%% file: index.tex
\documentclass[11pt]{article} 
\usepackage[utf8]{inputenc} 
\usepackage[a4paper, margin=1in]{geometry} 
\usepackage{titlesec} 
\titleformat{\paragraph}[hang]{\normalfont\bfseries\itshape}{}{0pt}{}
\titlespacing*{\paragraph}{0pt}{2.25ex plus 1ex minus .2ex}{1ex plus .2ex}
\usepackage{graphicx} 
\usepackage{caption} 
\usepackage{fancyhdr} 
\usepackage{amsmath, amssymb} 
\usepackage{booktabs} 
\usepackage{longtable} 
\usepackage{float} 
\usepackage{mathptmx} 

\usepackage{array} 
\usepackage[most]{tcolorbox} 
\usepackage{listings} 
\lstdefinelanguage{Rust}{
  keywords={fn,let,mut,as,impl,for,in,if,else,match,struct,enum,pub,use,self,Vec,f64,usize,return},
  morekeywords={assert,assert_eq,expect,map,collect,zip,iter},
  morecomment=[l]{//},
  morecomment=[s]{/*}{*/},
  morestring=[b]",
  sensitive=true,
}
\usepackage{tikz}
\usetikzlibrary{shapes.symbols,shapes.geometric,shadows,arrows.meta,
positioning,fit,backgrounds,calc}
\tikzset{>={Latex[width=1.5mm,length=2mm]}}
\tikzset{
  wfbox/.style   = {draw, rounded corners, align=center,
                    minimum width=3cm, minimum height=1cm, inner sep=4pt},
  wflabel/.style = {draw=none, align=center, font=\bfseries},
  wfflow/.style  = {->, thick},
  wfboth/.style  = {<->, thick},
  wfaux/.style   = {<->, thick, dashed},
  wfgap/.style   = {node distance=1.4cm},
}
\usepackage{pgfplots}
\pgfplotsset{compat=1.18}
\usepackage{pgfplotstable}
\usepackage{csvsimple-l3}

\usepackage{siunitx}

\usepackage{pgfplots}
\pgfplotsset{compat=1.18}
\usepackage{filecontents}

\newcolumntype{x}[1]{>{\centering\let\newline\\\arraybackslash\hspace{0pt}}p{#1}}

\usepackage[backend=bibtex,
style=ieee,
sorting=none,
isbn=true,
doi=true,
url=true]{biblatex}
\renewcommand{\thetable}{\Roman{table}} 

\usepackage[hidelinks]{hyperref} 

\AtBeginDocument{}

\makeatletter
\renewcommand{\@maketitle}{%
\begin{center}
    {\fontsize{14}{16}\selectfont \textbf{\@title} \par}
    \vskip 1em
    {\@author}
    \vskip 1em
    {\@date}
\end{center}%
\setcounter{footnote}{0}%
}
\makeatother

\title{Research Notes on Agentic Porting, Construction and Initial
Verification and Validation of Libraries within the
Open Source Unified TRAnsient Multi-Phase Advanced Reactor
simulation Kit (Outram Park) Part I: A Software Engineering Log for 
Thermal Hydraulics}
\author{\textbf{Theodore Kay Chen Ong, Sicong Xiao} \\ 
Singapore Nuclear Research and Safety Institute (SNRSI) \\ 
16 Prince George's Pk, Singapore 118415 \\ 
snrokct@nus.edu.sg \\ \vskip 2em
}
\date{}

\begin{document}


\begin{flushleft}
\textbf{ABSTRACT}
\end{flushleft}
\noindent 

Agentic porting of multiple open-source libraries into
Rust, with human in the loop, has been performed 
for construction of modules within the Open-source Unified TRAnsient 
Multi-Phase Advanced Reactor simulation Kit (Outram Park). With this 
new methodology, verification and validation with human expertise, 
rather than code generation has become the bottleneck 
in developing reliable simulation codes. In this research log, we present the 
porting of OpenFOAM libraries into the Outram-Foam Rust libraries, their 
preliminary verification and validation (V\&V) efforts, and their subsequent 
use in the development of open-source two-phase homogeneous-equilibrium (HEM)
choked-flow solvers for the Thermo-hydraulic AI Multi-Phase INtegrated 
Emulator System (TAMPINES) libraries within Outram Park such as 
tampines-steam-tables. V\&V efforts of Outram-Foam show that the cavity and
Sod shock tube cases agree with literature values. For the lid-driven cavity
at $Re=100$, a three-mesh grid-convergence study ($20\times20$, $40\times40$,
$80\times80$ at a constant refinement ratio of two) gives an observed order of
accuracy of 1.12, consistent with the first-order upwind convection the solver
uses, and Richardson extrapolation of that sequence recovers the benchmark
centerline velocity of Ghia et al.\ to within 0.64\%. The port therefore
converges at its formal rate and to the published reference value, rather than
merely agreeing to within a stated tolerance on one mesh. Moreover, the
preliminary development of tampines-steam-tables shows good agreement with
Moody's HEM charts. Thereafter, the 1D HEM 
solver, developed agentically, is presented and
preliminarily validated against the Edwards blowdown case. Productivity 
increases were observed with the use of Claude Code, but domain expertise
supplied by human experts remains critically required to ensure
the generated code can solve the problem effectively. Further work remains to be done in V\&V, but the agentic coding methodology
described here demonstrates considerable potential to speed up production and
development of open-source libraries such as Outram Park. These are presented
as research notes rather than as a finished study: the results are interim, the
development history is recorded as it happened, and the contribution is
principally one of software engineering applied to computational reactor
physics rather than one of new physics.

\vspace{1em}
\begin{flushright}
\textbf{KEYWORDS} \par
\vspace{1pt}
Open Source, TAMPINES, TUAS, OUTRAM PARK, Agentic Coding, LLM, AI
\end{flushright}

\maketitle{}
\section{INTRODUCTION}

Simulation of nuclear reactors is important for both research and education.
In Singapore's context, where Prime Minister Lawrence Wong announced 
that Singapore would embark on the IAEA Integrated Nuclear Infrastructure
Review (INIR) \cite{ema_singapore_inir_2026}, 
this is especially relevant to the nuclear safety and human 
resource development aspects of the INIR. Reliable reactor simulations are also
required for regulators to independently assess the safety claims of vendors. 
However, simulation of 
such reactors in a comprehensive manner often requires a suite of tools 
ranging from materials, to neutronics and thermal hydraulics. Such codes 
are also not readily available due to export control, thus reducing the 
reproducibility of research done using export controlled and proprietary 
codes \cite{ong2024digital}. Such barriers have historically made it hard
for these export controlled codes to be used in the arena of human resource 
development, where export controlled and proprietary codes are off limits 
to most students who may want to learn about nuclear power and nuclear safety.
Moreover, such data is difficult to visualise 
and often without a freely available integrated graphical user interface (GUI). 
Hence, its use was largely confined to technically competent audiences. 

This lack of open-source capability was apparent to the author
even before Singapore embarked on the INIR, and the incompressible open-source system
code TUAS \cite{ong2024tuasgithubrepo}, was developed in Rust along with its GUIs also written in Rust,
which could compile natively across Windows, Mac and Linux systems 
\cite{ong2023masters,ong2024digital,ong2024tuas}. TUAS was also used to build the
CIET Educational Simulator which was validated against forced and natural-circulation 
flows in CIET \cite{ong2024tuas,ong2025nureth}. Figure~\ref{fig:ciet_educational_simulator}
details the GUI of the CIET Educational Simulator, which shows its
capability to display the temperature distribution during
coupled natural-circulation, thus enabling the user to intuitively 
grasp the temperature profile one should roughly observe during natural-circulation:

\begin{figure}[H]
    \centering
	\includegraphics[width=0.95\textwidth]{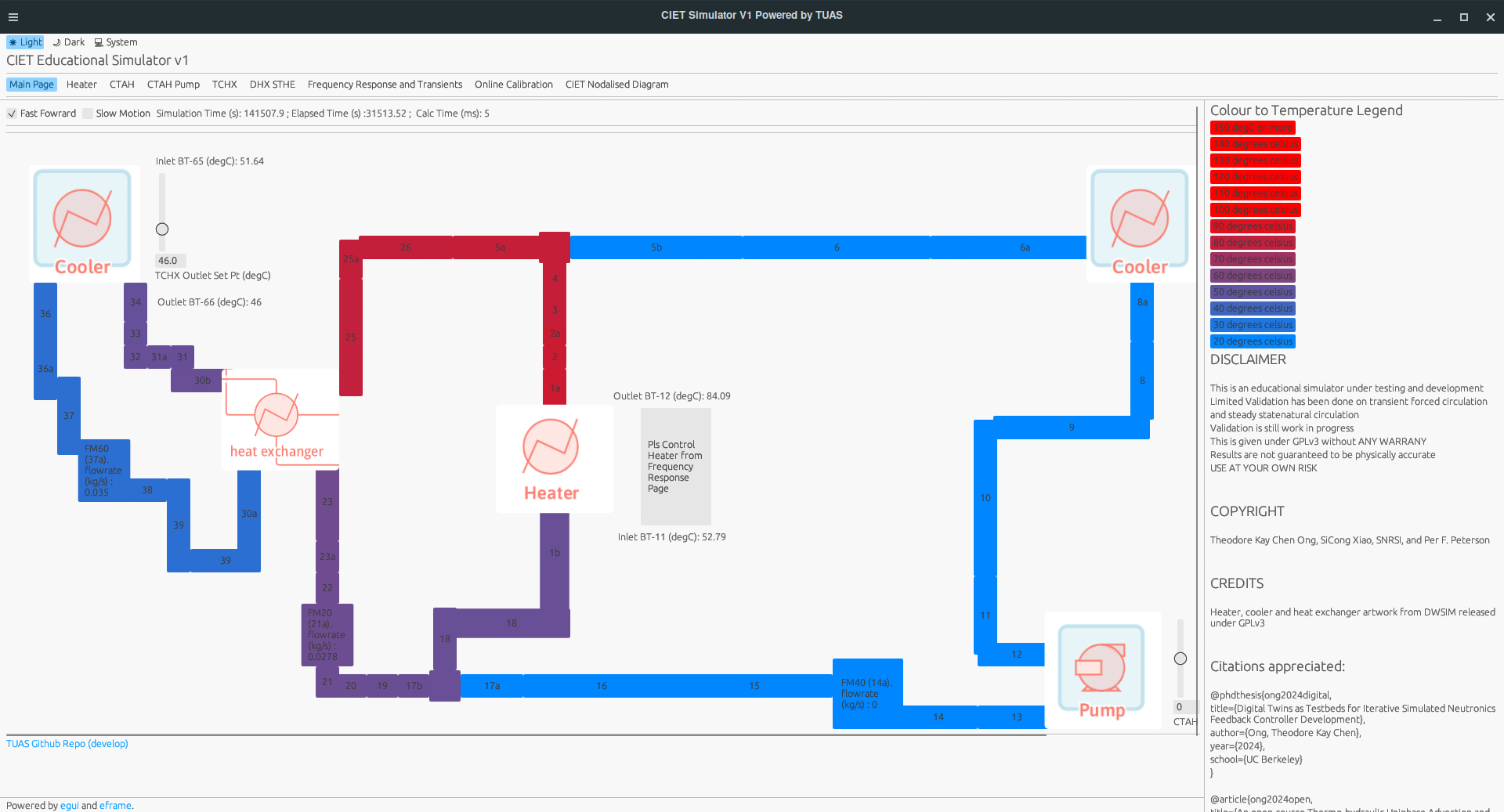}
    \caption{\textbf{CIET Educational Simulator}}
    \label{fig:ciet_educational_simulator}
\end{figure}

Nevertheless, TUAS was limited in its capabilities as it used the Boussinesq 
approximation and could only simulate single-phase, largely incompressible flow.
Density changes were not considered except where natural-circulation was 
computed. To serve the purposes of the INIR for Singapore, a larger 
suite of simulation codes would be required. The author envisioned this 
suite as the Open-source Unified TRAnsient Multi-Phase Advanced Reactor 
simulation Kit (OUTRAM PARK), named after the Outram Park MRT station 
in Singapore much like TUAS was named after the TUAS industrial 
region in Singapore. Outram Park would contain codes such as TUAS, in addition
to a neutronics suite, thermal-hydraulics simulation capabilities and
fission-product transport. Such a scope for a simulator suite would be vast
and cost many man-years to develop. Nevertheless, Outram Park was envisioned 
as a long term goal which could provide Singapore with tools required to 
serve its purposes, and as of July 2026 
it became more relevant because of the INIR announcement.

Such capabilities were traditionally fulfilled by a fragmented open-source 
ecosystem, with open-source frameworks and codes being written in various 
languages such as Python, C++, C and Fortran with various compilation configurations.
This made the multiphysics coupling required for reactor simulations difficult.
Nevertheless, the open-source ecosystem was still rich, and one could 
in theory, combine them to produce a coupled multiphysics simulation. 
These tools include 
Finite Volume codes such as OpenFOAM \cite{jasak2009openfoam} and GeN-Foam 
\cite{fiorina2015gen} helping fill thermal hydraulics, OpenMC \cite{romano2015openmc}
and NJOY \cite{njoy2016} fulfilling the part of nuclear data and Monte 
Carlo neutronics. System codes and simulators such as OpenModelica 
\cite{fritzson2020openmodelica} and DWSIM \cite{tangsriwong2020modeling}
were also used by researchers in evaluating system level phenomena. 
MOOSE is also a notable contribution with its open-source Finite 
Element Method \cite{gaston2009moose}, and this was used to construct 
Cardinal \cite{novak2023coupled} which coupled the spectral element 
solvers with OpenMC. However, certain codes built on MOOSE such as BISON 
and SAM \cite{zou2019sam} 
were closed source and not easily available to the general public, thus the 
results they produced were not easily reproducible by students and researchers 
without the clearance for such codes.

Within this open-source ecosystem,
the main challenge was that integration was cumbersome. With each 
C++, C and Fortran code requiring their own set of dependencies and build 
version, compiling them to work together and integrate in order to run 
a nuclear simulation would be a challenge. OpenFOAM has wmake while
OpenMC and MOOSE use cmake. Integrating OpenFOAM and OpenMC directly would 
therefore be a challenge. Even within MOOSE, an integrated system including a 
Python script of 20,000 lines long was created just to handle dependencies,
documentation and the ability to run the same code across various compilers 
and operating systems \cite{slaughter2021code}. Even MOOSE currently only 
runs on Linux or macOS, and virtual machines (including WSL) are required to set it
up.

The ideal of Outram Park would be to have an open-source 
Rust alternative of MOOSE where Cargo 
would be used as the single compiler across multiple platforms including, 
Windows, MacOS and Linux (including Android). These would enable Outram 
Park to build user friendly, performant and reasonably accurate graphical 
applications which run on all these platforms as tools for both education 
and research. Such an effort, a reinvention 
of the wheel you might say, would be impractical, monumental and 
expensive to do for a typical researcher in Singapore.
However, with the advent of artificial intelligence and agentic coding,
this vision of Outram Park is now realisable within a reasonable timeframe.

Agentic coding would allow the Singapore Nuclear Research and Safety 
Institute (SNRSI) to substantially improve its coding output and verification and
Validation (V\&V) efforts. This aligns closely with the Singapore government's 
AI agenda \cite{wong_singapore_budget_2026_ai}, where Prime Minister Lawrence
Wong highlighted the use of AI in the national budget. As part of this Agenda,
there was a major push for the Infocomm Media Development Authority to
train 40,000 tech professionals to use automated AI coding systems (agentic
coding) by 2029 \cite{imda_aixtech_2026}. In alignment with this agenda,
agentic coding was used to accelerate the development of Outram Park and
produce deliverables previously not possible. Of course, agentic coding 
is strictly limited to open-source and non-export controlled code, where 
Outram Park lies. 

Readers proficient in nuclear engineering may express a healthy scepticism
of the use of large language models, and this is warranted given the 
hallucinations risk of AI \cite{huang2025survey}. In nuclear engineering,
where quality control and nuclear quality assurance standards (eg. NQA-1) 
are of utmost importance, use of LLMs carry risks that 
must be mitigated \cite{sun2026quality} if they are ever used in nuclear 
engineering fields. Despite these risks, the benefits of AI and LLMs are 
apparent and such that the Nuclear Regulatory Commission has started considering 
its use \cite{nrc2024ai}. Moreover, in established nuclear engineering coding 
frameworks such as MOOSE, LLMs have also seen increased use such as in the 
MOOSE agent \cite{zhang2025mooseagent}. This is a testament to the usefulness 
of AI models in nuclear engineering. In fact, as of 15 Sep 2026, OpenAI has also 
claimed the breakthrough solutions in the Navier--Stokes 
equations \cite{cao2026distribution}, albeit with a fair amount of controversy 
generated in Sep 2026 \cite{zhao2026ai}. This controversy cannot be glossed 
over as it mentions that OpenAI could not rule out that de-identified 
product-usage data contributed to model improvement \cite{zhao2026ai}. This 
has given rise to concerns " among mathematicians around how AI 
models learn from interacting with their users — and whether the models, 
together with the developers and researchers using them, are 
giving due credit to previous work." \cite{nature2026attribution}. This 
is a textbook case testifying to both the rapidly evolving capabilities of AI,
as well as risks pertaining to data provenance, privacy, attribution, 
and academic integrity.

The question now, is what to do with these 2026 software engineering 
practices and LLM tools at our disposal given 
its risks and benefits. The answer is perhaps not to disregard 
use of AI completely, but use it in a careful manner where AI-assisted 
scientific development must include a framework for structured verification 
and validation (V\&V), traceability and transparency, which is especially 
important in the NQA-1 context \cite{bhave2026bridging}. Given these 
developments, and the overarching push to use AI 
models in Singapore and the tech industry, it 
is therefore important to investigate the use of AI in software 
engineering for nuclear engineering codes and perhaps attempt to use 
some of these frameworks for quality control. 

This research log details the agentic library porting and construction of various
libraries in Rust, as well as their initial V\&V efforts for the Outram Park
backend repository \cite{ong2026outramparkbackend}. We also examine the
productivity increase from using AI agents. For this research log, we confine
ourselves to thermal-hydraulics; nuclear data and neutronics are left to a
companion paper. Readers should note that this is not meant to be a full blown 
thermal-hydraulics paper, its bent towards software engineering and quality 
control is deliberately logged in hopes of it being useful to other software 
developers who may consider using agentic LLMs.

\section{METHODS}

\subsection{Origin of Discovery}
The author began by purchasing his own subscription of Claude Code, initially
for personal use and paid for out of his own pocket, but with experimentation,
he found that Claude Code (both
via the Opus and Sonnet) was able to translate codes
from C to Rust with exceptional ease, provided there were enough tokens.
Specifically, the author is an avid guitarist and used lingot \cite{cereijo_lingot_c_2020}
for tuning his guitar. Nevertheless, lingot development stalled, and it could 
not compile on Windows natively. Hence, there was motivation to translate 
it to Rust. This would involve translation of Fast Fourier Transforms, signal 
processing and front end GUI to Rust. The author was able to do so using 
Claude Code within one weekend \cite{ong_lingot_tuner_rust_2026,
ong_lingot_video_series_2026} and test the tuning GUI app on real guitars.
Given this exceptional capability of Claude 
Code, even Sonnet, the author then had the idea to apply this methodology 
to Outram Park. If agentic porting of signal processing, Fast Fourier 
transforms and GUI to Rust could be done in a single weekend, this 
could then imply that the Rust translation required for Outram Park was actually 
possible. Or at least more possible as compared to the daunting 
task of translation by hand. 
This hypothesis had to be tested, and the results were shown in 
this research log. To check if such practices were documented in literature, the author 
did literature searches and found similar efforts being done to translate legacy 
Fortran to C++ \cite{gupta2025legacy}. Thus the practice of agentic ports 
for legacy codes was performed in peer reviewed literature, and this gave the author
additional confidence to use this in Outram Park.
Once the utility of the tool to Outram Park development had been demonstrated,
the SNRSI CEO permitted the author to claim the Claude Code subscription as a
work expense. The subscription costs incurred thereafter were therefore borne
by SNRSI rather than by the author personally, though the initial exploratory
period described above was self-funded.

\subsection{Outram-Foam, Rust Port of OpenFOAM}

During the initial exploratory phase,
the next target was the rather refractory and unwieldy OpenFOAM library. This 
is because OpenFOAM could only be compiled on Linux systems with the use 
of a customised bashrc file and wmake script. Such manual compilation and 
editing of OpenFOAM was cumbersome as compared to Rust, where the build 
systems were modernised using a simple "cargo build --release" command.
and execution speeds were similar to C and Fortran
\cite{kailasa2022mostly}. Should the libraries be portable, then it would 
avail the entire open-source solver library to Rust for use in Outram Park
albeit under the GNU GPLv3 license. This is, however, an advantage for 
education and research as the strongly copyleft GPLv3 license forces code 
derivatives of OpenFOAM to be made open-source for students and researchers alike.
In the NQA-1 context, keeping source code open-source facilitates traceability,
transparency, reproducibility and auditability \cite{mangul2019recommendations,
barba2022open,spinellis2009evaluating}. In fact, there is a philosophy 
that exists where "given enough eyeballs, all bugs are shallow". This is 
known as Linus's law. It may not mean that open-source code is completely 
secure, but in some cases, having more independent reviewers scrutinize 
code may lead to more bugs being discovered and patched \cite{meneely2009secure}.
This has helped in the context 
of Linux kernel development \cite{linuxkernel2026cve,linuxkernel2026security}, 
where security vulnerabilities and other such defects are subject to public 
scrutiny, review and correction.

Porting OpenFOAM libraries to Rust was undertaken because the OpenFOAM library 
has many useful matrix operations and 
finite volume method libraries which, could in theory, be used for 
two phase flows, but they were impractical to humanly import over and 
translate to Rust. 
Claude Code was then used to port and debug the open-source OpenFOAM libraries
to Rust, including the many matrix solvers, discretisation methods, 
fvMatrix and source files. It was proven for use in the coding community 
\cite{liu2026dive} and proved to be extremely useful in porting the libraries
over. In doing so, of course, one has had to do unit testing to verify the 
code that was developed. For this, unit tests were imported 
over from OpenFOAM as well to verify that the matrix solvers were 
ported over correctly. 
In following naming convention of solvers based on OpenFOAM such as GeN-Foam 
\cite{fiorina2015gen} and HRMFoam \cite{schmidt2010hrmfoam,schmidt_hrmfoam_github},
the Outram Park Rust fork of OpenFOAM is henceforth named Outram-Foam.

\subsection{The role of AI in Tampines}
The goal, however, was not simply to port CFD software to Rust. 
At the point of porting, the author had conceived of a multi-phase version 
of TUAS, which was the Thermo-hydraulic Artificial-intelligence Multi 
Phase INtegrated Emulator System (TAMPINES), named after the Tampines 
MRT station on the east side of Singapore. TAMPINES was to simulate multi-phase
flows using various models which firstly included the Homogeneous 
Equilibrium Model (HEM), the simplest of multi-phase thermal hydraulics 
modelling for water and steam. From thereon, there was intent to move on 
to drift flux modelling, and eventually a six equation model if there 
was capacity. OpenFOAM solvers such as driftFluxFoam, rhoCentralFoam 
and rhoPimpleFoam would have been ideal for such cases. While impractical 
to translate to Rust quickly for a 1D solver, agentic porting now made 
this possible. 

Even with appropriate and accurate ports, this may not, in itself, 
be sufficient.
The goal was to use TAMPINES for real-time simulation such 
as in the case of the CIET Educational Simulator for reactor modelling and 
secondary loop modelling for various reactors. Knowing that the equations 
may not have been easy enough to solve in real-time, surrogate models and AI 
would have been part of the solution. Hence, the name in TAMPINES includes 
A for AI. Serendipitously, it would now also imply that TAMPINES was made 
with the help of agentic porting, an AI centric methodology.

\subsection{Tampines Steam Tables and Critical Flow}

Now, every OpenFOAM solver requires some form of thermophysical properties.
Even for the HEM model, multiphase compressible flow phenomena such as 
choked flow and critical heat flux models also had to be ported. To break 
this problem down, thermodynamics was to be done first, and heat transfer 
done later. Heat Transfer and Critical Heat Flux 
is the subject of future work, not in this research log.

As for the first step, thermodynamics would have to be implemented first. The 
author initially hand coded tampines-steam-tables from Wagner et al. 
\cite{wagner2008international} for the steam table thermodynamics. This initially 
took a number of months. However, for any real two-phase thermodynamics package 
to be complete, multi-phase choked flow would have to be tackled. However, 
multi-phase choked flow algorithms and formulae were harder to find than for single 
phase. Such algorithms 
and problems were absent in standard 
thermodynamics textbooks such as Cengel \cite{cengel2002thermodynamics}
which only included single phase choked flow. To the best of the author's knowledge,
no open-source library nuclear related numerical model for critical choked 
flow for LOCA exists for porting. Even ThermoSysPro \cite{ElHefni2019}
within the OpenModelica ecosystem and DWSIM were not found (with the help of Agentic 
AI scouring its source code) to have choked flow correlations programmed into 
it. Such flow phenomena are extremely important for modelling loss of coolant 
accidents and turbines. This reflects a lack of multi-phase critical choked flow code within 
the open-source community. For this, tampines-steam-tables would start 
to fill a gap by providing the HEM choked flow algorithms 
and contribute solvers to open-source code using agentic 
coding methodology.
For this, initial attempts were performed using human 
debugging, as even for a HEM model, there was a discontinuity where the speed 
of sound dropped drastically between subcooled liquid and bubble point.
However, this debugging process was then accelerated and automated 
using Claude Opus.

\subsection{Human in the Loop Agentic Porting methodology}

\paragraph{Reduction of Bad Engineering and Hallucinated Debugging Cycles}

Despite the benefits of agentic coding and porting,
agentic porting of nuclear simulation software is not a fully automated task.
This is because software has to be well organised and structured to ensure 
re-use. Moreover, AI agents, even the Opus models, were observed to apply
the wrong method of automated debugging and required explicit human intervention 
to prevent waste of time and AI tokens. 

One example is where a tampines-steam-tables function obtaining enthalpy 
from (p,T) for single phase water or steam flows was called for a 
steam generator simulation for which multi-phase flow 
was obviously expected. The function 
coded was specifically called "h\_tp\_eqm\_single\_phase" which signalled 
its use in single phase regions and would panic on two phase regions. Opus 
kept chasing these crashes incorrectly, adding unnecessary guards, and exhausting 
the token limit. For the author who hand coded these functions, the mistake 
was obvious, and he injected a prompt to ensure that Opus used (p,h) flashes 
rather than (p,T) flashes because the former already contained quality  
data. With this correction, Opus could then converge on the correct solution 
which prevented crashing.

\paragraph{Structuring Workflow with Claude Code and other AI tools}

\subparagraph{Outram-Foam Translation}
Early experiments were conducted using the National University of
Singapore's institutional AI platform, NUS AI Know. While suitable for
conventional conversational assistance, it proved insufficient for the
requirements of large-scale software porting. At the time of evaluation,
the system accepted only limited file uploads and could not recursively
inspect, modify, execute, and iterate over a local software workspace.
Consequently, repository-scale tasks required extensive manual context
management, resulting in high interaction overhead.

In contrast, agentic coding systems such as Anthropic Claude Code were
capable of operating across an entire software repository, autonomously
reading files, proposing modifications, executing tests, observing
failures, and iterating on solutions. The potential of this workflow
became apparent during an early experimental project involving the
translation of a guitar-tuner application into Rust. This exercise
demonstrated that repository-scale translation and iterative debugging
could be performed with substantially less manual intervention than was
possible using conversational interfaces alone.

Rust was selected as the target implementation language for several
practical reasons. The scientific software ecosystems evaluated during
this work, including OpenFOAM, OpenMC, NJOY, MOOSE, and PFLOTRAN,
employ heterogeneous build systems, compiler toolchains, dependency
management strategies, and language combinations. These include
\texttt{wmake}, \texttt{make}, \texttt{cmake}, C++, C, and Fortran
toolchains. Reproducing legacy software environments frequently required
substantial effort to resolve compiler and dependency incompatibilities.
In contrast, Rust provides a unified package-management and build system
through Cargo, reducing the operational complexity associated with
multi-language scientific software development \cite{veytsman2024rewrite}.
Historically, the scientific-computing capabilities in Rust were much 
less developed than in C++ or Fortran, but with today's agentic coding 
capabilities, this gap can be closed much more quickly than before.

Experience accumulated during the project also revealed significant
differences between model capabilities. Claude Sonnet proved effective
for localized translation tasks, individual modules, and relatively
well-bounded implementation requests. However, larger architectural
tasks, complex physics models, and new subsystem designs frequently
required the more capable Claude Opus models. Claude Haiku was generally
reserved for smaller utility tasks and isolated functions. In practice,
larger objectives were typically scaffolded using Opus before being
decomposed into smaller subtasks that could subsequently be delegated to
Sonnet.

A key methodological principle was hierarchical decomposition. OpenFOAM,
for example, consists of multiple layers of abstraction, including
fundamental matrix structures, \texttt{fvMatrix}, \texttt{fvMesh},
finite-volume operators, turbulence models, thermophysical models, and
application-level solvers. Rather than attempting direct translation of
entire applications, development proceeded from lower-level numerical
components toward higher-level physics modules. Matrix solvers, array
libraries, ordinary differential equation integrators, algebraic
equation solvers, and polynomial root-finding utilities were translated
and verified first before more complex finite-volume infrastructure was
constructed.

Every translated component was accompanied by an explicit verification
step. Following implementation, AI-generated code was subjected to a
lightweight verification procedure referred to within Claude Code as a
``smoke test.'' For example, when validating a linear-system solver, a
matrix $\mathbf{A}$ and solution vector $\mathbf{x}$ were chosen
\emph{a priori} and the right-hand-side vector computed according to

\begin{equation}
\mathbf{b} = \mathbf{A}\mathbf{x}.
\end{equation}

The translated solver was then required to recover $\mathbf{x}$ from
$\mathbf{A}$ and $\mathbf{b}$. Successful reconstruction within
acceptable numerical tolerances was treated as evidence that the
translation was functionally correct and free from obvious regressions
\cite{higham2002accuracy}. When a translated implementation failed its
smoke test, the agent was instructed to iteratively diagnose and correct
the defect until the verification case passed.

Initial verification often employed small synthetic matrices generated
during development, including the classical $5\times5$ Hilbert matrix.
However, such examples were regarded only as preliminary checks. Final
verification employed larger benchmark problems drawn from established
scientific-computing repositories. In particular, the Harwell--Boeing
matrix \texttt{nnc261}, originating from nuclear-engineering
applications, was used as a representative sparse-matrix benchmark
\cite{duff1989sparse,nnc261_matrixmarket}. Forward and backward solution tests were
performed using the same methodology described above. Successful
recovery of the original solution vector on both synthetic and
application-derived benchmark problems was considered a prerequisite for
integrating translated numerical components into higher-level software
modules.

Within the Outram Park architecture, the OpenFOAM-derived functionality
was decomposed into multiple Rust crates. Fundamental numerical
capabilities were implemented within the
\texttt{outram-foam-basic-lib} crate, which contained matrix structures,
linear solvers, array utilities, and related numerical infrastructure.
Turbulence modelling functionality was isolated within the
\texttt{outram-foam-turbulence-lib} crate, while application-level solver
logic was implemented within the
\texttt{outram-foam-appbuilder-lib} crate. The latter contained translated
solver implementations, including \texttt{rhoPimpleFoam} and other
application-specific finite-volume solvers.

This decomposition strategy reduced the complexity of individual
translation tasks and enabled each layer of functionality to be verified
before being used as a dependency by higher-level modules. As a result,
verification effort could be concentrated on well-defined software
components prior to full solver integration.

\subparagraph{Tampines Steam Tables}

Development of the \texttt{tampines-steam-tables} library began with the
implementation of the IAPWS-IF97 industrial formulation for water and
steam properties \cite{wagner2008international}. The Gibbs free-energy
formulations for pressure--temperature calculations, together with the
pressure--enthalpy (\textit{ph}), pressure--entropy (\textit{ps}), and
enthalpy--entropy (\textit{hs}) formulations, were implemented directly
from the published standard. Much of the implementation work was
performed manually using text-processing workflows within Neovim. Large
tabulated datasets from the standard were converted into intermediate
CSV-format representations before being translated into Rust data
structures. Although repetitive, this approach enabled the published
tables to be incorporated with minimal transcription logic and preserved
traceability to the original source document.

Verification proceeded in several stages. Initially, the implemented
equations were checked against the numerical examples provided in the
IAPWS-IF97 reference documentation \cite{wagner2008international}.
Subsequently, steam-table values were transcribed and used to validate
both forward and inverse thermodynamic-property calculations. Particular
attention was devoted to validating the \textit{hs} formulation, whose
domain of applicability is more restricted than other regions of the
standard. Saturation, superheated, and subcooled steam tables were
individually transcribed into Rust test fixtures and verified through
automated unit-test procedures. Verification was performed across entire
tables rather than isolated points wherever practical. Most of this work
predated the adoption of Claude Code and was completed using
conventional manual development workflows.

The implementation of the homogeneous-equilibrium-model (HEM)
two-phase critical-flow capability presented a substantially greater
challenge. Unlike the steam-table implementation, suitable open-source
reference implementations were not readily available. Existing
open-source process-simulation frameworks investigated during the
project, including ThermoSysPro and other Modelica-based libraries, did
not provide equivalent functionality suitable for direct reuse.
Consequently, the implementation required both literature review and
independent algorithm development.

The original motivation for the work arose from turbine-simulation
requirements within fluoride-salt-cooled high-temperature reactor
(FHR) studies. During this investigation it became apparent that
critical-flow phenomena play an important role in steam-turbine systems,
particularly under high-pressure conditions approaching sonic or
supersonic flow. More importantly, two-phase critical flow is also a
fundamental phenomenon in nuclear-engineering safety analysis,
particularly during loss-of-coolant accidents (LOCAs). As a result, the
development of an open-source critical-flow capability was considered an
important prerequisite for future thermal-hydraulic modelling efforts.

The initial development effort drew upon classical compressible-flow
treatment from \cite{cengel2002thermodynamics}. However, these methods
primarily address single-phase fluids and were insufficient for direct
application to two-phase steam-water systems. Validation was therefore
performed against established critical-flow literature, particularly the
critical-flow correlations of Moody \cite{moody1975maximum} and the
review of two-phase critical-flow models by Saha
\cite{saha1978review}. Experimental and published benchmark data were
digitized using GraphReader \cite{graphreader} to enable quantitative
comparison against the implementation.

An important practical difficulty arose from the form in which benchmark
data were reported. Published studies generally provide stagnation
properties and measured critical mass fluxes. Direct reproduction of
these results required iterative determination of the critical pressure
and critical-flow conditions, creating a numerically challenging
optimization problem. An alternative validation strategy was therefore
adopted. Critical mass-flux values and critical thermodynamic states
were digitized directly from published figures. Using the steam-table
infrastructure already developed within TAMPINES, the corresponding
thermodynamic properties at the choke point were reconstructed. Given
critical density and mass flux, flow velocity could be calculated
directly, allowing stagnation conditions to be recovered through
appropriate thermodynamic flash calculations. This substantially reduced
the complexity of the validation procedure by avoiding certain classes
of nested iterative calculations.

The forward problem remained more challenging. Starting from stagnation
conditions, isentropic pressure reduction was performed through a
pressure--entropy flash calculation. At each candidate pressure, local
thermodynamic properties, density, and sonic velocity were determined,
allowing a predicted mass flux to be calculated. The critical mass flux
was then identified through numerical maximization. AI-assisted
development tools proposed the use of golden-section search
\cite{price2012golden} to accelerate this optimization process and
reduce computational cost compared with exhaustive scanning.

During development, significant numerical difficulties were encountered
near saturation boundaries. In particular, the sonic velocity exhibits
strong discontinuities across the bubble point, transitioning from
liquid-phase values of the order of kilometres per second to
substantially lower values in two-phase regions. These discontinuities
degraded the behaviour of the optimization algorithm and complicated
identification of the true maximum mass-flux condition. Extensive
debugging was therefore required. Diagnostic quantities, including
pressure, density, sonic velocity, and predicted mass flux, were
systematically recorded and analyzed. AI-assisted tooling proved
particularly useful during this phase because it automated generation of
diagnostic output and accelerated investigation of numerical anomalies.

Several alternative optimization strategies were subsequently explored.
Separate optimization procedures were tested within and outside the
two-phase dome, and the resulting predictions were compared across a
wide range of conditions. Development records and Git history indicate
that much of this iterative refinement occurred during the latter half
of June 2026. Repeated testing and validation ultimately yielded a
stable implementation capable of reproducing the benchmark trends
reported in the literature. Quantitative validation results are
presented in a later section.

Integration studies followed, coupling the TAMPINES steam-table and
critical-flow infrastructure with the Rust implementation of
\texttt{rhoPimpleFoam} and forming the basis of the emerging
\texttt{Outram-Foam} validation effort. The resulting integration tests were
an important step toward applying the framework to nuclear-engineering
transient analysis.

The principal contribution of the \texttt{tampines-steam-tables}
project is therefore not merely the implementation of steam-property
calculations, but the creation of an open-source HEM-based critical-flow
capability suitable for nuclear-engineering applications. Although the
HEM approximation is known to have limitations relative to more advanced
non-equilibrium models, it provides a foundation upon which more
sophisticated delayed-equilibrium, homogeneous-relaxation, and
two-fluid approaches may subsequently be constructed.

\subparagraph{Refactoring into a Rust Workspace and Monorepo}

As development progressed, it became desirable to consolidate the
previously independent software components into a single Rust workspace
and monorepo. Earlier versions of the software stack suffered from
dependency-management challenges. For example, different projects relied
upon different versions of the \texttt{uom} (Units of Measurement)
library, requiring manual coordination whenever shared functionality was
introduced. Such version mismatches increased maintenance burden and
complicated cross-project integration.

AI-assisted development tools recommended migration to a Cargo workspace
architecture. Under this arrangement, multiple Rust crates share a
common dependency graph, build configuration, and version-management
strategy. As a result, all workspace members automatically use
consistent library versions, substantially reducing dependency conflicts
and simplifying maintenance.

The workspace architecture also provided an opportunity to eliminate
external numerical-library dependencies that had previously complicated
deployment. Earlier implementations relied upon OpenBLAS-based linear
algebra infrastructure, requiring platform-specific installation and
configuration procedures. This created practical difficulties on systems
such as Android devices and other environments where OpenBLAS was not
readily available. Following the introduction of OpenFOAM-derived sparse
and iterative solver infrastructure, small-to-medium-sized numerical
systems could instead be solved using pure-Rust implementations.

This capability enabled numerical infrastructure to be shared across
multiple Outram Park projects. The resulting solver implementations were
used within TUAS, the TAMPINES steam-table and critical-flow framework,
and the TEH-O-PRKE reactor-kinetics package. Although the latter is
outside the scope of the present thermal-hydraulics paper, it forms part
of the broader software ecosystem and is discussed in the companion
neutronics publication. The same numerical infrastructure also provides
the foundation for the \texttt{rhoPimpleFoam}-derived implementation
used in the Edwards--O'Brien blowdown integration studies.

The workspace architecture significantly improved software portability.
Because the resulting codebase relied primarily upon Rust-native
dependencies and Cargo-based build procedures, it could be compiled and
executed across a wide range of platforms without substantial
reconfiguration. Successful execution was demonstrated on desktop Linux
systems as well as Android-based devices.

AI-assisted development played a largely operational role during this
refactoring process. Tasks such as dependency alignment, version
migration, workspace configuration, compiler-error diagnosis, and crate
integration were effectively automated. Compared with the algorithmic
and physics-development challenges described elsewhere in this work,
workspace refactoring proved relatively straightforward. The principal
benefit of AI assistance was therefore a reduction in engineering effort
associated with software maintenance and ecosystem integration rather
than the development of new numerical methods.

\subsection{Use of AI to Track Provenance}

Generative AI was also employed as a provenance-tracing and project
archaeology tool. During the development of Outram Park, a substantial
amount of functionality was implemented over extended periods of time
across multiple repositories and software components. As development
progressed, it became increasingly difficult for the authors to recall
the precise sequence of implementation decisions, debugging efforts,
verification activities, and feature additions solely from memory.

AI-assisted repository analysis was therefore used to inspect Git commit
histories, summarize development timelines, identify relevant pull
requests and implementation milestones, and recover historical design
rationale recorded in commit messages and source-code comments. This
capability proved particularly useful when preparing the present
manuscript, as it enabled previously forgotten implementation details,
verification studies, and development decisions to be located rapidly and
cross-referenced against repository history.

Rather than serving as a source of technical content, the AI acted as an
interface to the project's historical record. Retrieved information was
subsequently verified by the authors against the underlying Git history,
source code, commit metadata, benchmark results, and documentation before
being incorporated into the manuscript. This workflow enabled specific
implementation details, verification activities, and development
milestones to be documented with greater accuracy than would have been
possible using author recollection alone.

The use of AI-assisted provenance tracing also reduced the cognitive
burden associated with navigating large software repositories. Instead of
manually inspecting extensive commit histories and source trees, the
authors were able to query implementation history at a higher semantic
level, allowing greater effort to be directed toward scientific
interpretation, verification activities, and manuscript preparation.

\subsection{Reduction of Cognitive Load in Prose Writing and Prompt 
Engineering}

A secondary use of generative AI within the project was the refinement of
informal technical notes into publication-ready academic prose. During
development, the primary author frequently recorded implementation
rationale, debugging observations, verification results, and design
decisions in abbreviated engineering prose rather than directly composing
journal-style text. This approach was adopted to reduce the cognitive
overhead associated with first-draft composition and to allow technical
ideas to be captured rapidly while attention remained focused on software
development and scientific reasoning.

For example, prose describing the translation and
verification of OpenFOAM numerical infrastructure was initially recorded
as informal Singlish (an English-based creole spoken in Singapore 
\cite{ningsih2023singlish}) engineering text (in \LaTeX{} format)
and subsequently transformed into structured
academic prose using AI-assisted language revision. 
The transformation was
restricted to language, organization, and readability. Technical content,
verification procedures, numerical results, interpretations, and
conclusions remained author-generated. The refined text was then reviewed
by the authors to ensure technical equivalence with the original notes.

This was the original text, which was then translated into proper journal
language:

\begin{tcolorbox}[breakable, enhanced, colback=blue!3, colframe=blue!40!black,
  fonttitle=\bfseries, title={Author's original note (verbatim LaTeX source 
	code, informal Singlish)}]
\begin{verbatim}
	
	OpenFOAM right, for example, got matrices, fvMatrix, fvmesh etc etc. 
The idea right is to break it down. First the basics, got matrix solvers,
arrays, ODEs, algebraic solver, quadratic and cubic equation solver. 
Quite basic lah. Ask Claude translate and test first. And make sure right,
every time translation comes, there should be some sanity check.  Claude 
calls it a smoke test lah. So for example, for solving Ax=b matrix, 
we can determine A and x, then calculate b. Then we try to back calculate 
x using the solver. if can hor, then okay lah. swee liao, solver verified.
\cite{higham2002accuracy}
If not, ask claude keep debugging until swee swee.

Ahh this one lah. nnc261 \cite{duff1989sparse}.

https://math.nist.gov/MatrixMarket/data/Harwell-Boeing/nucl/nnc261.html

Claude made smaller test like the hilbert 5x5 matrix. I mannaully do the 
nnc261. The forward backward test. Once that one done, ho seh liao.

\end{verbatim}
\end{tcolorbox}

Typing this less verbose form out and having it translated into proper academic 
text was valuable in reducing cognitive load for the author. This was because 
the author was more comfortable expressing ideas in Singlish, his colloquial
native tongue, than English.

This workflow was particularly valuable during large-scale agentic porting
activities, as well as prose writing, where implementation 
progress frequently outpaced the ability
to document design decisions and prose in formal publication style. 
By separating
technical idea capture from language refinement, cognitive resources could
be directed toward numerical verification, software architecture, and
scientific interpretation rather than grammatical editing and prose
construction. 
Such practices used and described in reputable 
literature \cite{lin2025supercharging} (Nature Biomedical engineering), 
were indeed valuable to the author's interaction with AI and the writing 
of prose from \LaTeX{} based Singlish prose to English. This technique 
should be useful for Singaporeans more accustomed to expressing ideas in 
colloquial language rather than English, but could also be useful for those 
writing ideas in any colloquial language and converting it to 
English suitable for a journal article.

The use of AI for language refinement is consistent with contemporary
publisher guidance permitting generative AI tools to assist with
organization, readability, and manuscript preparation, provided that
human authors retain responsibility for the scientific content, verify all
outputs, and disclose their use appropriately. 

Annals of Nuclear Energy states \cite{annals_nuclear_energy_guide}:

\begin{tcolorbox}[breakable, enhanced, colback=green!3, colframe=green!45!black,
  fonttitle=\bfseries, title={Annals of Nuclear Energy, Guide for Authors}]
Authors preparing a manuscript for this journal may use AI tools to support
them. However, these tools must never be used as a substitute for human
critical thinking, expertise and evaluation. AI tools may only be applied
with human oversight and control.

Please read Elsevier's detailed author policy on the use of AI tools, as
stated in Elsevier's generative AI policies for journals.
\end{tcolorbox}

Elsevier explicitly states \cite{elsevier2026genai}:

\begin{tcolorbox}[breakable, enhanced, colback=green!3, colframe=green!45!black,
  fonttitle=\bfseries, title={Elsevier, Generative AI Policies for Journals}]
Increasingly, these tools, including AI agents and deep research tools, are
helping researchers to synthesize complex literature, provide an overview of a
field or research question, identify research gaps, generate ideas and provide
tailored support for tasks such as content organization and improving language
and readability. Authors preparing a manuscript for an Elsevier journal can use
AI tools to support them in these tasks. However, these tools must never be used
as a substitute for human critical thinking, expertise and evaluation.

AI tools may be used to support the creation of data visualizations only when
the visual output is directly derived from underlying data.
\end{tcolorbox}

The authors reviewed all
AI-assisted revisions and retained full responsibility for the technical
accuracy and conclusions presented in the manuscript.

\subsection{Verification and Validation (V\&V)}

Verification and validation are particularly important for AI-assisted
software development. Large language models are known to generate
plausible but incorrect outputs, a phenomenon commonly referred to as
hallucination \cite{alansari2026large}. Consequently, code that compiles
successfully cannot automatically be assumed to be correct. The same
principle applies to conventionally developed scientific software:
implementation correctness and physical validity must be demonstrated
through systematic testing before software can be regarded as
trustworthy \cite{hook2009testing}. Within the Outram Park ecosystem,
all software components are regarded as experimental until they have
successfully passed an appropriate verification and validation
procedure.

The V\&V workflow employed throughout this work combined AI-assisted
automation with mandatory human review. Benchmark data were first
extracted from published figures using GraphReader
\cite{graphreader}. The resulting datasets were exported into CSV format
for subsequent use within the verification workflow.

Verification tests were then implemented in Rust and executed using:

\begin{verbatim}
cargo test --release
\end{verbatim}

The generated output was manually inspected by the authors to determine
whether the implemented functionality behaved as expected. When
anomalies were observed, additional diagnostics were added and the
tests were repeated until the behaviour of the implementation was
understood.

Verification outputs were written to CSV files within a Git-ignored
verification-and-validation directory. These files could then be
re-ingested into AI tools, including Claude Code, Microsoft Enterprise
Copilot, and NUS AI Know, for post-processing. Typical tasks included
automatic generation of Python matplotlib plotting scripts,
\LaTeX{} \texttt{pgfplotstable} tables, and visualization code.
The resulting figures and tables were reviewed by the authors before
being incorporated into the manuscript.

This workflow substantially reduced the effort required to transform
raw verification output into publication-quality figures and tables
while retaining human oversight of benchmark selection, numerical
verification, result interpretation, and scientific conclusions.

The workflow described above was intended primarily as a productivity
mechanism rather than as a complete substitute for conventional
verification practice. In its current form, the workflow provides a
substantial level of confidence in implementation correctness by
combining automated test generation, benchmark comparisons, numerical
inspection, and human review.

For applications requiring a higher degree of assurance, authors may
elect to independently derive and implement verification tests without
AI assistance. The resulting manually derived reference solutions can
then be compared against AI-generated implementations and against the
CSV-based validation workflow described above. Such independent
cross-checking provides an additional layer of verification by reducing
the possibility that errors in AI-generated code and AI-generated test
logic remain undetected.

Although this final verification step was not systematically applied to
all case studies reported in the present work, the workflow adopted here
was generally sufficient to identify the majority of implementation and
translation errors encountered during development. The authors therefore
regard the AI-assisted workflow as an effective first-pass verification
strategy, while recognising that independently derived validation tests
remain the preferred standard when the highest degree of confidence is
required.

\subsection{Use of Agents and Subagents in Claude Code Max}

This work made extensive use of the multi-agent capabilities available
within Claude Code Max \cite{anthropic2026pricing}. The larger token
allocation and extended usage limits permitted the simultaneous
execution of multiple agent and subagent workflows across different
repositories and development tasks.

A recurring pattern throughout the project was the decomposition of
large software-porting objectives into multiple parallel activities.
Rather than translating a single repository sequentially, separate
agents were assigned to independent numerical, thermophysical, and
software-engineering tasks. This enabled concurrent development across
multiple projects, including ports and reimplementations related to
OpenMC, NJOY, OpenFOAM-derived infrastructure, and associated Outram
Park components.

The multi-agent workflow proved particularly useful for repository-scale
translation tasks. Individual agents could be assigned responsibility
for understanding source-code structure, generating Rust
implementations, performing verification tests, resolving compiler
errors, and preparing documentation. Higher-level coordinating agents
were then used to review and integrate the outputs of these subordinate
agents into a unified codebase.

Practical experience revealed that model capabilities were not uniform.
Large architectural and repository-wide tasks generally benefited from
more capable models, while localized refactoring and implementation work
could often be delegated to smaller models. As a result, agentic
development frequently adopted a hierarchical structure, in which a
higher-level model generated implementation plans and task decomposition
strategies before dispatching narrower subtasks to subordinate agents.

The principal advantage of this approach was increased throughput.
Multiple repositories and software components could be developed,
reviewed, and validated concurrently rather than sequentially. In
practice, this significantly accelerated progress on projects spanning
multiple physics domains and software stacks.

However, increased development throughput also introduced new human
factors considerations. Although agent fleets reduced the effort
associated with implementation, they increased the monitoring burden on
the human developer. Simultaneous review of multiple streams of code
generation, verification output, debugging reports, and scientific
results introduced substantial cognitive demands. Consequently, the
limiting factor in development frequently became human review capacity
rather than code-generation speed.

The authors therefore regard multi-agent development as a powerful
productivity mechanism when combined with rigorous human oversight,
verification, and validation. While agent fleets can substantially
increase software-development throughput, scientific responsibility
remains with the human developer, who must review implementation
choices, validate numerical results, and assess the physical
correctness of generated code.

\section{Results}

\subsection{Speed Increase by Agentic Porting}

One of the most striking observations arising from this work was the
development velocity enabled by agentic software-development workflows.
The combination of repository-scale code understanding, automated
compiler-error resolution, dependency management, code generation,
testing, and iterative debugging substantially reduced the engineering
effort required to construct new scientific software components. Tasks
that would traditionally require extensive manual investigation and
software-engineering effort, particularly language-porting activities
and dependency management across heterogeneous code bases, could often
be completed significantly more rapidly through AI-assisted workflows.

Historically, software efforts of comparable breadth would often require
multiple developers working over extended periods of time. In contrast,
the Outram Park project demonstrated that a single researcher (or two),
supported by agentic development tools, could make meaningful progress
across multiple domains including computational fluid dynamics,
thermophysical property modelling, reactor kinetics, nuclear-data
processing, Monte Carlo transport, process-system simulation, and
digital-twin infrastructure within a relatively short period.

An additional observation was that development throughput increased
further following adoption of Claude Code Max in mid-July 2026. Larger
context windows, increased token availability, and multi-agent
execution capabilities enabled multiple repositories and software
components to be developed concurrently rather than sequentially.
Consequently, productivity became increasingly limited by human review,
verification, and scientific judgement rather than by code-generation
capacity itself.

The cumulative effect of these development activities is summarized in
Table~\ref{tab:outram_park_timeline}. Over approximately four weeks,
substantial progress was achieved across the Outram Park ecosystem,
ranging from finite-volume infrastructure and thermophysical property
libraries to nuclear-data processing, Monte Carlo transport, reactor
kinetics, and digital-twin foundations. Although only the
thermal-hydraulics-related developments are discussed in detail in the
present paper, the timeline provides context for the broader software
ecosystem that emerged during the same development period.

\begin{table}[H]
\centering
\caption{Development milestones of Outram Park, 19~June~--~16~July~2026.
Roughly four weeks of agentic-AI-assisted development (Claude Code)
delivered substantial productivity gains over conventional scientific
software engineering workflows. All commit hashes and full messages are
publicly available in the project repository under the develop branch.
Items marked \emph{in progress} are not yet validated; see the per-crate
\texttt{verification\_and\_validation/} records for validation status.}
\label{tab:outram_park_timeline}
\small
\begin{tabular}{@{}
  >{\raggedright\arraybackslash}p{0.11\linewidth}
  >{\raggedright\arraybackslash}p{0.45\linewidth}
  >{\raggedright\arraybackslash}p{0.37\linewidth}@{}}
\toprule
\textbf{Date} & \textbf{Milestone} & \textbf{Cumulative capability} \\
\midrule
19~Jun & Active development begins & Baseline: prior \texttt{tuas}, \texttt{teh-o-prke} \\
23--24~Jun & \texttt{outram-foam-basic-lib} v0.1.2--v0.1.5 & FV operators, thermophysics, matrix solvers \\
25~Jun & \texttt{outram-foam-appbuilder-lib} scaffolded & rhoCentralFoam, rhoPimpleFoam skeletons \\
26~Jun & Ghia benchmark; MUSCL + k-$\omega$ SST added & Ghia cavity 1.9\% $L^\infty$$^{a}$; Sod $L^1$ 3.7\% \\
26~Jun & vanLeer MUSCL reconstruction & rhoCentralFoam Sod $L^1$ $\to$ 0.7\% \\
28~Jun & \texttt{njoy-outram-park-fork} scaffolded & ENDF tape reader; RECONR Phase 1 \\
28~Jun & RECONR Phase 2b SLBW/MLBW & U-235 thermal cross sections validated \\
2~Jul & WMP + fast MGXS baked (ENDF/B-VIII.0) & Offline nuclear-data blob (\SI{4.7}{\mega\byte}) \\
3~Jul & First end-to-end Godiva $k_{\text{eff}}$ & $k_{\text{eff}} = 1.129$ (+12852 pcm), coupled physics live \\
3~Jul & Anisotropic elastic scatter (MF=4) & Godiva HIGH $k_{\text{eff}} = 0.996$ ($-373$ pcm) \\
3--4~Jul & MF=5 fission spectrum $\chi(E)$; HEATR H1--H7 & Full continuous-energy transport physics \\
6~Jul & Sod shock tube validation documented & LaTeX-ready V\&V infrastructure \\
7~Jul & U-238 capture wing pedestal fix & RRR $L^1$ vs OpenMC: 0.30 $\to$ 0.0007 (400$\times$) \\
7~Jul & Godiva HIGH-fidelity V\&V document & $k_{\text{eff}} = 1.00094 \pm 0.00198$ (+94 pcm) \\
8~Jul & CoolProp fork: 137-fluid coverage & Helmholtz EOS + transport + VLE \\
10~Jul & CoolProp: HumidAir, incompressibles, mixtures (codegen) & 126 incompressibles + 840 binary pairs; V\&V scaffold across 13 crates \\
13~Jul & DWSIM balance-of-plant ports + \texttt{tampines} crate & Valve/HX/pump/expander, pipe correlations; digital-twin GUI scaffold \\
13~Jul & Real IAPWS-IF97 $(p,h)$ flash in \texttt{correct\_thermo} & Two-phase-safe array thermodynamics \\
14~Jul & Nordheim--Fuchs prompt-excursion kinetics & teh-o-prke $\to$ nee\_soon $\to$ digital twin; 909 tests pass \\
14~Jul & Two-phase boiling unblocked; OpenFOAM pressure bounding & Boiling stable; V\&V log + student teaching doc \\
15~Jul & \texttt{outram-mc} Monte Carlo transport (CSG + surface tracking) & Thermal LWR pincell $k_\infty = 1.398$ (OpenMC comparison \emph{in progress}) \\
15~Jul & TRISO random packing + delta-tracking; depletion (CRAM) & Doubly-heterogeneous $k_\infty$; burnup driver; HexLattice \\
15~Jul & GeN-Foam multiphysics port begun (SP3/S$_N$/multiRegion) & \emph{In progress}: single-phase TH core pending \\
16~Jul & First crates.io release prep & coolprop 0.1.0 (GPL-3.0), tuas 0.1.4, tampines 0.2.2, outram-foam \\
16~Jul & Current tip & Multi-physics suite; validation ongoing \\
\bottomrule
\end{tabular}

\vspace{2pt}
{\footnotesize
\begin{flushleft}
$^{a}$~This row records the cavity result \emph{as it stood on 26~June~2026},
obtained on a single refined mesh. The three-mesh grid-convergence study
reported in Section~\ref{sec:cavity} --- which adds the $80\times80$ mesh and
the observed order of accuracy, Tables~\ref{tab:ghia-validation-finest}
and~\ref{tab:ghia-grid-convergence} --- was performed on 17~September~2026 and
therefore postdates this table's 19~June~--~16~July window. It is reported
here because it re-measures work described in this table, not because it
belongs to it. Note also that the MUSCL reconstruction named in this row was
added to \texttt{rhoCentralFoam} (next row) and is not used by the cavity
case, whose convection scheme is first-order upwind throughout.
\end{flushleft}}
\end{table}

\subsection{Outram-Foam Verification and Validation}

As discussed previously, Outram-Foam is a Rust implementation derived
from the numerical methods and software architecture of OpenFOAM.
However, because Outram-Foam is a newly developed codebase, systematic
verification and validation activities were necessary to establish
confidence in the correctness of its implementation.

The verification strategy adopted in this work proceeded in several
stages. First, the underlying matrix and linear-algebra libraries were
verified using established benchmark problems. Second, code-to-code
verification was performed against OpenFOAM using the classical
lid-driven cavity benchmark. Finally, validation against the published
reference solution of Ghia et al.\cite{ghia1982high} was carried out,
including a mesh-refinement study to assess discretization error and
solution convergence.

\subsubsection{Matrix Libraries}

The first verification exercise focused on the matrix and linear-algebra
infrastructure that forms the numerical foundation of Outram-Foam.
For this purpose, the Harwell--Boeing benchmark matrix
\texttt{nnc261} was obtained from the Matrix Market repository
\cite{nnc261_matrixmarket,duff1989sparse}. A method of manufactured
solutions was employed, in which a known reference solution vector was
first prescribed and subsequently used to generate the corresponding
right-hand-side vector. The solver was then required to recover the
original solution vector from the resulting linear system.

The resulting verification test is shown here:

\begin{lstlisting}[caption={Verification test for the dense LU solver (\texttt{SquareMatrix::solve}, Crout LU with scaled partial pivoting) against the NNC261 Harwell--Boeing matrix, using the method of manufactured solutions.},label={lst:nnc261}]
#[test]
fn nnc261_lu_roundtrip() {
    let mtx_path = std::path::Path::new(env!("CARGO_MANIFEST_DIR"))
        .join("tests").join("nnc261.mtx");
    let contents = std::fs::read_to_string(&mtx_path)
        .expect("could not read tests/nnc261.mtx");

    let a = load_matrix_market(&contents);
    assert_eq!(a.n(), 261, "expected a 261x261 matrix");

    // Manufactured reference solution: x_ref[i] = i + 1  (1, 2, ..., 261)
    let n = a.n();
    let x_ref: Vec<f64> = (1..=n).map(|i| i as f64).collect();

    // Forward pass: b = A * x_ref
    let b = mat_vec(&a, &x_ref);

    // Solve A * x = b
    let x_solved = a.solve(&b).expect("NNC261 must be non-singular");

    // Check 1 (primary): relative residual ||A*x - b||_inf / ||b||_inf < 1e-10
    let ax = mat_vec(&a, &x_solved);
    let residual: Vec<f64> = ax.iter().zip(b.iter())
        .map(|(a, b)| a - b).collect();
    let rel_residual = norm_inf(&residual) / norm_inf(&b);
    assert!(rel_residual < 1e-10,
        "relative residual {rel_residual:.3e} exceeds 1e-10");

    // Check 2 (secondary): relative solution error
    // ||x - x_ref||_inf / ||x_ref||_inf < 0.05
    let err: Vec<f64> = x_solved.iter().zip(x_ref.iter())
        .map(|(a, b)| a - b).collect();
    let rel_err = norm_inf(&err) / norm_inf(&x_ref);
    assert!(rel_err < 0.05,
        "relative solution error {rel_err:.3e} exceeds 0.05");
}
\end{lstlisting}

Residuals for the iterative calculations were evaluated using two
separate criteria.

The first criterion was the relative residual, which measures the extent
to which the computed solution satisfies the original linear system.
Using the infinity norm,\footnote{The infinity norm is defined as the
maximum absolute value among all components of a vector.} the relative
residual was calculated as

\begin{equation}
r_{\mathrm{rel}}
=
\frac{\left\lVert A\mathbf{x}-\mathbf{b}\right\rVert_{\infty}}
     {\left\lVert \mathbf{b}\right\rVert_{\infty}}.
\end{equation}

This quantity represents the normalized discrepancy between the computed
right-hand side $A\mathbf{x}$ and the prescribed vector $\mathbf{b}$.
A value close to zero indicates that the numerical solution satisfies
the linear system to high accuracy.

The second criterion was the relative solution error. Unlike the
residual, which measures how well the equations are satisfied, the
relative solution error directly compares the recovered solution vector
against a known reference solution,

\begin{equation}
e_{\mathrm{rel}}
=
\frac{\left\lVert \mathbf{x}-\mathbf{x}_{\mathrm{ref}}
\right\rVert_{\infty}}
     {\left\lVert \mathbf{x}_{\mathrm{ref}}
\right\rVert_{\infty}}.
\end{equation}

For the \texttt{nnc261} benchmark, the measured solution error was

\begin{equation}
e_{\mathrm{rel}}
=
1.184\times10^{-2},
\end{equation}

corresponding to approximately $1.18\%$ relative error. Although this
may initially appear large compared with the residual, the benchmark
matrix is highly ill-conditioned ($\kappa \approx 10^{13}$). The
observed error therefore remains consistent with expected numerical
behaviour for such a challenging test problem and lies comfortably
within the prescribed acceptance criterion of $5\%$.

Having demonstrated satisfactory performance on the initial 
benchmark problems, the author proceeded to more challenging 
verification and validation studies.

\subsubsection{Cavity Case}
\label{sec:cavity}

The lid-driven cavity problem was selected as the next verification
exercise. This benchmark is one of the most widely used validation
cases within the OpenFOAM community and is distributed as part of the
official OpenFOAM tutorial suite
\cite{openfoam_cavity_tutorial}. As a consequence, it provides a
convenient and well-understood baseline for assessing the correctness
of finite-volume implementations.

The reference cavity case was first executed using OpenFOAM. The
resulting mesh (\texttt{polyMesh}), boundary conditions,
\texttt{controlDict}, and velocity-field data were retained as
reference information. The same case was subsequently executed using
Outram-Foam's \texttt{pimpleFoam} solver.

To enable a direct comparison, Outram-Foam's \texttt{pimpleFoam}
solver was configured to operate in a PISO-like mode equivalent to that
of OpenFOAM's \texttt{icoFoam}. Under these conditions, differences
between the two solutions can be attributed primarily to implementation
details rather than differences in solution methodology.

Centerline velocity profiles were then compared between OpenFOAM
\texttt{icoFoam} and Outram-Foam's \texttt{pimpleFoam}. Following this
code-to-code verification step, comparison with the benchmark data of
Ghia et al.\cite{ghia1982high} was also performed. Benchmarking
against peer-reviewed literature was considered preferable to solely
comparing against another software implementation, since agreement with
published benchmark data provides stronger evidence of correctness.

The benchmark data reported by Ghia et al.\cite{ghia1982high} were
manually reviewed and verified before use. Initial comparison suggested
that further improvement might be obtained through mesh refinement.
Consequently, a mesh-refinement study was also conducted in order to
assess the influence of discretization error on the solution.

The resulting comparison between OpenFOAM \texttt{icoFoam} and
Outram-Foam's \texttt{pimpleFoam} is shown below. Excellent agreement
was observed between the two solvers, indicating that the finite-volume
operators, pressure--velocity coupling procedures, and associated
numerical infrastructure were translated correctly. The detailed data
are provided to facilitate independent reproduction of the verification
exercise.

The agreement between OpenFOAM \texttt{icoFoam} and Outram-Foam's
\texttt{pimpleFoam} was excellent, as evidenced by the small residual
differences reported in Table~\ref{tab:ux-centerline}.

\begin{table}[H]
  \centering
  \caption{Comparison of dimensionless $x$-velocity $u_x$ along the
           vertical centerline ($y/L$) between the Outram-Foam's 
		   pimpleFoam and OpenFOAM's icoFoam reference.}
  \label{tab:ux-centerline}
  \sisetup{
    table-format        = +1.5,
    table-number-alignment = center,
    table-align-text-post = false,
  }
  \begin{tabular}{
    S[table-format=1.4]
    S[table-format=+1.5]
    S[table-format=+1.5]
    S[table-format=1.2e+1]
  }
    \toprule
    {$y/L$} & {$u_x^{\text{Rust}}$} & {$u_x^{\text{icoFoam}}$} & {$|\Delta u_x|$} \\
    \midrule
    0.0250 & -0.01770 & -0.01768 & 2.00e-5 \\
    0.0750 & -0.04609 & -0.04609 & 0.00e+0 \\
    0.1250 & -0.07024 & -0.07035 & 1.10e-4 \\
    0.1750 & -0.09194 & -0.09222 & 2.80e-4 \\
    0.2250 & -0.11222 & -0.11275 & 5.30e-4 \\
    0.2750 & -0.13165 & -0.13250 & 8.50e-4 \\
    0.3250 & -0.15033 & -0.15155 & 1.22e-3 \\
    0.3750 & -0.16783 & -0.16947 & 1.64e-3 \\
    0.4250 & -0.18317 & -0.18523 & 2.06e-3 \\
    0.4750 & -0.19468 & -0.19713 & 2.45e-3 \\
    0.5250 & -0.19991 & -0.20261 & 2.70e-3 \\
    0.5750 & -0.19546 & -0.19818 & 2.72e-3 \\
    0.6250 & -0.17689 & -0.17930 & 2.41e-3 \\
    0.6750 & -0.13869 & -0.14035 & 1.66e-3 \\
    0.7250 & -0.07439 & -0.07479 & 4.00e-4 \\
    0.7750 & +0.02303 & +0.02442 & 1.39e-3 \\
    0.8250 & +0.16035 & +0.16400 & 3.65e-3 \\
    0.8750 & +0.34340 & +0.34892 & 5.52e-3 \\
    0.9250 & +0.57470 & +0.58034 & 5.64e-3 \\
    0.9750 & +0.84923 & +0.85202 & 2.79e-3 \\
    \bottomrule
  \end{tabular}
\end{table}

\begin{tikzpicture}
  \begin{axis}[
      name=main,
      width=0.85\linewidth, height=0.45\linewidth,
      xlabel={}, xticklabels={},
      ylabel={$u_x$},
      xmin=0, xmax=1,
      grid=both, legend pos=north west, legend cell align=left,
  ]
    \addplot+[thick, mark=none, color=black]
      table[x=y_over_L, y=icofoam_ux] {ux_centerline.dat};
    \addlegendentry{icoFoam from OpenFOAM}
    \addplot+[only marks, mark=o, mark size=1.4pt, color=red!75!black]
      table[x=y_over_L, y=rust_ux] {ux_centerline.dat};
    \addlegendentry{pimpleFoam from Outram-Foam}
  \end{axis}

  \begin{axis}[
      at={(main.below south west)}, anchor=north west,
      width=0.85\linewidth, height=0.30\linewidth,
      xlabel={$y/L$},
      ylabel={$u_x^{\text{Rust}}-u_x^{\text{icoFoam}}$},
      xmin=0, xmax=1,
      grid=both,
      yticklabel style={/pgf/number format/fixed, /pgf/number format/precision=3},
  ]
    \addplot+[thick, mark=*, mark size=1.2pt, color=blue!70!black]
      table[x=y_over_L,
            y expr=\thisrow{rust_ux}-\thisrow{icofoam_ux}]
            {ux_centerline.dat};
  \end{axis}
\end{tikzpicture}

The agreement between Outram-Foam's \texttt{pimpleFoam} and the
benchmark data of Ghia et al.~\cite{ghia1982high} is also encouraging,
as evidenced by the relatively small discrepancies reported in
Table~\ref{tab:ghia-validation}.

Following successful code-to-code verification against OpenFOAM,
comparison against the benchmark data of Ghia et al.~\cite{ghia1982high}
was performed. This recommendation initially arose from the
AI-assisted workflow. The suggestion was accepted because comparison
against peer-reviewed benchmark data provides stronger validation than
agreement with a single software implementation alone. While
code-to-code verification establishes consistency with OpenFOAM,
agreement with benchmark data from the literature provides additional
confidence that the solver is correctly reproducing the underlying
physics.

The benchmark values were obtained from Table~I of
Ghia et al.~\cite{ghia1982high} for the lid-driven cavity problem at a
Reynolds number of $Re=100$. The benchmark data were manually reviewed
against the original publication prior to their use in the validation
study.

The resulting comparison between Ghia et al.~and Outram-Foam's
\texttt{pimpleFoam} is presented in Table~\ref{tab:ghia-validation}.

\begin{table}[H]
  \centering
  \caption{Validation of the Rust port against the Ghia et al.\ (1982)
           lid-driven cavity benchmark: $x$-velocity $u_x$ along the
           vertical centerline ($y/L$) at $Re=100$.}
  \label{tab:ghia-validation}
  \begin{tabular}{
    S[table-format=1.4]
    S[table-format=+1.5]
    S[table-format=+1.5]
    S[table-format=1.5]
  }
    \toprule
    {$y/L$} & {$u_x^{\text{Ghia}}$} & {$u_x^{\text{Rust}}$} & {$|\Delta u_x|$} \\
    \midrule
    0.0000 & +0.00000 & +0.00000 & 0.00000 \\
    0.0547 & -0.03717 & -0.03348 & 0.00369 \\
    0.0625 & -0.04192 & -0.03775 & 0.00417 \\
    0.0703 & -0.04775 & -0.04201 & 0.00574 \\
    0.1016 & -0.06434 & -0.05678 & 0.00756 \\
    0.1719 & -0.10150 & -0.08652 & 0.01498 \\
    0.2813 & -0.15662 & -0.12566 & 0.03096 \\
    0.4531 & -0.21090 & -0.16510 & 0.04580 \\
    0.5000 & -0.20581 & -0.16520 & 0.04061 \\
    0.6172 & -0.13641 & -0.12885 & 0.00756 \\
    0.7344 & +0.00332 & -0.02215 & 0.02547 \\
    0.8516 & +0.23151 & +0.19253 & 0.03898 \\
    0.9531 & +0.68717 & +0.62802 & 0.05915 \\
    0.9609 & +0.73722 & +0.67634 & 0.06088 \\
    0.9688 & +0.78871 & +0.72529 & 0.06342 \\
    0.9766 & +0.84123 & +0.77883 & 0.06240 \\
    1.0000 & +1.00000 & +1.00000 & 0.00000 \\
    \midrule
    \multicolumn{3}{r}{\textbf{Max $|\Delta u_x|$}} & \bfseries 0.06342 \\
    \multicolumn{3}{r}{\textbf{Mean $|\Delta u_x|$}} & \bfseries 0.02773 \\
    \multicolumn{3}{r}{\textbf{RMS $|\Delta u_x|$}}  & \bfseries 0.03630 \\
    \bottomrule
  \end{tabular}
\end{table}

An initial comparison against the benchmark data of
Ghia et al.~\cite{ghia1982high} was performed using a relatively coarse
$20\times20$ computational mesh.

At this stage, the observed discrepancies were not unexpected because
the mesh resolution was relatively low. The maximum pointwise error was
approximately $6.3\%$ of the lid velocity, with the largest deviations
occurring in regions of strong velocity gradients near the moving lid.

It is important to state what discretization produced that figure, because
it sets the accuracy that can be expected. Convection in Outram-Foam's
\texttt{pimpleFoam} is assembled with first-order upwind differencing
(\texttt{DivScheme::GaussUpwind}, the solver default), while the Laplacian is
second order. The formal spatial order of the cavity solution is therefore
\textbf{one}, not two, and the gap to Ghia et al.~on a coarse mesh is
dominated by first-order numerical diffusion. Ghia et al.~themselves used a
$129\times129$ grid. A $6.3\%$ maximum deviation is thus the expected
behaviour of this scheme at this resolution rather than an indication of an
implementation defect --- a proposition that the grid-convergence study below
tests directly.

Following inspection of the validation results, a mesh-refinement study was
undertaken to quantify the effect of spatial resolution on solution accuracy.
Two grids can only show that the error became smaller; they cannot establish an
observed order of accuracy. A third grid was therefore added, giving the
sequence $20\times20$, $40\times40$ and $80\times80$ at a constant refinement
ratio $r=2$.

All three meshes are generated from the same \texttt{blockMeshDict} with only
the cell count substituted, so they are identical in geometry, patch naming and
patch type, and all three have an even cell count so that the centerline
sampling is performed identically on each. This matters: the sampling line
$x=L/2$ falls between two columns of cell centres on an even mesh and exactly on
a column on an odd one, so a sequence mixing the two would place a different
sampling error on different members of it and corrupt the measured order.

The per-station results are given in Tables~\ref{tab:ghia-validation-fine}
and~\ref{tab:ghia-validation-finest}, and the convergence quantities derived
from them in Table~\ref{tab:ghia-grid-convergence}.

\begin{table}[H]
  \centering
  \caption{Validation of the Rust port against the Ghia et al.\ (1982)
           lid-driven cavity benchmark at $Re=100$ on the refined mesh:
           $x$-velocity $u_x$ along the vertical centerline ($y/L$).}
  \label{tab:ghia-validation-fine}
  \begin{tabular}{
    S[table-format=1.4]
    S[table-format=+1.5]
    S[table-format=+1.5]
    S[table-format=1.5]
  }
    \toprule
    {$y/L$} & {$u_x^{\text{Ghia}}$} & {$u_x^{\text{Rust}}$} & {$|\Delta u_x|$} \\
    \midrule
    0.0000 & +0.00000 & +0.00000 & 0.00000 \\
    0.0547 & -0.03717 & -0.03669 & 0.00048 \\
    0.0625 & -0.04192 & -0.04142 & 0.00050 \\
    0.0703 & -0.04775 & -0.04576 & 0.00199 \\
    0.1016 & -0.06434 & -0.06265 & 0.00169 \\
    0.1719 & -0.10150 & -0.09654 & 0.00496 \\
    0.2813 & -0.15662 & -0.14359 & 0.01303 \\
    0.4531 & -0.21090 & -0.19058 & 0.02032 \\
    0.5000 & -0.20581 & -0.18883 & 0.01698 \\
    0.6172 & -0.13641 & -0.13781 & 0.00140 \\
    0.7344 & +0.00332 & -0.01337 & 0.01669 \\
    0.8516 & +0.23151 & +0.21428 & 0.01723 \\
    0.9531 & +0.68717 & +0.66985 & 0.01732 \\
    0.9609 & +0.73722 & +0.71930 & 0.01792 \\
    0.9688 & +0.78871 & +0.77433 & 0.01438 \\
    0.9766 & +0.84123 & +0.82990 & 0.01133 \\
    1.0000 & +1.00000 & +1.00000 & 0.00000 \\
    \midrule
    \multicolumn{3}{r}{\textbf{Max $|\Delta u_x|$}}  & \bfseries 0.02032 \\
    \multicolumn{3}{r}{\textbf{RMS $|\Delta u_x|$}}  & \bfseries 0.01197 \\
    \bottomrule
  \end{tabular}
\end{table}

\begin{table}[H]
  \centering
  \caption{Validation of the Rust port against the Ghia et al.\ (1982)
           lid-driven cavity benchmark at $Re=100$ on the finest
           ($80\times80$) mesh: $x$-velocity $u_x$ along the vertical
           centerline ($y/L$).}
  \label{tab:ghia-validation-finest}
  \begin{tabular}{
    S[table-format=1.4]
    S[table-format=+1.5]
    S[table-format=+1.5]
    S[table-format=1.5]
  }
    \toprule
    {$y/L$} & {$u_x^{\text{Ghia}}$} & {$u_x^{\text{Rust}}$} & {$|\Delta u_x|$} \\
    \midrule
    0.0000 & +0.00000 & +0.00000 & 0.00000 \\
    0.0547 & -0.03717 & -0.03730 & 0.00013 \\
    0.0625 & -0.04192 & -0.04198 & 0.00006 \\
    0.0703 & -0.04775 & -0.04659 & 0.00116 \\
    0.1016 & -0.06434 & -0.06404 & 0.00030 \\
    0.1719 & -0.10150 & -0.09973 & 0.00177 \\
    0.2813 & -0.15662 & -0.15101 & 0.00561 \\
    0.4531 & -0.21090 & -0.20229 & 0.00861 \\
    0.5000 & -0.20581 & -0.19916 & 0.00665 \\
    0.6172 & -0.13641 & -0.13910 & 0.00269 \\
    0.7344 & +0.00332 & -0.00536 & 0.00868 \\
    0.8516 & +0.23151 & +0.22560 & 0.00591 \\
    0.9531 & +0.68717 & +0.68185 & 0.00532 \\
    0.9609 & +0.73722 & +0.73228 & 0.00494 \\
    0.9688 & +0.78871 & +0.78455 & 0.00416 \\
    0.9766 & +0.84123 & +0.83811 & 0.00312 \\
    1.0000 & +1.00000 & +1.00000 & 0.00000 \\
    \midrule
    \multicolumn{3}{r}{\textbf{Max $|\Delta u_x|$}}  & \bfseries 0.00868 \\
    \multicolumn{3}{r}{\textbf{RMS $|\Delta u_x|$}}  & \bfseries 0.00455 \\
    \bottomrule
  \end{tabular}
\end{table}

\begin{table}[H]
  \centering
  \caption{Grid-convergence quantities for the lid-driven cavity at $Re=100$,
           derived from the three meshes above at a constant refinement ratio
           $r=2$. The functional is $u_x/U_{\mathrm{lid}}$ interpolated at the
           fixed station $y/L=0.4531$, Ghia's tabulated point nearest the
           recirculation peak and the location of the largest coarse-mesh
           error. The observed order is
           $p=\ln(|f_1-f_2|/|f_2-f_3|)/\ln r$; Richardson extrapolation is
           $f_{h\to0}=f_3+(f_3-f_2)/(r^p-1)$; and
           $\mathrm{GCI}_{23}=F_s|(f_3-f_2)/f_3|/(r^p-1)$ with $F_s=1.25$.}
  \label{tab:ghia-grid-convergence}
  \footnotesize
  \begin{tabular}{@{}lr@{}}
    \toprule
    \textbf{Quantity} & \textbf{Value} \\
    \midrule
    $f_1$, $20\times20$ ($\Delta x = \SI{5.00e-3}{m}$)   & $-0.165103$ \\
    $f_2$, $40\times40$ ($\Delta x = \SI{2.50e-3}{m}$)   & $-0.190580$ \\
    $f_3$, $80\times80$ ($\Delta x = \SI{1.25e-3}{m}$)   & $-0.202290$ \\
    \addlinespace[2pt]
    Observed order of accuracy $p$                        & $1.121$ \\
    Richardson extrapolation $f_{h\to0}$                  & $-0.21225$ \\
    Ghia et al.\ (1982) at the same station               & $-0.21090$ \\
    Difference, extrapolated vs.\ benchmark               & $0.64\,\%$ \\
    $\mathrm{GCI}_{23}$ ($F_s=1.25$)                      & $6.16\,\%$ \\
    \addlinespace[2pt]
    RMS $|\Delta u_x|$ vs.\ Ghia, $20\to40\to80$          & $0.03630 \to 0.01197 \to 0.00455$ \\
    Trend in the benchmark error, $\log_2(e_i/e_{i+1})$   & $1.600$, $1.396$ \\
    \bottomrule
  \end{tabular}
\end{table}


Each refinement reduces the maximum pointwise error: $6.3\%$ of the lid
velocity on $20\times20$, $2.0\%$ on $40\times40$ and $0.87\%$ on
$80\times80$. Increasing agreement is therefore achieved with the lid-driven
cavity flow of Ghia et al.~\cite{ghia1982high}, and the rate at which it is
achieved is quantified below.

\begin{figure}[H]
  \centering
  \begin{tikzpicture}
    \begin{axis}[
        name=main,
        width=0.85\linewidth, height=0.45\linewidth,
        xticklabels={},
        ylabel={$u_x$},
        xmin=0, xmax=1,
        grid=both, minor tick num=1,
        legend pos=north west, legend cell align=left,
        tick label style={font=\small}, label style={font=\small},
        legend style={font=\small},
    ]
      \addplot+[only marks, mark=square*, mark size=2.2pt, color=black]
        table[x=y_over_L, y=ghia_ux] {ghia_validation_fine.dat};
      \addlegendentry{Ghia et al.\ (1982)}

      \addplot+[thick, mark=o, mark size=1.6pt,
                color=orange!80!black, densely dashed]
        table[x=y_over_L, y=rust_ux] {ghia_validation_coarse.dat};
      \addlegendentry{Outram-Foam, $20\times20$}

      \addplot+[thick, mark=triangle, mark size=1.8pt,
                color=red!75!black, dashed]
        table[x=y_over_L, y=rust_ux] {ghia_validation_fine.dat};
      \addlegendentry{Outram-Foam, $40\times40$}

      \addplot+[thick, mark=diamond, mark size=1.8pt,
                color=blue!70!black]
        table[x=y_over_L, y=rust_ux] {ghia_validation_finest.dat};
      \addlegendentry{Outram-Foam, $80\times80$}
    \end{axis}

    \begin{axis}[
        at={(main.below south west)}, anchor=north west,
        width=0.85\linewidth, height=0.30\linewidth,
        xlabel={$y/L$},
        ylabel={$|\Delta u_x|$},
        xmin=0, xmax=1, ymin=0,
        ymode=log,
        log basis y=10,
        ymin=1e-4, ymax=1e-1,
        grid=both,
        tick label style={font=\small}, label style={font=\small},
        legend style={font=\small, at={(0.02,0.95)}, anchor=north west},
        legend cell align=left,
    ]
      \addplot+[thin, mark=o, mark size=1.4pt,
                color=orange!80!black, densely dashed]
        table[x=y_over_L, y=abs_err] {ghia_validation_coarse.dat};
      \addlegendentry{$20\times20$}
      \addplot+[thin, mark=triangle*, mark size=1.4pt,
                color=red!75!black, dashed]
        table[x=y_over_L, y=abs_err] {ghia_validation_fine.dat};
      \addlegendentry{$40\times40$}
      \addplot+[thin, mark=diamond*, mark size=1.4pt,
                color=blue!70!black]
        table[x=y_over_L, y=abs_err] {ghia_validation_finest.dat};
      \addlegendentry{$80\times80$}
    \end{axis}
  \end{tikzpicture}
  \caption{Grid convergence of the Rust port against the Ghia et al.\ (1982)
           lid-driven cavity benchmark at $Re=100$, on the three-mesh sequence
           $20\times20$, $40\times40$, $80\times80$ ($r=2$).
           Top: $u_x$ profiles along the vertical centerline; each refinement
           tracks the benchmark more closely, most visibly in the shear layer
           near $y/L\approx1$ and at the recirculation peak near
           $y/L\approx0.45$.
           Bottom: pointwise absolute error (log scale). RMS error over the 17
           stations falls $0.0363 \to 0.0120 \to 0.0046$.}
  \label{fig:ghia-grid-convergence}
\end{figure}
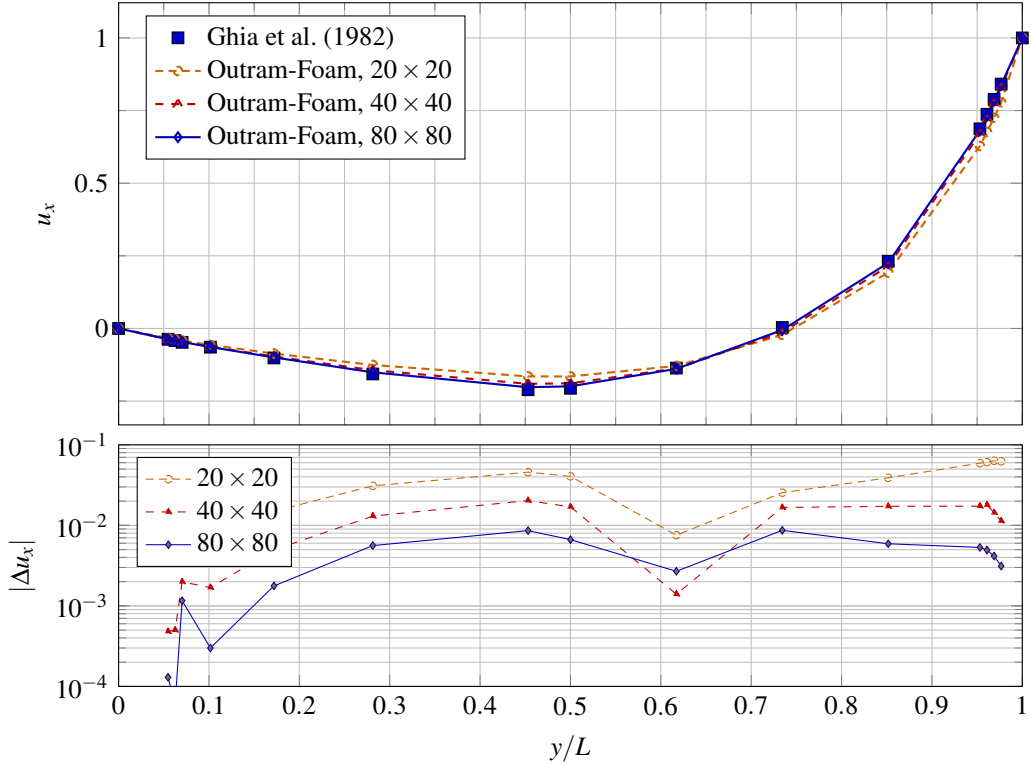

\begin{figure}[H]
  \centering
  \begin{tikzpicture}
    \begin{axis}[
        width=0.72\linewidth, height=0.48\linewidth,
        xlabel={cell width $\Delta x$ [m]},
        ylabel={RMS $|\Delta u_x|$ vs.\ Ghia},
        xmode=log, ymode=log,
        log basis x=10, log basis y=10,
        grid=both,
        legend pos=north west, legend cell align=left,
        tick label style={font=\small}, label style={font=\small},
        legend style={font=\small},
    ]
      \addplot+[thick, only marks, mark=*, mark size=2.4pt,
                color=blue!70!black]
        table[x=dx, y=rms_err] {ghia_grid_convergence.dat};
      \addlegendentry{measured}

      \addplot[thick, dashed, domain=1.25e-3:5.0e-3, samples=2,
               color=black]
        {0.03630*(x/5.0e-3)^1};
      \addlegendentry{slope 1 (first order)}

      \addplot[thick, dotted, domain=1.25e-3:5.0e-3, samples=2,
               color=gray!60!black]
        {0.03630*(x/5.0e-3)^2};
      \addlegendentry{slope 2 (second order)}
    \end{axis}
  \end{tikzpicture}
  \caption{Convergence of the centerline RMS error against cell width, with
           first- and second-order reference slopes anchored at the coarsest
           mesh. The measured points sit between the two references, which is
           the expected signature of a scheme combining first-order upwind
           convection with a second-order Laplacian. The verification order
           obtained by Richardson extrapolation of the solution itself is
           $p=1.121$ (Table~\ref{tab:ghia-grid-convergence}); the steeper
           apparent slope of the benchmark error --- $\log_2(e_i/e_{i+1})$ of
           $1.600$ and $1.396$ --- includes Ghia's own discretization error and
           is not an order of accuracy.}
  \label{fig:ghia-convergence-order}
\end{figure}
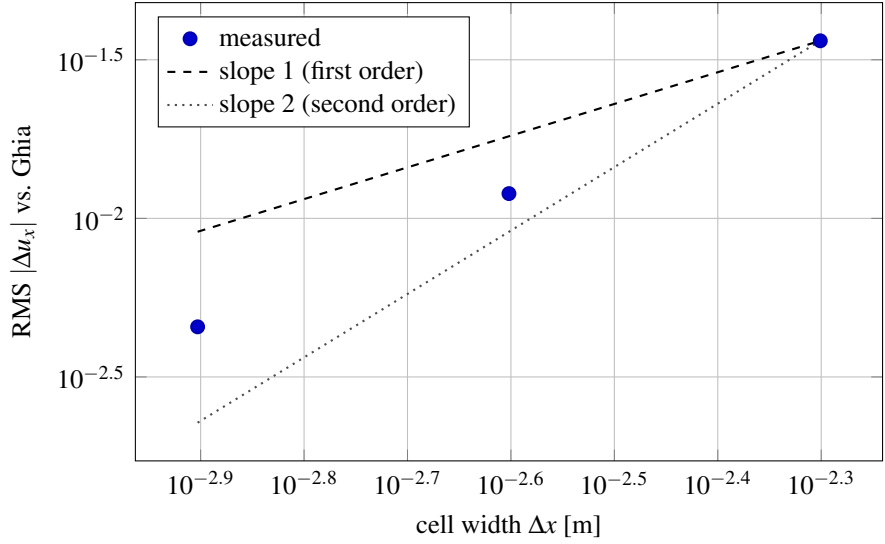

Figure~\ref{fig:ghia-grid-convergence} shows that each refinement improves
agreement with the benchmark solution of Ghia et al.~\cite{ghia1982high}. The
maximum pointwise error falls from $6.3\%$ of the lid velocity on the
$20\times20$ mesh to $2.0\%$ on $40\times40$ and $0.87\%$ on $80\times80$, and
the centerline RMS error from $0.0363$ to $0.0120$ to $0.0046$.

That the error decreases is necessary but not sufficient: it shows only that
the discretization is doing something sensible, not that it is converging at
the rate the scheme should deliver, nor that it is converging to the right
answer. The three-mesh sequence permits both to be measured, and
Table~\ref{tab:ghia-grid-convergence} reports them.

The observed order of accuracy, obtained by Richardson extrapolation of the
centerline velocity at $y/L=0.4531$, is $p=1.121$. This is the verification
result, and it sits essentially on the formal first order of the upwind
convection scheme in use. Richardson extrapolation of the same sequence gives
$u_x/U_{\mathrm{lid}} = -0.21225$ as $\Delta x \to 0$, against Ghia et al.'s
tabulated $-0.21090$ at that station --- a difference of $0.64\%$. The
discretization therefore converges at its formal rate, and converges to the
benchmark value rather than merely to some grid-independent value of its own.
The fine-grid convergence index on the $40\to80$ pair is
$\mathrm{GCI}_{23}=6.16\%$, which is the uncertainty that should be attached to
the $80\times80$ result.

One distinction is worth drawing explicitly, because conflating the two
overstates the result. The trend in the error \emph{against Ghia et al.} is
steeper than the order of accuracy: $\log_2(e_i/e_{i+1})$ gives $1.600$ for
$20\to40$ and $1.396$ for $40\to80$. Those figures include Ghia et al.'s own
discretization error and the fact that the benchmark error is not a clean power
law, and they are reported here as a validation trend only. The order of
accuracy of this implementation is the $1.121$ obtained from the solution
sequence itself, not the $1.4$--$1.6$ obtained from the benchmark comparison.

Taken together, these results are consistent with Outram-Foam's
\texttt{pimpleFoam} correctly reproducing the classical lid-driven cavity
benchmark: it agrees with OpenFOAM \texttt{icoFoam} under matched PISO
settings, converges at its formal order, and extrapolates to the published
reference value. Two limitations should be stated alongside. First, the study
is at $Re=100$, the lowest of the seven Reynolds numbers tabulated by
Ghia et al., where the flow is diffusion-dominated and the convective
discretization is correspondingly lightly exercised; agreement here does not
establish behaviour in convection-dominated regimes. Second, reaching benchmark
accuracy on a practical mesh is a question of scheme rather than of
implementation --- a second-order or limited convection scheme would reach the
same accuracy at far lower cost --- and that comparison has not been run.
Both are left to future work, readers should note that this is an incomplete, 
initial verficiation effort. 

\subsubsection{Sod Shock Tube Case for rhoCentralFoam Port}

The lid-driven cavity case is verified using literature data, but 
it only covers incompressible regimes. However, real fluid simulation 
in reactors would need compressible flow.
For that, we need to consider classic use cases for compressible flow 
such as shock formation.

OpenFOAM provides a Sod shock tube case, a classic benchmark \cite{toro2013riemann,sod1978survey}.
This case is commonly used as a benchmark for finite-volume methods 
\cite{marzouk2020sod}.
This OpenFOAM case used rhoCentralFoam; as a compressible solver, 
rhoCentralFoam is suitable for transonic flows, and the Sod shock tube is 
well established in the literature \cite{toro2013riemann,sod1978survey}
The Sod shock tube case was therefore adopted. 

As with the cavity case, OpenFOAM was used to construct the Sod shock 
tube geometry using its meshing tools.
Once the geometry was complete, it was provided to Claude Code. After asking 
Claude Code to implement rhoCentralFoam, it was asked to produce a tutorial 
example for V\&V, namely this Sod shock tube case, and Claude Code 
compared it to the exact Riemann solution by generating its own code.

The exact Riemann solution Claude Code used for V\&V was computed by
implementing the solver from Toro chapter 4 \cite{toro2013riemann}.
Results are here Fig.~\ref{fig:sod-shock-tube-riemann-solution}:

\begin{figure}[H]
    \centering
	\includegraphics[width=0.95\textwidth]{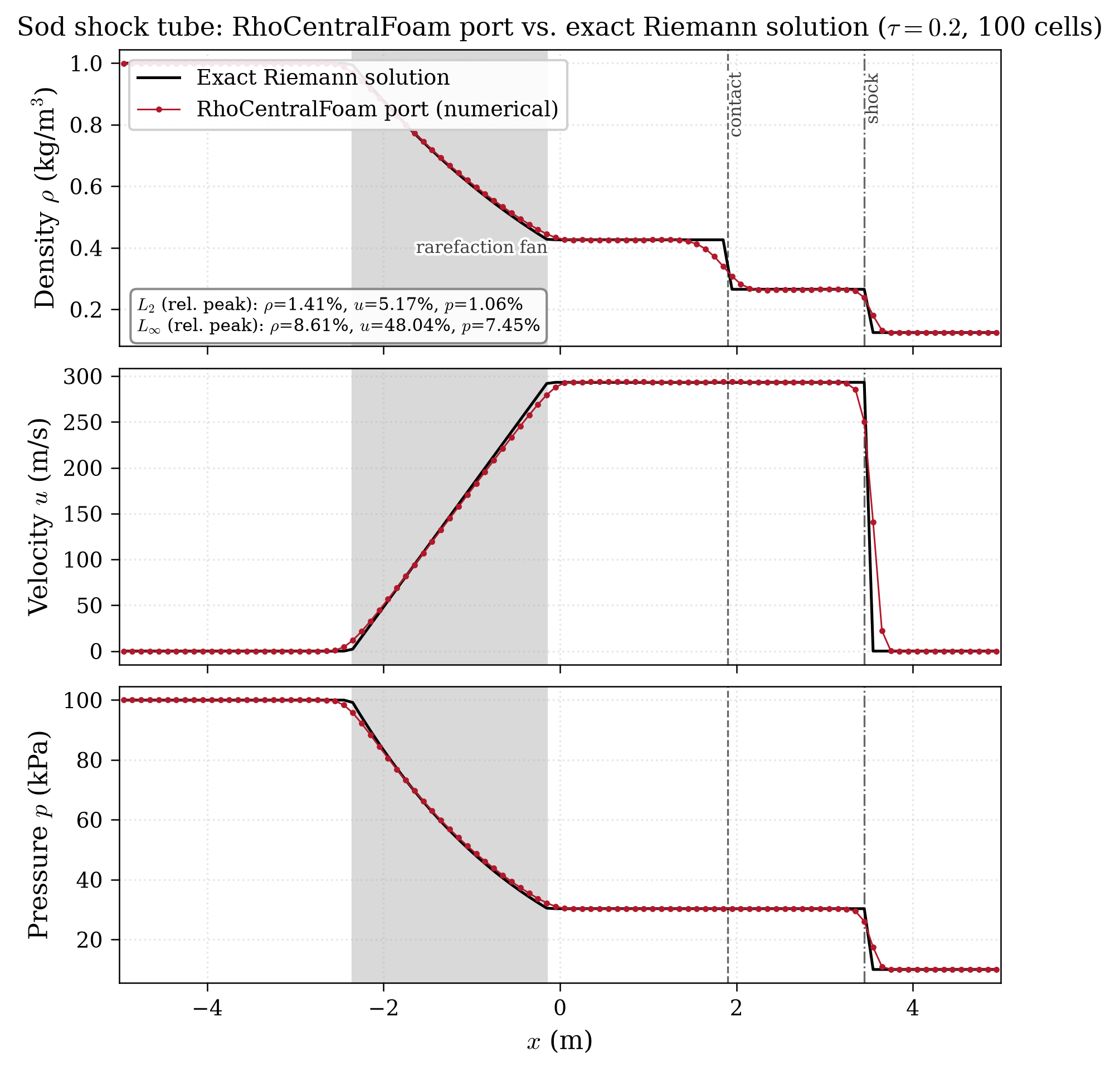}
	\caption{\textbf{Outram-Foam V\&V against the exact Riemann solution} \cite{toro2013riemann}}
    \label{fig:sod-shock-tube-riemann-solution}
\end{figure}

The residuals are acceptable, as seen in Fig~\ref{fig:sod-shock-tube-residual} for exact 
solution:

\begin{figure}[H]
    \centering
	\includegraphics[width=0.95\textwidth]{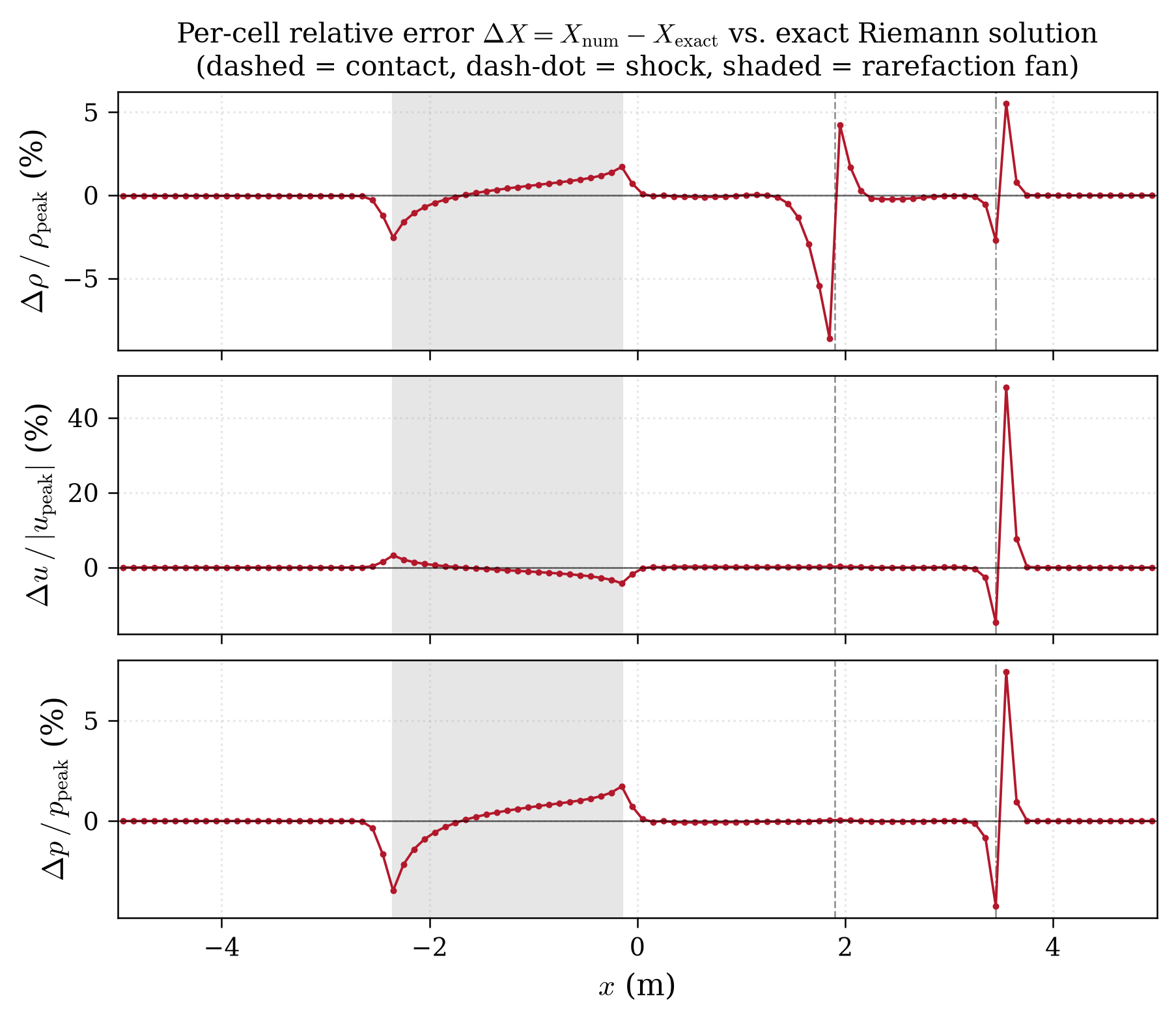}
	\caption{\textbf{Outram-Foam V\&V Residuals} \cite{toro2013riemann}}
    \label{fig:sod-shock-tube-residual}
\end{figure}

But how can one be sure it is correct? Claude Code generated both 
the coding results and the V\&V case, it effectively checked its own work. One must 
be meticulous and independently locate the V\&V data manually.

For this,
The port was compared against Sod's Table II \cite{sod1978survey}. Matches decently 
except at the rarefaction wave, as shown in 
Figure~\ref{fig:sod-shock-tube-table-ii}:

\begin{figure}[H]
    \centering
	\includegraphics[width=0.95\textwidth]{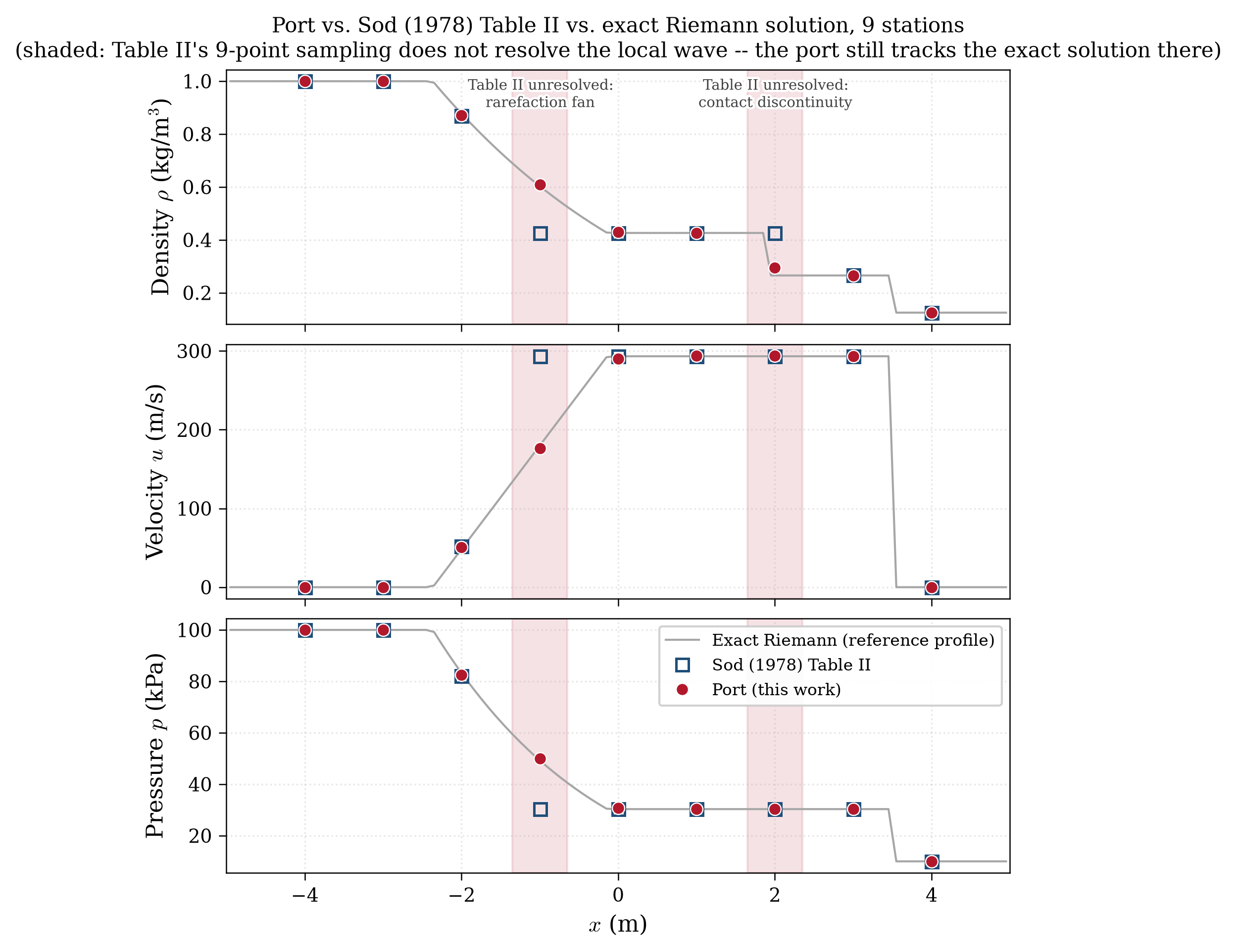}
	\caption{\textbf{Outram-Foam V\&V using Sod's Table II} \cite{sod1978survey}}
    \label{fig:sod-shock-tube-table-ii}
\end{figure}

The station-by-station values behind Figure~\ref{fig:sod-shock-tube-table-ii}
are given in Table~\ref{tab:sod-table-ii-comparison}.

\begin{table}[H]
  \centering
  \caption{\textbf{Sod shock tube: station-by-station comparison against the
  coarse 9-point Sod (1978) Table~II.} The Outram-Foam \texttt{rhoCentralFoam}
  port (100 cells) and the exact Riemann solution are the quantitative
  references; the 9-point tabulated column ($\rho_{\text{Sod~II}}$ etc.) is a
  coarse literature sanity check only and is \emph{not} treated as
  authoritative. Where the port departs from it (e.g.\ $x/L=0.4$ and $0.7$, in
  the rarefaction fan and at the contact) it still tracks the exact solution ---
  the 9-point table simply cannot resolve those features. Densities are
  in kg/m$^3$, velocity in m/s and pressure in Pa}
  \label{tab:sod-table-ii-comparison}
  \footnotesize
  \setlength{\tabcolsep}{4pt}
  \resizebox{\textwidth}{!}{%
  \begin{filecontents*}{results_and_discussion/outram-foam-sod-shock-tube/sod_table_ii_3way.csv}
x_over_L,rho_port,rho_tab,rho_exact,u_port,u_tab,u_exact,p_port,p_tab,p_exact
0.1,0.9996,1.0000,1.0000,0.00,0.00,0.00,100000.0,100000.0,100000.0
0.2,0.9996,1.0000,1.0000,0.00,0.00,0.00,100000.0,100000.0,100000.0
0.3,0.8717,0.8690,0.8775,50.61,51.86,48.28,82560.8,82200.0,83274.7
0.4,0.6090,0.4260,0.6029,176.65,293.14,180.04,49971.7,30300.0,49247.2
0.5,0.4303,0.4260,0.4263,290.31,293.14,293.29,30726.9,30300.0,30313.0
0.6,0.4262,0.4260,0.4263,293.71,293.14,293.29,30265.7,30300.0,30313.0
0.7,0.2953,0.4260,0.2656,293.90,293.14,293.29,30351.5,30300.0,30313.0
0.8,0.2652,0.2660,0.2656,293.48,293.14,293.29,30321.1,30300.0,30313.0
0.9,0.1250,0.1250,0.1250,0.00,0.00,0.00,10000.0,10000.0,10000.0
\end{filecontents*}
\pgfplotstabletypeset[
    col sep=comma,
    columns={x_over_L,rho_port,rho_tab,rho_exact,u_port,u_tab,u_exact,p_port,p_tab,p_exact},
    columns/x_over_L/.style={column name={$x/L$}, fixed, precision=2, zerofill},
    columns/rho_port/.style={column name={$\rho_{\text{port}}$}, fixed, precision=4, zerofill},
    columns/rho_tab/.style={column name={$\rho_{\text{Sod~II}}$}, fixed, precision=4, zerofill},
    columns/rho_exact/.style={column name={$\rho_{\text{exact}}$}, fixed, precision=4, zerofill},
    columns/u_port/.style={column name={$u_{\text{port}}$}, fixed, precision=2, zerofill},
    columns/u_tab/.style={column name={$u_{\text{Sod~II}}$}, fixed, precision=2, zerofill},
    columns/u_exact/.style={column name={$u_{\text{exact}}$}, fixed, precision=2, zerofill},
    columns/p_port/.style={column name={$p_{\text{port}}$}, fixed, precision=1, zerofill},
    columns/p_tab/.style={column name={$p_{\text{Sod~II}}$}, fixed, precision=1, zerofill},
    columns/p_exact/.style={column name={$p_{\text{exact}}$}, fixed, precision=1, zerofill},
    every head row/.style={before row=\toprule, after row=\midrule},
    every last row/.style={after row=\bottomrule},
  ]{results_and_discussion/outram-foam-sod-shock-tube/sod_table_ii_3way.csv}%
  }
\end{table}

This is a reasonable result for a first attempt.
Sod Table II was used because it was readily available as a first bound 
check. The tabulated data are provided for 9 grid points as mentioned 
in \cite{sod1978survey}. This is a relatively coarse mesh solution, as typical 
solutions can have as many as 100 points \cite{sod1978survey}. It is 
never as accurate as the exact solution, but still 
useful as a sanity check that Claude Code was not fabricating results.

Naturally, when a coarse mesh is compared to the Outram-Foam mesh of 100 
cells, the coarseness can explain the discrepancy.
Is there anything more concrete? For this, one may examine Fig 
4 of Marzouk's paper \cite{marzouk2020sod}. In his Fig 4 caption, Marzouk
cites the Castro solver's verification page \cite{castro_verification}.
That page also cites chapter 4 of Riemann
solver \cite{toro2013riemann}, the same chapter that Claude Code implemented.
The method Claude Code used is therefore the standard one, not one it fabricated.

More usefully, Marzouk solves the same Sod
problem at the same dimensionless time $\tau = 0.2$, and compares it against
the Castro solver, which is an entirely different code developed by a different group.
So if Outram-Foam also matches Marzouk, then it is not being self-validated
anymore. Furthermore, Marzouk states $\tau = 0.2$ explicitly \cite{marzouk2020sod}.
Sod's paper \cite{sod1978survey} never actually gives the time, neither for
Table II nor for his figures. Marzouk is therefore the citable source for
the comparison time.

The manual check of the reference is valuable, but it is also worth doing
a check that does not require complicated programming.
Marzouk's Figure~4 was digitised approximately. Pressure, velocity and
density were read off using graphreader \cite{graphreader} at dimensionless
time $\tau = 0.2$ as mentioned by Marzouk \cite{marzouk2020sod}:

\begin{figure}[H]
    \centering
	\includegraphics[width=0.95\textwidth]{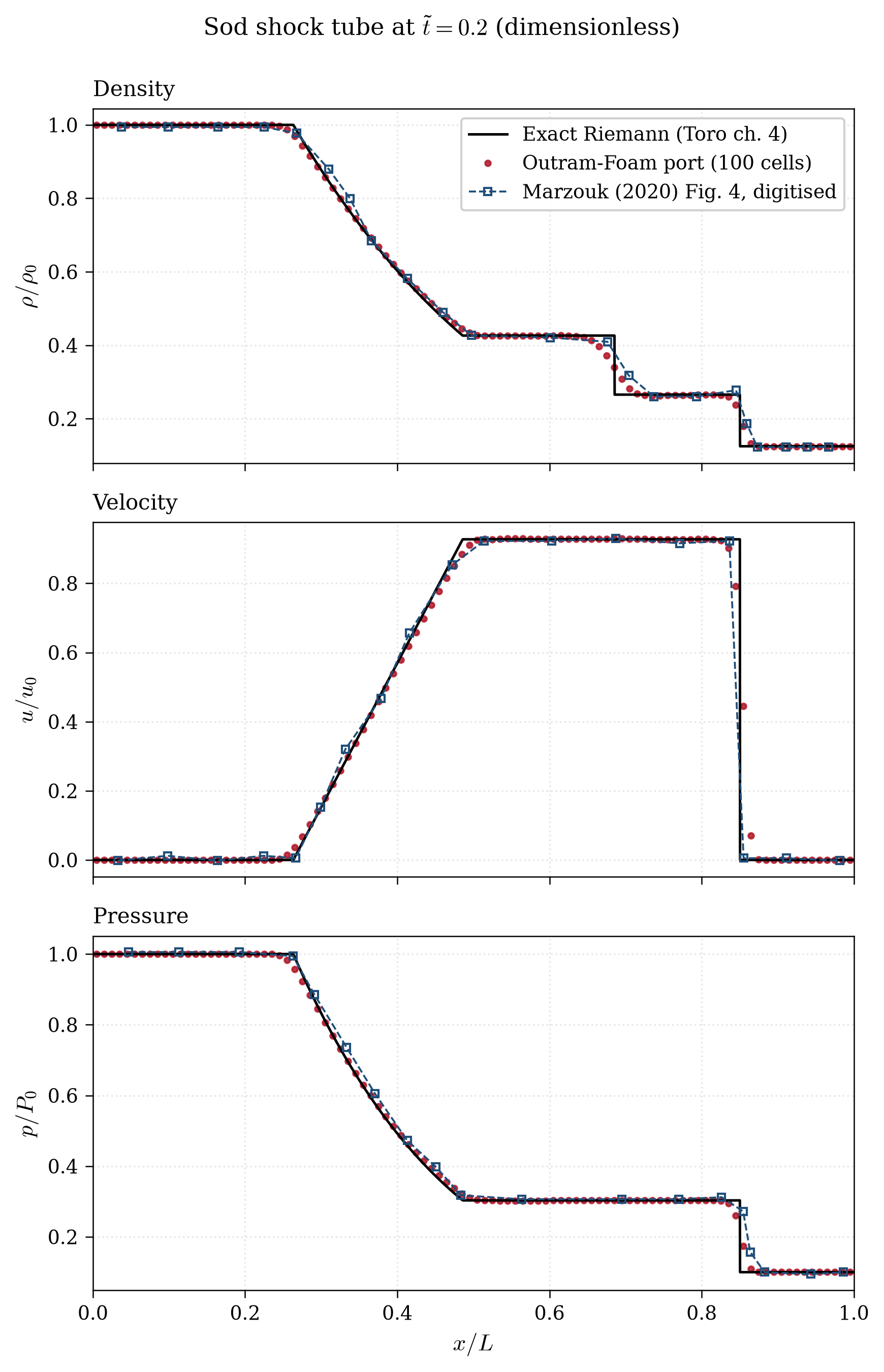}
	\caption{\textbf{Outram-Foam V\&V using Marzouk's Fig 4 Digitised 
	using GraphReader (approx only)} \cite{marzouk2020sod}}
    \label{fig:sod-shock-marzouk}
\end{figure}

A representative slice of the digitised pressure comparison is tabulated in
Table~\ref{tab:sod-marzouk-pressure}.

\begin{table}[H]
  \centering
  \caption{\textbf{Sod shock tube pressure: Outram-Foam port vs.\ Marzouk (2020)
  Fig.~4, digitised.} Ten representative stations of the pressure channel at
  $\tau=0.2$, dimensionless ($p/P_0$). Marzouk's points are read off a printed
  figure (approximate); the exact Riemann solution is the quantitative
  reference and $\Delta = p_{\text{Marzouk}} - p_{\text{exact}}$ measures
  digitisation agreement, \emph{not} solver error. The single large residual at
  $x/L\approx0.85$ is one digitised point straddling the shock jump. The density
  and velocity channels agree comparably over the full digitised sets
  (max$|\Delta|$: $\rho$ 0.061, $u$ 0.035; RMS 0.021 and 0.012).}
  \label{tab:sod-marzouk-pressure}
  \footnotesize
  \setlength{\tabcolsep}{6pt}
  \begin{filecontents*}{results_and_discussion/outram-foam-sod-shock-tube/sod_marzouk_pressure_10row.csv}
x_over_L,marzouk,exact,port,marz_minus_exact
0.0469,1.0057,1.0000,1.0000,+0.0057
0.1925,1.0057,1.0000,1.0000,+0.0057
0.2911,0.8854,0.8710,0.8602,+0.0144
0.3709,0.6045,0.5762,0.5812,+0.0283
0.4507,0.3981,0.3714,0.3823,+0.0267
0.5634,0.3064,0.3031,0.3024,+0.0033
0.7700,0.3064,0.3031,0.3030,+0.0033
0.8545,0.2720,0.1000,0.1788,+0.1720
0.8826,0.1000,0.1000,0.1000,+0.0000
0.9859,0.1000,0.1000,0.1000,+0.0000
\end{filecontents*}
\pgfplotstabletypeset[
    col sep=comma,
    columns={x_over_L,marzouk,exact,port,marz_minus_exact},
    columns/x_over_L/.style={column name={$x/L$}, fixed, precision=4, zerofill},
    columns/marzouk/.style={column name={$p_{\text{Marzouk}}$}, fixed, precision=4, zerofill},
    columns/exact/.style={column name={$p_{\text{exact}}$}, fixed, precision=4, zerofill},
    columns/port/.style={column name={$p_{\text{port}}$}, fixed, precision=4, zerofill},
    columns/marz_minus_exact/.style={column name={$\Delta$ (M$-$E)}, fixed, precision=4, zerofill},
    every head row/.style={before row=\toprule, after row=\midrule},
    every last row/.style={after row=\bottomrule},
  ]{results_and_discussion/outram-foam-sod-shock-tube/sod_marzouk_pressure_10row.csv}%
\end{table}

The CSV file was obtained manually. Then
Claude Code was asked to write a Python script to produce the PNG. This became 
Figure ~\ref{fig:sod-shock-marzouk}.

Overall, given how the Outram-Foam solver compares with  the exact Riemann
solution, the coarse Sod Table~II sanity check, and the independent Marzouk
digitised data, the port is qualitatively validated for the Sod shock tube,
with pressure and density agreeing to within 5\%, provided they are 
distanced from the discontinuities.  There, the errors can be as high as 
20\%. However, in the discontinuities, it is difficult to have lower error 
anyhow. Taking all this into account, the Sod shock case can be considered 
ported rather satisfactorily for something as difficult to solve as a shock 
front.

\subsection{TAMPINES Steam Tables}

\subsubsection{Steam Tables Testing Summary}

The tampines-steam-tables library took some time to develop. The commit
history is long, so it was summarised with AI assistance in
Table~\ref{tab:tampines-summary} which
summarises the two development phases; the fuller AI-generated summary 
history is in
Table~\ref{tab:tampines-history} in Appendix~\ref{app:tampines}.

\begin{table}[ht]
\centering
\caption{AI-generated Summary of \texttt{tampines-steam-tables} development. Phase~I spans 947
commits in the standalone repository (2025-01 to 2026-06); Phase~II covers 74
commits on the migrated crate within \texttt{outram-park-backend} (2026-06 to
2026-07). Full per-commit detail is given in Table~\ref{tab:tampines-history},
Appendix~\ref{app:tampines}.}
\label{tab:tampines-summary}
\footnotesize
\setlength{\tabcolsep}{5pt}
\begin{tabular}{@{}l>{\raggedright\arraybackslash}p{0.20\linewidth}>{\raggedright\arraybackslash}p{0.48\linewidth}@{}}
\toprule
\textbf{Period} & \textbf{Focus} & \textbf{Key milestones} \\
\midrule
\multicolumn{3}{@{}l}{\textit{Phase~I --- standalone IAPWS-IF97 property library}}\\
2025-01 & Forward equations & Regions 1--5 Gibbs/Helmholtz forward properties, B23 boundary and saturation line; dimensioned with \texttt{uom}, verified vs.\ IST sets A--C \\
2025-01/04 & Backward flashes & $(p,h)$, $(p,T)$, $(p,s)$ and $(h,s)$ inversions across the IF97 surface \\
2025-02 & Transport properties & Dynamic viscosity (2008), thermal conductivity (2011), surface tension and dielectric constant \\
2026-03/06 & Choked-flow scaffolding & Nozzle and converging--diverging solvers; Marviken blowdown harness; Moody and Zaloudek datasets digitised. In-dome solver passing but near-bubble-point artifact unresolved at migration; validation completed in Phase~II \\
\addlinespace[2pt]
\multicolumn{3}{@{}l}{\textit{Phase~II --- migrated into \texttt{outram-park-backend}}}\\
2026-06 & Migration & Vendored into the workspace; \texttt{uom}/\texttt{ndarray}/\texttt{egui} upgrades; BLAS decoupled \\
2026-06 & Choked-flow completion & Near-bubble-point artifact resolved; unified HEM dispatcher; Moody isobars pass (v0.2.1) \\
2026-07 & Array solver & \texttt{TampinesSteamArray} 1-D PIMPLE pipe solver and \texttt{tampines} framework crate; steam-generator integration \\
2026-07 & Blowdown V\&V & Edwards--O'Brien blowdown (v0.2.3); \texttt{HybridAllMach} solver stabilised (v0.2.4) \\
\bottomrule
\end{tabular}
\end{table}

As seen in Table~\ref{tab:tampines-history} and \ref{tab:tampines-summary},
Phase I concerned Tampines Steam tables as a standalone IAPWS-IF97 
library. This was meant for future development of Tampines and for the 
steam cycle simulation in an FHR Educational Simulator.

At that time, no known Rust library fully provided IAPWS-IF97 equations 
with (p,T), (p,s), (p,h) and (h,s) flashes. So this was a first, unpublished 
work.
At that time, all of the steam tables were copied from
\cite{wagner2008international}. Each flash was tested against the entire 
steam table. Only for (h,s) flash, there were parts in the saturation dome that 
could not be tested due to the limitations of the (h,s) equations. Those regimes were  
initially left as to-do items. But as of today, they fall back
to an iterative (p,s) equation to solve the parts where the (h,s) equation does not
apply in the saturation dome. Table~\ref{tab:tampines-flash-limits} summarises the
per-flash limitations found across the whole steam table.

\begin{table}[ht]
\centering
\caption{AI-generated summary of the property-flash limitations found when each
backward flash was tested against the full IAPWS-IF97 steam table of
\cite{wagner2008international} (triple point to 1000~bar, Regions~1--5).
Choked-flow / two-phase critical-flow limitations are excluded here and treated
separately. This table was generated with AI assistance from a survey of the
\texttt{tampines-steam-tables} test suite and source comments; the author
reviewed and verified its contents.}
\label{tab:tampines-flash-limits}
\footnotesize
\setlength{\tabcolsep}{4pt}
\begin{tabular}{@{}l>{\raggedright\arraybackslash}p{0.17\linewidth}>{\raggedright\arraybackslash}p{0.37\linewidth}>{\raggedright\arraybackslash}p{0.20\linewidth}@{}}
\toprule
\textbf{Flash} & \textbf{Region / condition} & \textbf{Limitation} & \textbf{Current handling} \\
\midrule
$(p,T)$ & Saturation dome (Region~4) & $(p,T)$ is non-unique inside the dome, so mixture $h,s,c_p,c_v,v,w$ cannot be resolved without a quality input & Not attempted (expected degeneracy) \\
\addlinespace[2pt]
$(p,T)$ & Single phase, 240--1000~bar & High-pressure single-phase table not implemented & Deferred to next major version \\
\addlinespace[2pt]
$(p,h)$ & Single phase, 1000~bar & Flash goes out of bounds at 1000~bar & Accepted Caveat, near max pressure \\
\addlinespace[2pt]
$(p,s)$ & Single phase, 240--1000~bar & High-pressure single-phase table not implemented & Deferred to next major version \\
\addlinespace[2pt]
$(h,s)$ & Triple-point pressure (0.006112127~bar) & $(h,s)$ equations cannot resolve triple-point liquid/vapour accurately & Accepted Caveat \\
\addlinespace[2pt]
$(h,s)$ & Single phase, 1000~bar & Flash goes out of bounds at 1000~bar & Accepted Caveat, near max pressure \\
\addlinespace[2pt]
$(h,s)$ & Saturation dome & $(h,s)$ backward equations invalid over parts of the dome & Falls back to iterative $(p,s)$ solve \\
\addlinespace[2pt]
$(h,s)$ & Near critical point & $T$ off by up to 5~$^\circ$C, $v$ up to 10\%, $h_{fg}$ up to 5\% & Accepted accuracy caveat \\
\addlinespace[2pt]
$(h,s)$ & Low pressure & Pressure back-equation error $\sim$20\% at 0.1--1~bar, falling to $\sim$2\% by 10--20~bar; dew point can fail (e.g.\ at 8~bar) & Accepted accuracy caveat \\
\bottomrule
\end{tabular}
\end{table}

For those, the flash equations agree well with those values in 
the table for areas outside the saturation dome. 
Of these, the (h,s) flash seems to be the most problematic around certain regions such 
as triple-point and critical point. Some parts of $V\&V$ are left out for the 
high pressure regions, but these are not particularly important for LWR-type 
scenarios or steam turbines, where maximum pressure is almost 150 bar for 
PWRs and 70-75 bar for BWRs. Hence, these are left out, as full coverage is not necessary here. 
This can be addressed in future work. SCWR conditions, which involve pressures
of around 22~MPa and above \cite{wu2022scwr}, cannot yet be handled.
Moreover, Singapore is not currently surveying SCWR at the moment. This is of little
concern.

For completeness, TAMPINES eventually had to implement 
compressible-flow solvers. From basic thermodynamics 
references such as Cengel \cite{cengel2002thermodynamics}, compressible-flow 
must include the critical flow phenomena. Therefore, choked flow was added 
and there were plans to do blowdown tests with V\&V using Zaloudek, Moody 
and Marviken tests. However, without the critical-flow solver machinery,
these were only scaffolds/harnesses, not implemented at all.

For critical flow, initial placeholder 
values were given for the speed of sound, compressibility, etc., so that it 
linearly interpolates against the saturated-liquid and saturated-vapour 
values. This was not accurate, but it was deferred to future work.
It will be addressed if necessary. Note that such values are linearly 
interpolated unless otherwise stated.

At this point, minimal AI was used, except to check the speed of sound, 
which was critical for choked flow. NUS-AI initially 
suggested that the author's linear interpolation was incorrect, and provided a formula 
for Wood--Wallis speed of sound (frozen speed of sound).

Of course, on further investigation, there were many kinds of speed of sound 
leading to different levels of choked flow in the two-phase dome.
This was very troublesome when surveying the NRC literature.

Moreover, no algorithm was found in the open-source domain 
to the best of the author's knowledge. Most of the codes which could do 
this were closed source. After initial debugging, the authors found that 
two-phase choked flow was difficult. 

However with move to agentic coding, the debugging process was sped up,
though it still needed plenty of human intervention.

With agentic coding, however, CoolProp was also translated to Rust (not
covered in this article, for length) which provided Helmholtz equations 
for both steam and helium. 

Of course, reproducing a steam table is not novel, and the distinction from
CoolProp should be stated carefully rather than overclaimed. CoolProp does
accept an $(h,s)$ input pair through its generic flash routine, so what
tampines-steam-tables adds there is a direct IAPWS-IF97 backward $(h,s)$
formulation rather than a capability CoolProp lacks outright.

The substantive difference is critical flow. For nuclear applications,
two-phase critical flow is important, and to the best of the authors'
knowledge --- based on a survey of CoolProp, ThermoSysPro and DWSIM, assisted
by agentic search of their source --- no open-source thermophysical property
library ships a verified HEM critical-flow solver. That survey is not
exhaustive, and the claim should be read as such. Subject to it,
tampines-steam-tables appears to be the first open-source steam-table library
with critical-flow thermodynamics and solvers embedded in it and verified
against published charts.

\subsubsection{Steady State homogeneous-equilibrium-model for Two Phase 
Choked Flow}

Initially, to break up the problem, the authors required understanding of 
the flash equations and their limitations, and to structure a plan towards 
obtaining a critical-flow solver. In doing so, the authors would know how best 
to prompt Claude Code to solve the problem. Without this guidance, Claude Code 
tended to go in circles, wasting tokens implementing the wrong solutions.

In any critical flow, we go from stagnation (p,h) to arrive at choked flow 
(p,h) using an isentropic process, wherein we normally use an iterative (p,s) 
flash to find the speed of sound. Now, due to the kink in the speed of sound 
at the bubble point, a specialised dispatch algorithm is required in 
order to find the choked flow as the mass flux was not unimodal and had 
two local maxima, usually one before the bubble point and one after.

It was important to know which local maxima to take.

The best approach to break down this problem was to first solve the inverse 
problem. Given choked flow conditions, what would the stagnation conditions 
be? 

A deliberate decision to scope down the choked flow to HEM. Additionally, we did not 
wish to address the other formulations, as HEM is the most basic. The remaining cases can be handled 
later.

This was the simpler problem to solve as it was not iterative. For this,
it was convenient to use data from Saha et al. \cite{saha1978review} 
especially since it is from the NRC, a gold-standard source. Moreover, it had graphs 
where data was in terms of choked flow conditions rather than stagnation 
conditions. This made it ideal for the choked flow to stagnation condition 
graph.
Results are shown in Figures~\ref{fig:zaloudek-critical-flow}:

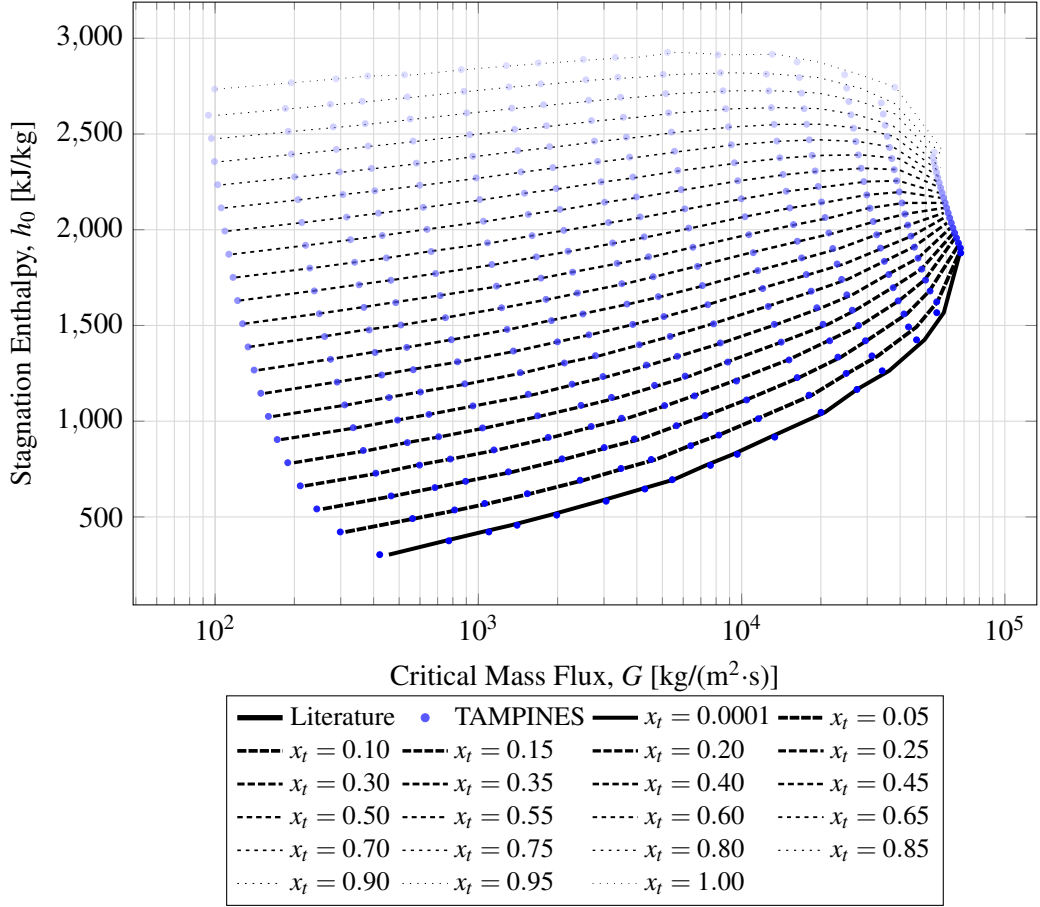
\begin{figure}[H]
  \centering
  \begin{tikzpicture}
  \begin{axis}[
      xlabel={Critical Mass Flux, $G$ [kg/(m$^2\cdot$s)]},
      ylabel={Stagnation Enthalpy, $h_0$ [kJ/kg]},
      xmode=log,
      legend style={at={(0.5,-0.15)}, anchor=north, font=\small, cells={anchor=west}},
      grid=both,
      grid style={gray!30},
      width=0.85\linewidth,
      height=0.6\linewidth,
      legend columns=4,
  ]

  \addlegendimage{thick, line width=2pt, black, mark=none}
  \addlegendentry{Literature}
  \addlegendimage{only marks, mark=*, mark size=1.5pt, color=blue!65}
  \addlegendentry{TAMPINES}

  \addplot[black, mark=none, line width=1.50pt] table[x=G_lit_kgm2s, y=h0_calc_kJkg, col sep=comma]{results_and_discussion/zaloudek_validation_xt_0001.csv};
  \addlegendentry{$x_t = 0.0001$}
  \addplot[color=blue!100, only marks, mark=*, mark size=1.1pt, opacity=1.00, forget plot] table[x=G_solver_kgm2s, y=h0_calc_kJkg, col sep=comma]{results_and_discussion/zaloudek_validation_xt_0001.csv};

  \addplot[black, mark=none, line width=1.44pt, dash pattern=on 3.62pt off 0.69pt] table[x=G_lit_kgm2s, y=h0_calc_kJkg, col sep=comma]{results_and_discussion/zaloudek_validation_xt_0p05.csv};
  \addlegendentry{$x_t = 0.05$}
  \addplot[color=blue!96, only marks, mark=*, mark size=1.1pt, opacity=0.97, forget plot] table[x=G_solver_kgm2s, y=h0_calc_kJkg, col sep=comma]{results_and_discussion/zaloudek_validation_xt_0p05.csv};

  \addplot[black, mark=none, line width=1.38pt, dash pattern=on 3.45pt off 0.78pt] table[x=G_lit_kgm2s, y=h0_calc_kJkg, col sep=comma]{results_and_discussion/zaloudek_validation_xt_0p10.csv};
  \addlegendentry{$x_t = 0.10$}
  \addplot[color=blue!93, only marks, mark=*, mark size=1.1pt, opacity=0.95, forget plot] table[x=G_solver_kgm2s, y=h0_calc_kJkg, col sep=comma]{results_and_discussion/zaloudek_validation_xt_0p10.csv};

  \addplot[black, mark=none, line width=1.31pt, dash pattern=on 3.27pt off 0.87pt] table[x=G_lit_kgm2s, y=h0_calc_kJkg, col sep=comma]{results_and_discussion/zaloudek_validation_xt_0p15.csv};
  \addlegendentry{$x_t = 0.15$}
  \addplot[color=blue!90, only marks, mark=*, mark size=1.1pt, opacity=0.92, forget plot] table[x=G_solver_kgm2s, y=h0_calc_kJkg, col sep=comma]{results_and_discussion/zaloudek_validation_xt_0p15.csv};

  \addplot[black, mark=none, line width=1.25pt, dash pattern=on 3.10pt off 0.96pt] table[x=G_lit_kgm2s, y=h0_calc_kJkg, col sep=comma]{results_and_discussion/zaloudek_validation_xt_0p20.csv};
  \addlegendentry{$x_t = 0.20$}
  \addplot[color=blue!86, only marks, mark=*, mark size=1.1pt, opacity=0.89, forget plot] table[x=G_solver_kgm2s, y=h0_calc_kJkg, col sep=comma]{results_and_discussion/zaloudek_validation_xt_0p20.csv};

  \addplot[black, mark=none, line width=1.19pt, dash pattern=on 2.92pt off 1.05pt] table[x=G_lit_kgm2s, y=h0_calc_kJkg, col sep=comma]{results_and_discussion/zaloudek_validation_xt_0p25.csv};
  \addlegendentry{$x_t = 0.25$}
  \addplot[color=blue!82, only marks, mark=*, mark size=1.1pt, opacity=0.86, forget plot] table[x=G_solver_kgm2s, y=h0_calc_kJkg, col sep=comma]{results_and_discussion/zaloudek_validation_xt_0p25.csv};

  \addplot[black, mark=none, line width=1.12pt, dash pattern=on 2.75pt off 1.14pt] table[x=G_lit_kgm2s, y=h0_calc_kJkg, col sep=comma]{results_and_discussion/zaloudek_validation_xt_0p30.csv};
  \addlegendentry{$x_t = 0.30$}
  \addplot[color=blue!79, only marks, mark=*, mark size=1.1pt, opacity=0.83, forget plot] table[x=G_solver_kgm2s, y=h0_calc_kJkg, col sep=comma]{results_and_discussion/zaloudek_validation_xt_0p30.csv};

  \addplot[black, mark=none, line width=1.06pt, dash pattern=on 2.57pt off 1.23pt] table[x=G_lit_kgm2s, y=h0_calc_kJkg, col sep=comma]{results_and_discussion/zaloudek_validation_xt_0p35.csv};
  \addlegendentry{$x_t = 0.35$}
  \addplot[color=blue!76, only marks, mark=*, mark size=1.1pt, opacity=0.81, forget plot] table[x=G_solver_kgm2s, y=h0_calc_kJkg, col sep=comma]{results_and_discussion/zaloudek_validation_xt_0p35.csv};

  \addplot[black, mark=none, line width=1.00pt, dash pattern=on 2.40pt off 1.32pt] table[x=G_lit_kgm2s, y=h0_calc_kJkg, col sep=comma]{results_and_discussion/zaloudek_validation_xt_0p40.csv};
  \addlegendentry{$x_t = 0.40$}
  \addplot[color=blue!72, only marks, mark=*, mark size=1.1pt, opacity=0.78, forget plot] table[x=G_solver_kgm2s, y=h0_calc_kJkg, col sep=comma]{results_and_discussion/zaloudek_validation_xt_0p40.csv};

  \addplot[black, mark=none, line width=0.94pt, dash pattern=on 2.23pt off 1.41pt] table[x=G_lit_kgm2s, y=h0_calc_kJkg, col sep=comma]{results_and_discussion/zaloudek_validation_xt_0p45.csv};
  \addlegendentry{$x_t = 0.45$}
  \addplot[color=blue!68, only marks, mark=*, mark size=1.1pt, opacity=0.75, forget plot] table[x=G_solver_kgm2s, y=h0_calc_kJkg, col sep=comma]{results_and_discussion/zaloudek_validation_xt_0p45.csv};

  \addplot[black, mark=none, line width=0.88pt, dash pattern=on 2.05pt off 1.50pt] table[x=G_lit_kgm2s, y=h0_calc_kJkg, col sep=comma]{results_and_discussion/zaloudek_validation_xt_0p50.csv};
  \addlegendentry{$x_t = 0.50$}
  \addplot[color=blue!65, only marks, mark=*, mark size=1.1pt, opacity=0.73, forget plot] table[x=G_solver_kgm2s, y=h0_calc_kJkg, col sep=comma]{results_and_discussion/zaloudek_validation_xt_0p50.csv};

  \addplot[black, mark=none, line width=0.81pt, dash pattern=on 1.87pt off 1.59pt] table[x=G_lit_kgm2s, y=h0_calc_kJkg, col sep=comma]{results_and_discussion/zaloudek_validation_xt_0p55.csv};
  \addlegendentry{$x_t = 0.55$}
  \addplot[color=blue!62, only marks, mark=*, mark size=1.1pt, opacity=0.70, forget plot] table[x=G_solver_kgm2s, y=h0_calc_kJkg, col sep=comma]{results_and_discussion/zaloudek_validation_xt_0p55.csv};

  \addplot[black, mark=none, line width=0.75pt, dash pattern=on 1.70pt off 1.68pt] table[x=G_lit_kgm2s, y=h0_calc_kJkg, col sep=comma]{results_and_discussion/zaloudek_validation_xt_0p60.csv};
  \addlegendentry{$x_t = 0.60$}
  \addplot[color=blue!58, only marks, mark=*, mark size=1.1pt, opacity=0.67, forget plot] table[x=G_solver_kgm2s, y=h0_calc_kJkg, col sep=comma]{results_and_discussion/zaloudek_validation_xt_0p60.csv};

  \addplot[black, mark=none, line width=0.69pt, dash pattern=on 1.52pt off 1.77pt] table[x=G_lit_kgm2s, y=h0_calc_kJkg, col sep=comma]{results_and_discussion/zaloudek_validation_xt_0p65.csv};
  \addlegendentry{$x_t = 0.65$}
  \addplot[color=blue!54, only marks, mark=*, mark size=1.1pt, opacity=0.64, forget plot] table[x=G_solver_kgm2s, y=h0_calc_kJkg, col sep=comma]{results_and_discussion/zaloudek_validation_xt_0p65.csv};

  \addplot[black, mark=none, line width=0.62pt, dash pattern=on 1.35pt off 1.86pt] table[x=G_lit_kgm2s, y=h0_calc_kJkg, col sep=comma]{results_and_discussion/zaloudek_validation_xt_0p70.csv};
  \addlegendentry{$x_t = 0.70$}
  \addplot[color=blue!51, only marks, mark=*, mark size=1.1pt, opacity=0.61, forget plot] table[x=G_solver_kgm2s, y=h0_calc_kJkg, col sep=comma]{results_and_discussion/zaloudek_validation_xt_0p70.csv};

  \addplot[black, mark=none, line width=0.56pt, dash pattern=on 1.18pt off 1.95pt] table[x=G_lit_kgm2s, y=h0_calc_kJkg, col sep=comma]{results_and_discussion/zaloudek_validation_xt_0p75.csv};
  \addlegendentry{$x_t = 0.75$}
  \addplot[color=blue!48, only marks, mark=*, mark size=1.1pt, opacity=0.59, forget plot] table[x=G_solver_kgm2s, y=h0_calc_kJkg, col sep=comma]{results_and_discussion/zaloudek_validation_xt_0p75.csv};

  \addplot[black, mark=none, line width=0.50pt, dash pattern=on 1.00pt off 2.04pt] table[x=G_lit_kgm2s, y=h0_calc_kJkg, col sep=comma]{results_and_discussion/zaloudek_validation_xt_0p80.csv};
  \addlegendentry{$x_t = 0.80$}
  \addplot[color=blue!44, only marks, mark=*, mark size=1.1pt, opacity=0.56, forget plot] table[x=G_solver_kgm2s, y=h0_calc_kJkg, col sep=comma]{results_and_discussion/zaloudek_validation_xt_0p80.csv};

  \addplot[black, mark=none, line width=0.44pt, dash pattern=on 0.83pt off 2.13pt] table[x=G_lit_kgm2s, y=h0_calc_kJkg, col sep=comma]{results_and_discussion/zaloudek_validation_xt_0p85.csv};
  \addlegendentry{$x_t = 0.85$}
  \addplot[color=blue!40, only marks, mark=*, mark size=1.1pt, opacity=0.53, forget plot] table[x=G_solver_kgm2s, y=h0_calc_kJkg, col sep=comma]{results_and_discussion/zaloudek_validation_xt_0p85.csv};

  \addplot[black, mark=none, line width=0.38pt, dash pattern=on 0.65pt off 2.22pt] table[x=G_lit_kgm2s, y=h0_calc_kJkg, col sep=comma]{results_and_discussion/zaloudek_validation_xt_0p90.csv};
  \addlegendentry{$x_t = 0.90$}
  \addplot[color=blue!37, only marks, mark=*, mark size=1.1pt, opacity=0.51, forget plot] table[x=G_solver_kgm2s, y=h0_calc_kJkg, col sep=comma]{results_and_discussion/zaloudek_validation_xt_0p90.csv};

  \addplot[black, mark=none, line width=0.31pt, dash pattern=on 0.48pt off 2.31pt] table[x=G_lit_kgm2s, y=h0_calc_kJkg, col sep=comma]{results_and_discussion/zaloudek_validation_xt_0p95.csv};
  \addlegendentry{$x_t = 0.95$}
  \addplot[color=blue!34, only marks, mark=*, mark size=1.1pt, opacity=0.48, forget plot] table[x=G_solver_kgm2s, y=h0_calc_kJkg, col sep=comma]{results_and_discussion/zaloudek_validation_xt_0p95.csv};

  \addplot[black, mark=none, line width=0.25pt, dash pattern=on 0.30pt off 2.40pt] table[x=G_lit_kgm2s, y=h0_calc_kJkg, col sep=comma]{results_and_discussion/zaloudek_validation_xt_1p00.csv};
  \addlegendentry{$x_t = 1.00$}
  \addplot[color=blue!30, only marks, mark=*, mark size=1.1pt, opacity=0.45, forget plot] table[x=G_solver_kgm2s, y=h0_calc_kJkg, col sep=comma]{results_and_discussion/zaloudek_validation_xt_1p00.csv};

  \end{axis}
  \end{tikzpicture}
  \caption{Zaloudek critical-flow validation: stagnation enthalpy vs.\ critical mass flux (log scale) for each quality curve.}
  \label{fig:zaloudek-critical-flow}
\end{figure}

With that done, we then tackle the forward iteration problem since the 
backward function is solved.

For this, Claude Code's Opus models were better, though they consume more tokens.
The reasoning and debugging in Opus 4.8 were far superior to that of Sonnet 
4.5 used during June-early July 2026. This enabled it to be better at debugging. 

It was used to check the shape of the mass-flux curves before and after the
bubble point. Claude Code proposed and implemented the Golden Section algorithm 
which was then checked against literature \cite{price2012golden} and accepted.

The idea was to apply golden-section search before and after the bubble point

However, considerable debugging was required around the bubble point, because
the kink is genuine and throws numerical solvers off. 
Depending on how deeply subcooled the solvers were, slightly differing 
algorithms and adjustments were needed. Claude Opus helped to debug 
this by suggesting solutions and performing iterative debugging. 

Claude Code was monitored from the author's phone, observing what Opus was
doing; for the most part it was left to fix issues automatically. It is
difficult to recall in detail which fixes Opus made and which the author
contributed. 

Nevertheless, Moody's graphs were digitised into CSV, and Claude was asked to
confirm that the graph was reproduced to within reasonable bounds.
This resulted in Figure~\ref{fig:moody-critical-flow}:

\begin{figure}[H]
  \centering
  \begin{tikzpicture}
  \begin{axis}[
      xlabel={Stagnation Enthalpy, $h_0$ [kJ/kg]},
      ylabel={Critical Mass Flux, $G$ [kg/(m$^2\cdot$s)]},
      ymode=log,
      legend style={at={(0.5,-0.15)}, anchor=north, font=\small, cells={anchor=west}},
      grid=both,
      grid style={gray!30},
      width=0.85\linewidth,
      height=0.6\linewidth,
      legend columns=3,
  ]

  \addlegendimage{thick, line width=2pt, black, mark=none}
  \addlegendentry{Literature}
  \addlegendimage{only marks, mark=*, mark size=1.5pt, color=blue!65}
  \addlegendentry{TAMPINES}

  \addplot[black, mark=none, line width=1.50pt] table[x=h0_kJkg, y=G_lit_kgm2s, col sep=comma]{results_and_discussion/moody_validation_pref_0p25.csv};
  \addlegendentry{$p_{\mathrm{ref}} = 0.25$ MPa}
  \addplot[color=blue!100, only marks, mark=*, mark size=1.1pt, opacity=1.00, forget plot] table[x=h0_kJkg, y=G_solver_kgm2s, col sep=comma]{results_and_discussion/moody_validation_pref_0p25.csv};

  \addplot[black, mark=none, line width=1.40pt, dash pattern=on 3.51pt off 0.75pt] table[x=h0_kJkg, y=G_lit_kgm2s, col sep=comma]{results_and_discussion/moody_validation_pref_0p50.csv};
  \addlegendentry{$p_{\mathrm{ref}} = 0.50$ MPa}
  \addplot[color=blue!94, only marks, mark=*, mark size=1.1pt, opacity=0.95, forget plot] table[x=h0_kJkg, y=G_solver_kgm2s, col sep=comma]{results_and_discussion/moody_validation_pref_0p50.csv};

  \addplot[black, mark=none, line width=1.29pt, dash pattern=on 3.22pt off 0.90pt] table[x=h0_kJkg, y=G_lit_kgm2s, col sep=comma]{results_and_discussion/moody_validation_pref_1p00.csv};
  \addlegendentry{$p_{\mathrm{ref}} = 1.00$ MPa}
  \addplot[color=blue!88, only marks, mark=*, mark size=1.1pt, opacity=0.91, forget plot] table[x=h0_kJkg, y=G_solver_kgm2s, col sep=comma]{results_and_discussion/moody_validation_pref_1p00.csv};

  \addplot[black, mark=none, line width=1.19pt, dash pattern=on 2.92pt off 1.05pt] table[x=h0_kJkg, y=G_lit_kgm2s, col sep=comma]{results_and_discussion/moody_validation_pref_2p00.csv};
  \addlegendentry{$p_{\mathrm{ref}} = 2.00$ MPa}
  \addplot[color=blue!82, only marks, mark=*, mark size=1.1pt, opacity=0.86, forget plot] table[x=h0_kJkg, y=G_solver_kgm2s, col sep=comma]{results_and_discussion/moody_validation_pref_2p00.csv};

  \addplot[black, mark=none, line width=1.08pt, dash pattern=on 2.63pt off 1.20pt] table[x=h0_kJkg, y=G_lit_kgm2s, col sep=comma]{results_and_discussion/moody_validation_pref_4p00.csv};
  \addlegendentry{$p_{\mathrm{ref}} = 4.00$ MPa}
  \addplot[color=blue!77, only marks, mark=*, mark size=1.1pt, opacity=0.82, forget plot] table[x=h0_kJkg, y=G_solver_kgm2s, col sep=comma]{results_and_discussion/moody_validation_pref_4p00.csv};

  \addplot[black, mark=none, line width=0.98pt, dash pattern=on 2.34pt off 1.35pt] table[x=h0_kJkg, y=G_lit_kgm2s, col sep=comma]{results_and_discussion/moody_validation_pref_6p00.csv};
  \addlegendentry{$p_{\mathrm{ref}} = 6.00$ MPa}
  \addplot[color=blue!71, only marks, mark=*, mark size=1.1pt, opacity=0.77, forget plot] table[x=h0_kJkg, y=G_solver_kgm2s, col sep=comma]{results_and_discussion/moody_validation_pref_6p00.csv};

  \addplot[black, mark=none, line width=0.88pt, dash pattern=on 2.05pt off 1.50pt] table[x=h0_kJkg, y=G_lit_kgm2s, col sep=comma]{results_and_discussion/moody_validation_pref_8p00.csv};
  \addlegendentry{$p_{\mathrm{ref}} = 8.00$ MPa}
  \addplot[color=blue!65, only marks, mark=*, mark size=1.1pt, opacity=0.73, forget plot] table[x=h0_kJkg, y=G_solver_kgm2s, col sep=comma]{results_and_discussion/moody_validation_pref_8p00.csv};

  \addplot[black, mark=none, line width=0.77pt, dash pattern=on 1.76pt off 1.65pt] table[x=h0_kJkg, y=G_lit_kgm2s, col sep=comma]{results_and_discussion/moody_validation_pref_10p00.csv};
  \addlegendentry{$p_{\mathrm{ref}} = 10.00$ MPa}
  \addplot[color=blue!59, only marks, mark=*, mark size=1.1pt, opacity=0.68, forget plot] table[x=h0_kJkg, y=G_solver_kgm2s, col sep=comma]{results_and_discussion/moody_validation_pref_10p00.csv};

  \addplot[black, mark=none, line width=0.67pt, dash pattern=on 1.47pt off 1.80pt] table[x=h0_kJkg, y=G_lit_kgm2s, col sep=comma]{results_and_discussion/moody_validation_pref_12p00.csv};
  \addlegendentry{$p_{\mathrm{ref}} = 12.00$ MPa}
  \addplot[color=blue!53, only marks, mark=*, mark size=1.1pt, opacity=0.63, forget plot] table[x=h0_kJkg, y=G_solver_kgm2s, col sep=comma]{results_and_discussion/moody_validation_pref_12p00.csv};

  \addplot[black, mark=none, line width=0.56pt, dash pattern=on 1.18pt off 1.95pt] table[x=h0_kJkg, y=G_lit_kgm2s, col sep=comma]{results_and_discussion/moody_validation_pref_14p00.csv};
  \addlegendentry{$p_{\mathrm{ref}} = 14.00$ MPa}
  \addplot[color=blue!48, only marks, mark=*, mark size=1.1pt, opacity=0.59, forget plot] table[x=h0_kJkg, y=G_solver_kgm2s, col sep=comma]{results_and_discussion/moody_validation_pref_14p00.csv};

  \addplot[black, mark=none, line width=0.46pt, dash pattern=on 0.88pt off 2.10pt] table[x=h0_kJkg, y=G_lit_kgm2s, col sep=comma]{results_and_discussion/moody_validation_pref_16p00.csv};
  \addlegendentry{$p_{\mathrm{ref}} = 16.00$ MPa}
  \addplot[color=blue!42, only marks, mark=*, mark size=1.1pt, opacity=0.54, forget plot] table[x=h0_kJkg, y=G_solver_kgm2s, col sep=comma]{results_and_discussion/moody_validation_pref_16p00.csv};

  \addplot[black, mark=none, line width=0.35pt, dash pattern=on 0.59pt off 2.25pt] table[x=h0_kJkg, y=G_lit_kgm2s, col sep=comma]{results_and_discussion/moody_validation_pref_20p00.csv};
  \addlegendentry{$p_{\mathrm{ref}} = 20.00$ MPa}
  \addplot[color=blue!36, only marks, mark=*, mark size=1.1pt, opacity=0.50, forget plot] table[x=h0_kJkg, y=G_solver_kgm2s, col sep=comma]{results_and_discussion/moody_validation_pref_20p00.csv};

  \addplot[black, mark=none, line width=0.25pt, dash pattern=on 0.30pt off 2.40pt] table[x=h0_kJkg, y=G_lit_kgm2s, col sep=comma]{results_and_discussion/moody_validation_pref_30p00.csv};
  \addlegendentry{$p_{\mathrm{ref}} = 30.00$ MPa}
  \addplot[color=blue!30, only marks, mark=*, mark size=1.1pt, opacity=0.45, forget plot] table[x=h0_kJkg, y=G_solver_kgm2s, col sep=comma]{results_and_discussion/moody_validation_pref_30p00.csv};

  \addplot[red, mark=none, line width=2pt] table[x expr=\thisrow{x}*232.6, y expr=\thisrow{y}*4882, col sep=comma]{results_and_discussion/moody_bubble_point_curve.csv};
  \addlegendentry{Bubble-point locus (Moody)}

  \end{axis}
  \end{tikzpicture}
  \caption{Moody critical-flow validation: critical mass flux (log scale) vs.\ stagnation enthalpy for each isobar pressure.}
  \label{fig:moody-critical-flow}
\end{figure}
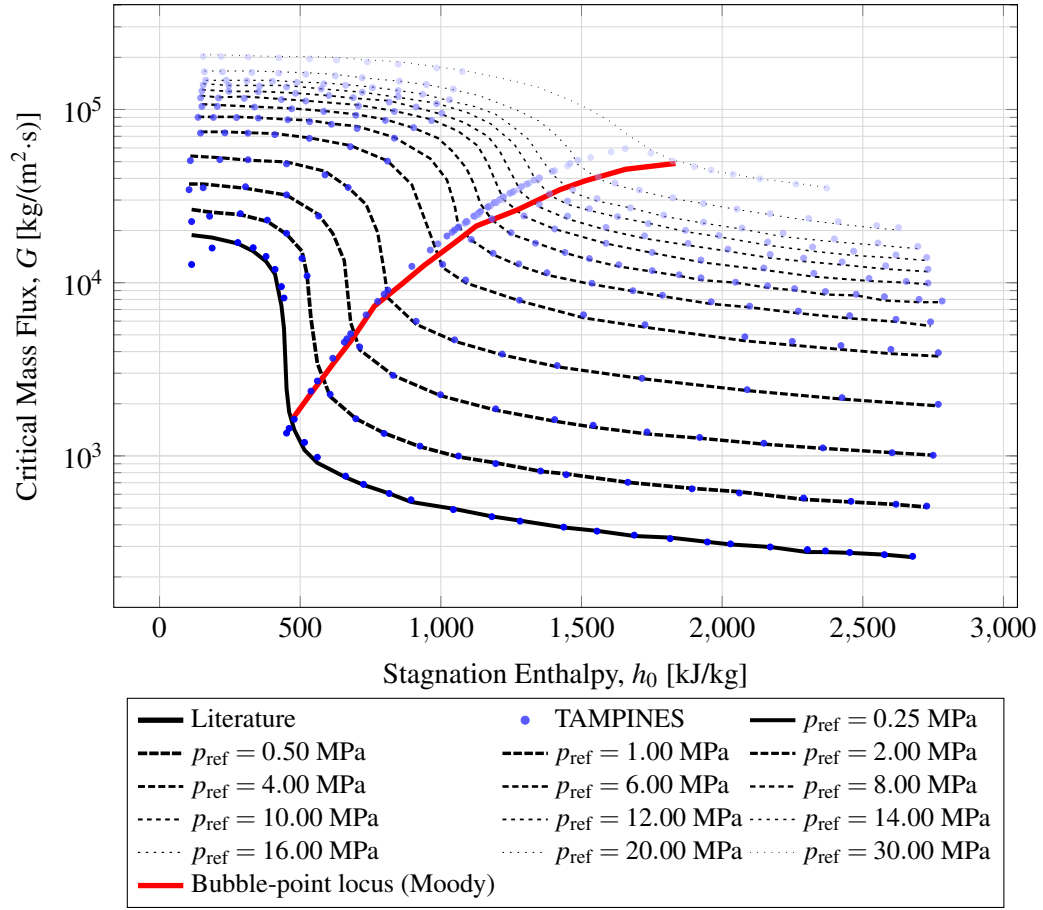

The agreement is acceptable except for the kink region at the bubble point. 
This is difficult because of the discontinuity. 

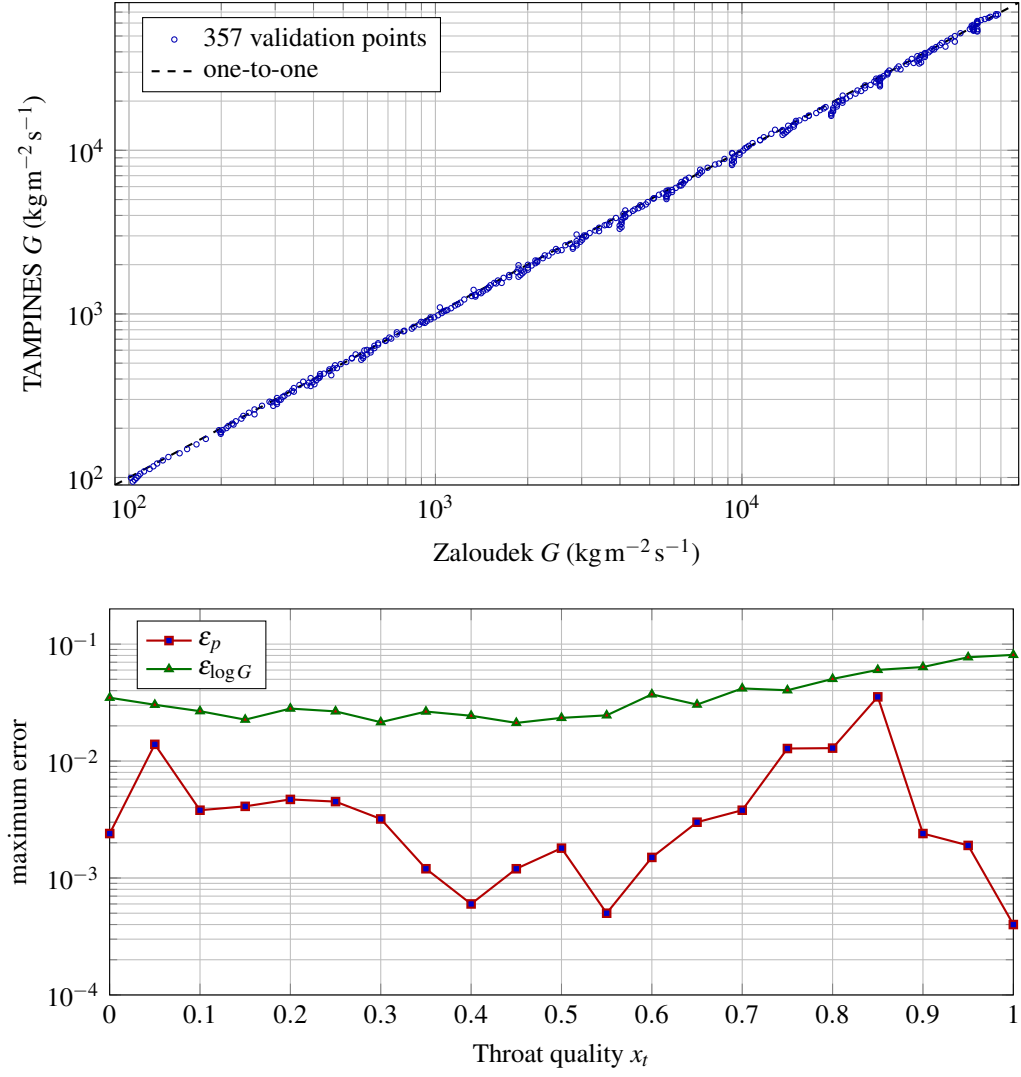
\begin{figure}[H]
  \centering
  \begin{tikzpicture}
    \begin{axis}[
        name=zaloudekparity,
        width=0.85\linewidth,
        height=0.50\linewidth,
        xlabel={Zaloudek $G$ (\si{kg.m^{-2}.s^{-1}})},
        ylabel={TAMPINES $G$ (\si{kg.m^{-2}.s^{-1}})},
        xmode=log,
        ymode=log,
        xmin=90, xmax=80000,
        ymin=90, ymax=80000,
        grid=both,
        legend pos=north west,
        legend cell align=left,
        tick label style={font=\small},
        label style={font=\small},
        legend style={font=\small},
    ]
      \addplot+[only marks, mark=o, mark size=1.0pt, color=blue!70!black]
        table[x=G_lit_kgm2s, y=G_solver_kgm2s, col sep=comma]
        {results_and_discussion/zaloudek_validation_full.csv};
      \addlegendentry{357 validation points}

      \addplot+[thick, mark=none, color=black, dashed]
        coordinates {(90,90) (80000,80000)};
      \addlegendentry{one-to-one}
    \end{axis}
  \end{tikzpicture}

  \vspace{1em}

  \begin{tikzpicture}
    \begin{axis}[
        width=0.85\linewidth,
        height=0.42\linewidth,
        xlabel={Throat quality $x_t$},
        ylabel={maximum error},
        xmin=0, xmax=1,
        ymin=1e-4, ymax=2e-1,
        ymode=log,
        grid=both,
        legend pos=north west,
        legend cell align=left,
        tick label style={font=\small},
        label style={font=\small},
        legend style={font=\small},
    ]
      \addplot+[thick, mark=square*, mark size=1.4pt, color=red!70!black]
        table[x=x_t, y=max_p_rel_err, col sep=comma]
        {results_and_discussion/zaloudek_validation_summary.csv};
      \addlegendentry{$\epsilon_p$}

      \addplot+[thick, mark=triangle*, mark size=1.6pt, color=green!45!black]
        table[x=x_t, y=max_G_log10_err, col sep=comma]
        {results_and_discussion/zaloudek_validation_summary.csv};
      \addlegendentry{$\epsilon_{\log G}$}
    \end{axis}
  \end{tikzpicture}
  \caption{Zaloudek critical-flow validation of the TAMPINES HEM dispatcher.
           Top: mass-flux parity over all extracted points. Bottom: maximum
           pressure and mass-flux errors by throat quality.}
  \label{fig:zaloudek-validation}
\end{figure}

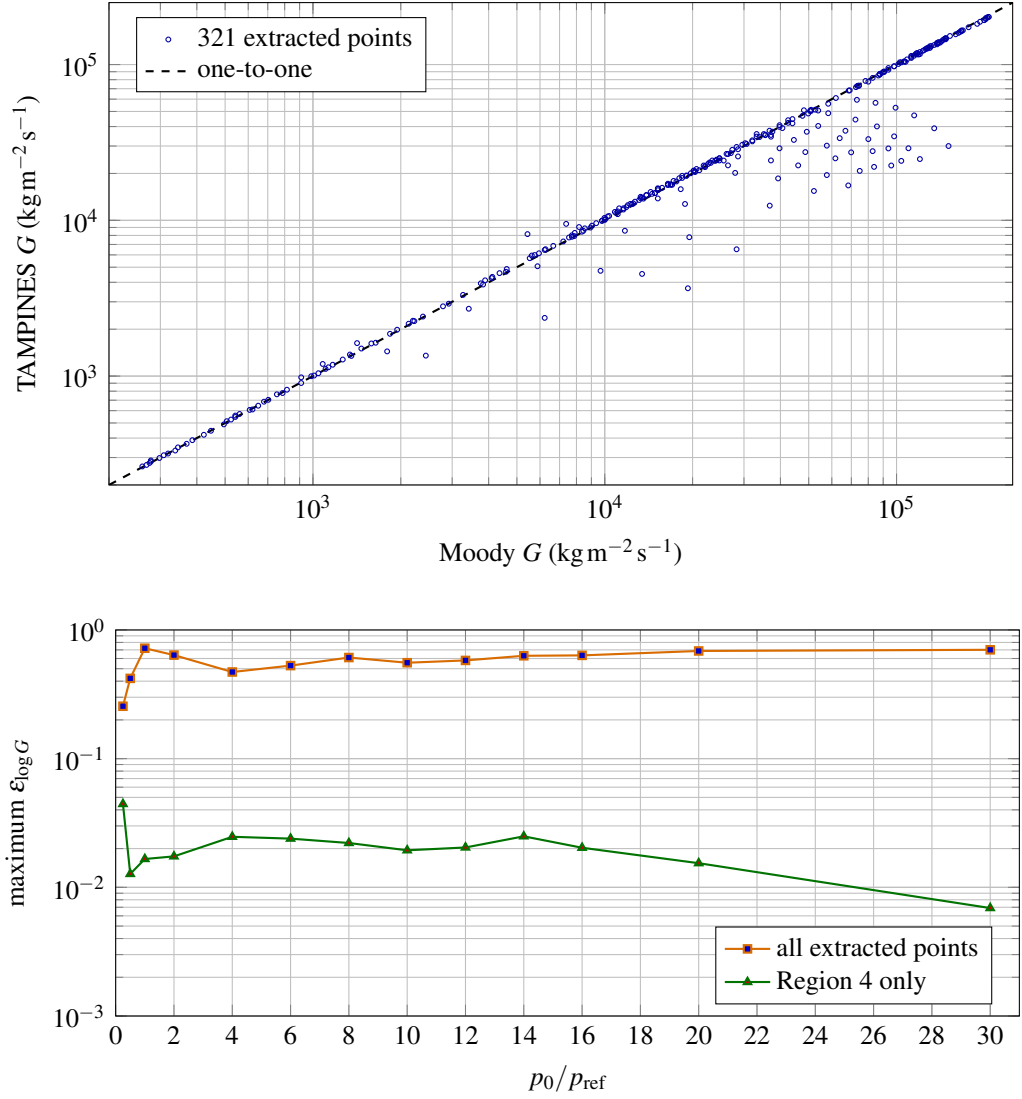
\begin{figure}[H]
  \centering
  \begin{tikzpicture}
    \begin{axis}[
        width=0.85\linewidth,
        height=0.50\linewidth,
        xlabel={Moody $G$ (\si{kg.m^{-2}.s^{-1}})},
        ylabel={TAMPINES $G$ (\si{kg.m^{-2}.s^{-1}})},
        xmode=log,
        ymode=log,
        xmin=200, xmax=250000,
        ymin=200, ymax=250000,
        grid=both,
        legend pos=north west,
        legend cell align=left,
        tick label style={font=\small},
        label style={font=\small},
        legend style={font=\small},
    ]
      \addplot+[only marks, mark=o, mark size=0.9pt, color=blue!65!black]
        table[x=G_lit_kgm2s, y=G_solver_kgm2s, col sep=comma]
        {results_and_discussion/moody_validation_full.csv};
      \addlegendentry{321 extracted points}

      \addplot+[thick, mark=none, color=black, dashed]
        coordinates {(200,200) (250000,250000)};
      \addlegendentry{one-to-one}
    \end{axis}
  \end{tikzpicture}

  \vspace{1em}

  \begin{tikzpicture}
    \begin{axis}[
        width=0.85\linewidth,
        height=0.42\linewidth,
        xlabel={$p_0/p_{\mathrm{ref}}$},
        ylabel={maximum $\epsilon_{\log G}$},
        xmin=0, xmax=31,
        ymin=1e-3, ymax=1,
        ymode=log,
        grid=both,
        legend pos=south east,
        legend cell align=left,
        tick label style={font=\small},
        label style={font=\small},
        legend style={font=\small},
    ]
      \addplot+[thick, mark=square*, mark size=1.4pt, color=orange!85!black]
        table[x=p0_over_pref, y=max_log10_err_all, col sep=comma]
        {results_and_discussion/moody_validation_summary.csv};
      \addlegendentry{all extracted points}

      \addplot+[thick, mark=triangle*, mark size=1.6pt, color=green!45!black]
        table[x=p0_over_pref, y=max_log10_err_region4, col sep=comma]
        {results_and_discussion/moody_validation_summary.csv};
      \addlegendentry{Region 4 only}
    \end{axis}
  \end{tikzpicture}
  \caption{Moody critical-flow validation of the TAMPINES HEM stagnation
           mass-flux routine. Top: mass-flux parity over all extracted
           isobar points. Bottom: maximum log-scale mass-flux error by
           stagnation-pressure isobar, comparing the full diagnostic dataset
           with the asserted in-dome Region 4 subset.}
  \label{fig:moody-validation}
\end{figure}

The errors are summarized in Table \ref{tab:zaloudek-validation-summary},
\ref{tab:moody-validation-summary}, Figure \ref{fig:zaloudek-validation},
and Figure \ref{fig:moody-validation}:

\begin{table}[H]
  \centering
  \caption{Summary of the Zaloudek critical-flow validation dataset for
           the TAMPINES HEM dispatcher, benchmarked against the Zaloudek
           curves digitised from Fig.~2 of Saha \cite{saha1978review}. The
           pressure error $\epsilon_p=|p_{\mathrm{crit,solver}}-
           p_{\mathrm{crit,lit}}|/\max(|p_{\mathrm{crit,solver}}|,
           |p_{\mathrm{crit,lit}}|)$ is the relative pressure difference as
           evaluated by the regression tests. The mass-flux columns report the
           \emph{absolute} log-scale error $\epsilon_{\log G}=
           |\log_{10}G_{\mathrm{solver}}-\log_{10}G_{\mathrm{lit}}|$ and its
           linear-scale equivalent $\epsilon_G=|G_{\mathrm{solver}}-
           G_{\mathrm{lit}}|/G_{\mathrm{lit}}=1-10^{-\epsilon_{\log G}}$;
           $\epsilon_G$ is a derived diagnostic and is not directly asserted.
           Note that the regression-test pass/fail gate for mass flux is a
           \emph{relative} tolerance on $\log_{10}G$ rather than the absolute
           $\epsilon_{\log G}$ tabulated here.}
  \label{tab:zaloudek-validation-summary}
  \scriptsize
  \resizebox{\textwidth}{!}{%
  \begin{filecontents*}{results_and_discussion/zaloudek_validation_summary.csv}
x_t,n_points,region_counts,max_p_rel_err,mean_p_rel_err,max_G_log10_err,mean_G_log10_err,max_G_rel_err,mean_G_rel_err
0.0001,17,Region1:16;Region3:1,0.0024,0.0012,0.0348,0.0187,0.0771,0.0430
0.05,17,Region1:3;Region3:1;Region4:13,0.0139,0.0051,0.0303,0.0113,0.0673,0.0255
0.10,17,Region1:1;Region3:2;Region4:14,0.0038,0.0017,0.0267,0.0092,0.0597,0.0208
0.15,17,Region3:2;Region4:15,0.0041,0.0014,0.0226,0.0088,0.0508,0.0201
0.20,17,Region3:2;Region4:15,0.0047,0.0009,0.0281,0.0121,0.0628,0.0274
0.25,17,Region3:1;Region4:16,0.0045,0.0006,0.0266,0.0104,0.0593,0.0235
0.30,17,Region3:1;Region4:16,0.0032,0.0006,0.0215,0.0084,0.0484,0.0192
0.35,17,Region3:1;Region4:16,0.0012,0.0004,0.0265,0.0091,0.0593,0.0208
0.40,17,Region3:1;Region4:16,0.0006,0.0002,0.0244,0.0083,0.0547,0.0189
0.45,17,Region3:1;Region4:16,0.0012,0.0004,0.0212,0.0074,0.0477,0.0171
0.50,17,Region3:1;Region4:16,0.0018,0.0003,0.0234,0.0078,0.0525,0.0180
0.55,17,Region3:1;Region4:16,0.0005,0.0003,0.0246,0.0079,0.0550,0.0180
0.60,17,Region3:1;Region4:16,0.0015,0.0003,0.0372,0.0102,0.0820,0.0231
0.65,17,Region3:2;Region4:15,0.0030,0.0003,0.0304,0.0104,0.0675,0.0234
0.70,17,Region3:2;Region4:15,0.0038,0.0003,0.0418,0.0151,0.0918,0.0340
0.75,17,Region3:2;Region4:15,0.0128,0.0008,0.0403,0.0173,0.0885,0.0388
0.80,17,Region3:2;Region4:15,0.0129,0.0008,0.0504,0.0254,0.1096,0.0566
0.85,17,Region2:1;Region3:2;Region4:14,0.0354,0.0021,0.0602,0.0283,0.1293,0.0626
0.90,17,Region2:2;Region3:2;Region4:13,0.0024,0.0002,0.0637,0.0366,0.1364,0.0805
0.95,17,Region2:6;Region3:1;Region4:10,0.0019,0.0002,0.0772,0.0421,0.1628,0.0918
1.00,17,Region2:16;Region3:1,0.0004,0.0001,0.0809,0.0331,0.1699,0.0718
\end{filecontents*}
\pgfplotstabletypeset[
    col sep=comma,
    columns={x_t,n_points,max_p_rel_err,mean_p_rel_err,max_G_log10_err,mean_G_log10_err,max_G_rel_err,mean_G_rel_err},
    columns/x_t/.style={column name={$x_t$}, fixed, precision=4, zerofill},
    columns/n_points/.style={column name={$N$}, int detect},
    columns/max_p_rel_err/.style={column name={max $\epsilon_p$}, fixed, precision=4, zerofill},
    columns/mean_p_rel_err/.style={column name={mean $\epsilon_p$}, fixed, precision=4, zerofill},
    columns/max_G_log10_err/.style={column name={max $\epsilon_{\log G}$}, fixed, precision=4, zerofill},
    columns/mean_G_log10_err/.style={column name={mean $\epsilon_{\log G}$}, fixed, precision=4, zerofill},
    columns/max_G_rel_err/.style={column name={max $\epsilon_G$}, fixed, precision=4, zerofill},
    columns/mean_G_rel_err/.style={column name={mean $\epsilon_G$}, fixed, precision=4, zerofill},
    every head row/.style={before row=\toprule, after row=\midrule},
    every last row/.style={after row=\bottomrule},
  ]{results_and_discussion/zaloudek_validation_summary.csv}%
  }
\end{table}

\begin{table}[H]
  \centering
  \caption{Summary of the Moody critical-flow validation dataset for the
           TAMPINES HEM stagnation mass-flux routine, benchmarked against the
           Moody maximum-discharge curve \cite{moody1975maximum}. The mass-flux
           error is the \emph{absolute} log-scale difference
           $\epsilon_{\log G}=|\log_{10}G_{\mathrm{solver}}-
           \log_{10}G_{\mathrm{lit}}|$, which is the metric asserted by the
           regression tests; $\epsilon_G=|G_{\mathrm{solver}}-
           G_{\mathrm{lit}}|/G_{\mathrm{lit}}=1-10^{-\epsilon_{\log G}}$ is its
           linear-scale equivalent, reported as a diagnostic. The ``all''
           columns include diagnostic Region 1 and Region 3 points that are not
           asserted, while the Region 4 columns isolate the in-dome two-phase
           points asserted at the standard Moody log-scale tolerance
           ($\epsilon_{\log G}\le 0.06$).}
  \label{tab:moody-validation-summary}
  \scriptsize
  \resizebox{\textwidth}{!}{%
  \begin{filecontents*}{results_and_discussion/moody_validation_summary.csv}
p0_over_pref,n_points,region_counts,max_log10_err_all,mean_log10_err_all,max_log10_err_region4,mean_log10_err_region4,max_G_rel_err_all,mean_G_rel_err_all,max_G_rel_err_region4,mean_G_rel_err_region4
0.25,32,Region1:11;Region4:21,0.2554,0.0364,0.0444,0.0074,0.5010,0.0810,0.1076,0.0174
0.50,24,Region1:9;Region4:15,0.4204,0.0310,0.0127,0.0043,0.6202,0.0552,0.0297,0.0099
1.00,21,Region1:9;Region4:12,0.7212,0.0817,0.0166,0.0062,0.8100,0.1171,0.0391,0.0145
2.00,18,Region1:10;Region4:8,0.6377,0.0751,0.0174,0.0078,0.7697,0.1143,0.0408,0.0182
4.00,18,Region1:8;Region4:10,0.4712,0.0381,0.0247,0.0182,0.6621,0.0647,0.0584,0.0428
6.00,26,Region1:13;Region4:13,0.5284,0.0451,0.0239,0.0102,0.7038,0.0740,0.0566,0.0239
8.00,30,Region1:14;Region4:16,0.6105,0.0605,0.0221,0.0091,0.7548,0.0940,0.0522,0.0213
10.00,23,Region1:13;Region4:10,0.5563,0.0641,0.0194,0.0079,0.7222,0.1021,0.0457,0.0184
12.00,25,Region1:14;Region4:11,0.5798,0.0655,0.0204,0.0155,0.7368,0.1056,0.0480,0.0363
14.00,25,Region1:14;Region4:11,0.6299,0.0684,0.0249,0.0138,0.7655,0.1051,0.0590,0.0324
16.00,27,Region1:17;Region4:10,0.6345,0.0763,0.0203,0.0112,0.7680,0.1149,0.0480,0.0263
20.00,28,Region1:18;Region4:10,0.6866,0.0894,0.0154,0.0082,0.7942,0.1301,0.0360,0.0192
30.00,24,Region1:15;Region3:4;Region4:5,0.7011,0.0965,0.0069,0.0044,0.8010,0.1430,0.0161,0.0101
\end{filecontents*}
\pgfplotstabletypeset[
    col sep=comma,
    columns={p0_over_pref,n_points,max_log10_err_all,mean_log10_err_all,max_log10_err_region4,mean_log10_err_region4,max_G_rel_err_region4},
    columns/p0_over_pref/.style={column name={$p_0/p_{\mathrm{ref}}$}, fixed, precision=2, zerofill},
    columns/n_points/.style={column name={$N$}, int detect},
    columns/max_log10_err_all/.style={column name={max $\epsilon_{\log G}$, all}, fixed, precision=4, zerofill},
    columns/mean_log10_err_all/.style={column name={mean $\epsilon_{\log G}$, all}, fixed, precision=4, zerofill},
    columns/max_log10_err_region4/.style={column name={max $\epsilon_{\log G}$, R4}, fixed, precision=4, zerofill},
    columns/mean_log10_err_region4/.style={column name={mean $\epsilon_{\log G}$, R4}, fixed, precision=4, zerofill},
    columns/max_G_rel_err_region4/.style={column name={max $\epsilon_G$, R4}, fixed, precision=4, zerofill},
    every head row/.style={before row=\toprule, after row=\midrule},
    every last row/.style={after row=\bottomrule},
  ]{results_and_discussion/moody_validation_summary.csv}%
  }
\end{table}

From these, one can see that the data matches well. Error was 
quantified using log scale, because digitisation was in log scale.

Also, it is interesting to note that the HEM solver managed to reproduce 
the shape of the bubble point curve due to the kink of 
Moody's critical-flow diagram under Figure 1 of Saha \cite{saha1978review}.
This shows that the trends produced match those in the literature.
This is also shown in Figure \ref{fig:moody-critical-flow} which shows that 
the solver can roughly produce Moody's bubble point locus at least for lower 
quality steam.

Overall, the trend is broadly there. Although there are some points 
where the solver fails to produce the critical mass flux, on the whole 
it appears acceptable for this attempt. Future refinements can be made. 

The main issue now is whether this will impact critical flow in a benchmarked 
test. Hence, we proceed to the Edwards pipe demonstration.

\subsubsection{Transient Edwards Pipe Blowdown Test}

Since we do not know if the steady-state errors are too large,
it is best to test this in a transient case. This is Edwards blowdown.

Only preliminary solver results for the Edwards pipe blowdown test are
presented here, based on agentic coding. Rigorous derivation is 
reserved for future work, as it is out of scope for this research log.

The Edwards pipe is a good choice of test, as both RELAP and TRACE use it; the
nodalisation and initial-condition data used here are taken from the RELAP
paper \cite{edwards1970depressurization}. 

Here there is both code-to-code verification and validation, so it can be
determined whether the behaviour at the bubble point causes serious numerical
difficulty.

\paragraph{Summary of Debugging}

The authors subsequently investigated whether the SIMPLE or PISO algorithms
would be more appropriate for the present application, although using either
approach would increase the computational cost.
Table~\ref{tab:edwards-debug-timeline} summarises the issues encountered and
the corresponding stages of the investigation. Readers should note that
\texttt{rhoCentralFoam} falls under the Kurganov--Noelle--Petrova (KNP)
classification, and that the two names are sometimes used interchangeably in
this prose.

\begin{table}[H]
\centering
\caption{AI-generated summary of the Edwards--O'Brien blowdown debugging timeline for
the \texttt{TampinesSteamArray} PIMPLE / \texttt{HybridAllMach} Rust solver.
Reconstructed with AI assistance from the case READMEs and the solver debugging
notes; the authors reviewed and verified its contents. RMSE
values are gauge-station GS-1 pressure against the Edwards data
\cite{edwards1970depressurization}.}
\label{tab:edwards-debug-timeline}
\footnotesize
\setlength{\tabcolsep}{4pt}
\begin{tabular}{@{}>{\raggedright\arraybackslash}p{0.15\linewidth}>{\raggedright\arraybackslash}p{0.37\linewidth}>{\raggedright\arraybackslash}p{0.38\linewidth}@{}}
\toprule
\textbf{Stage} & \textbf{Issue encountered} & \textbf{Fix and outcome} \\
\midrule
\multicolumn{3}{@{}l}{\textbf{Rust \texttt{TampinesSteamArray} PIMPLE solver (16 July 2026)}}\\
16 Jul --- Energy conservation (A) & Plateau collapses to 17.4~psia (vs.\ $\sim$350); the non-conservative $\rho\,\partial h/\partial t$ form over-drains enthalpy during the flash & Conservative $\mathrm{ddt}$ with continuity-consistent $\rho_{\mathrm{cont}}$ and compressibility $\psi=\partial\rho/\partial p|_h$; plateau recovers to 392.7~psia (RMSE 59.8), no clamps \\
\addlinespace[2pt]
16 Jul --- Shock capturing (B) & High-frequency ringing at the near-sonic flashing front from the low-dissipation pressure solver & Mach-weighted KNP (\texttt{rhoCentralFoam}) dissipation in a \texttt{HybridAllMach} mode; dissipate static enthalpy, gate on $\min(\mathrm{Ma})$; ringing $-55.5\%$ \\
\addlinespace[2pt]
16 Jul --- Tail stability (C) & Hybrid crashes at $t\approx0.18$~s: a rarefied cell's enthalpy diverges and trips the $(p,h)$ 273.15~K validity edge & Rarefied-tail density taper (KNP off below 50, full above 100~kg/m$^3$); stable over the full 600~ms (RMSE 30.6) \\
\bottomrule
\end{tabular}
\end{table}

The Edwards validation required sustained, human-led iterative development:
the solver reached a stable 600~ms run only after the three targeted fixes
above, each of which depended on understanding the underlying physics rather
than on generating more code. Domain knowledge and practical experience remain
essential and are not replaced by AI models, including highly capable ones such
as Opus, when developing new numerical solvers.

This table was AI-generated, and the author was initially unfamiliar with the
details. Claude was therefore asked to explain the concepts from the energy
balance upward, so that the steps taken could be checked for reasonableness
against the literature, or at least against first principles. That checking is
the substantive point rather than an aside: an AI-generated summary that has
not been independently understood is not a result.

What follows is an account of the fixes Claude Code made, presented as interim
work. Fuller solver derivation and quantification is reserved for a later
paper, the scope being too large for this one.

\paragraph{Enthalpy Drain Fix}

First, this is common sense energy balance:

\begin{equation*}
	\rho \frac{D h}{D t} = S
\end{equation*}

$S$ represents all the source and sink terms.
This comes from the full conservative form:

\begin{equation*}
	\frac{\partial (\rho h)}{\partial t} + \nabla \cdot (\rho u h) = S
\end{equation*}

When taking product rule:
\begin{equation*}
	\rho \frac{D h}{D t} + h \left( 	
	\frac{\partial \rho}{\partial t} + \nabla \cdot (\rho u)
	\right) = S
\end{equation*}

Normally the mass conservation equations necessitate that:
\begin{equation*}
	\frac{\partial \rho}{\partial t} + \nabla \cdot (\rho u) = 0
\end{equation*}

Though in approximate form, discretisation and after we apply Gauss's theorem and 
approximate face fluxes as per normal finite volume method
\cite{greenshields2022notes,jasak1996openfoamthesisphd}:
\begin{equation*}
	V_\text{cell} \frac{\rho^\text{new} - \rho^\text{old}}{\Delta t} 
	+ \sum_f (\rho u)_f \cdot A_f = 0
\end{equation*}

However, due to discretisation or other such errors, this term may not 
drop to zero. So, under numerical solution:

\begin{equation*}
	V_\text{cell} \frac{\rho^\text{new} - \rho^\text{old}}{\Delta t} 
	+ \sum_f (\rho u)_f \cdot A_f \neq 0
\end{equation*}

This caused the overcooling due to the spurious 
$h \left(V_\text{cell} \frac{\rho^\text{new} - \rho^\text{old}}{\Delta t} 
+ \sum_f (\rho u)_f \cdot A_f\right)$ term. 

So using the $\rho^\text{new}$ was not suitable. If we want the continuity 
to hold, we had to force the continuity equation to zero.

\begin{equation*}
	V_\text{cell} \frac{\rho^\text{cont} - \rho^\text{old}}{\Delta t} 
	+ \sum_f (\rho u)_f \cdot A_f = 0
\end{equation*}

In doing so: 

\begin{equation*}
	\rho^\text{cont} \frac{D h}{D t} + h \left( 	
	0
	\right) = S
\end{equation*}

\begin{equation*}
	\rho^\text{cont} \frac{D h}{D t} = S
\end{equation*}

When $\rho^\text{cont}$ is used, the continuity error used for the energy 
balance equation is solved, preventing the overdrain of enthalpy.

The mass balance itself is not written in terms of pressure, whereas 
rhoPimpleFoam is essentially based on the pimpleFoam algorithm which is 
a pressure based solver rather than a density based solver like rhoCentralFoam.

The mass balance itself needs to be written in terms of pressure.

Starting with:
\begin{equation*}
	\frac{\partial \rho}{\partial t} + \nabla \cdot (\rho u) = 0
\end{equation*}

In differential form, we can use thermodynamic calculus, for pure steam

\begin{equation*}
	d \rho (p, T) = \left( \frac{\partial \rho}{\partial T} \right)_p dT +
	\left( \frac{\partial \rho}{\partial p} \right)_T dp
\end{equation*}

Or if one desires the use of enthalpy:

\begin{equation*}
	d \rho (p, h) = \left( \frac{\partial \rho}{\partial h} \right)_p dh +
	\left( \frac{\partial \rho}{\partial p} \right)_h dp
\end{equation*}

Now, in the saturation dome, then the $\rho(p,T)$ equation doesn't quite 
apply, because in saturation, under thermodynamic equilibrium, then T and p 
are not enough to describe the thermodynamic state of the system. One 
needs to include quality. rhoPimpleFoam traditionally deals with single phase 
flows as in for the Sod shock tube, but for two phase flashing, this algorithm 
(the closure part specifically) fails without proper 
correction. Again, the PIMPLE algorithm still works, but 
one cannot use a rhoPimpleFoam translation in this case as it uses the 
wrong thermodynamics. With AI-assisted debugging, the 
latter form was used instead.

Now, under constant enthalpy (which translates to constant temperature 
in the single phase case for ideal gas), the derivative is reduced to:

\begin{equation*}
	d \rho (p, h) = \left( \frac{\partial \rho}{\partial p} \right)_h dp
\end{equation*}

\begin{equation*}
	d \rho (p, h)|_h = \psi_h dp
\end{equation*}

Equivalently, the constant temperature version used in original rhoPimpleFoam 
is:
\begin{equation*}
	d \rho (p, t)|_T = \psi_T dp
\end{equation*}

From hereon, we define $\psi_h \equiv 
\left( \frac{\partial \rho}{\partial p} \right)_h$.

With discretisation, we obtain:

\begin{equation*}
	\rho^\text{new} = \rho^{old} + \psi_h (p - p^{old})
\end{equation*}

Now, $\psi_T$ used for rhoPimpleFoam was wrong, so in agentic debugging,
$\psi_h$ was used instead. This ensured that the plateau at 392.7 psia 
was recovered for the Edwards case, and that the solver was stable. 
Moreover, using $\psi_h$ would ensure that the solver can work in both 
single and two phase regimes. This resulted in Figure~\ref{fig:enthalpy-drain-fix}:

\begin{figure}[H]
    \centering
	\includegraphics[width=0.65\textwidth]{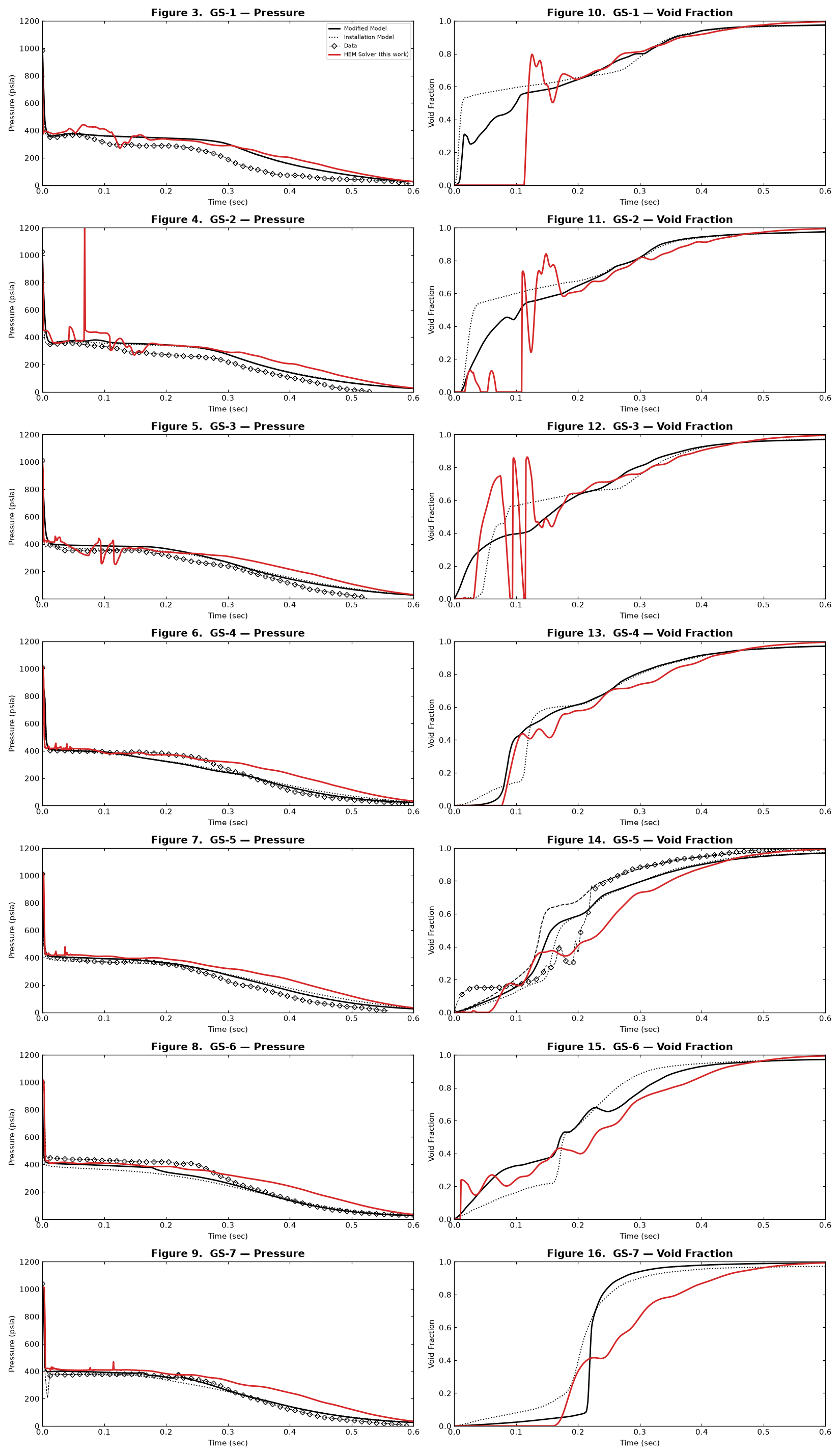}
	\caption{Enthalpy Drain Fix Results with Ringing}
    \label{fig:enthalpy-drain-fix}
\end{figure}

Nevertheless, there was noticeable ringing and oscillation through the system 
even with these fixes.

\paragraph{Shock Capturing Fix}

To remove ringing and oscillation, it was initially decided to use rhoCentralFoam 
type solver to reduce the shocks. Unfortunately, these were quite 
futile as they reduced ringing, but re-introduced the instability issues
as seen in Figure~\ref{fig:edwards-rhocentralfoam-fix}:
\begin{figure}[H]
    \centering
	\includegraphics[width=0.65\textwidth]{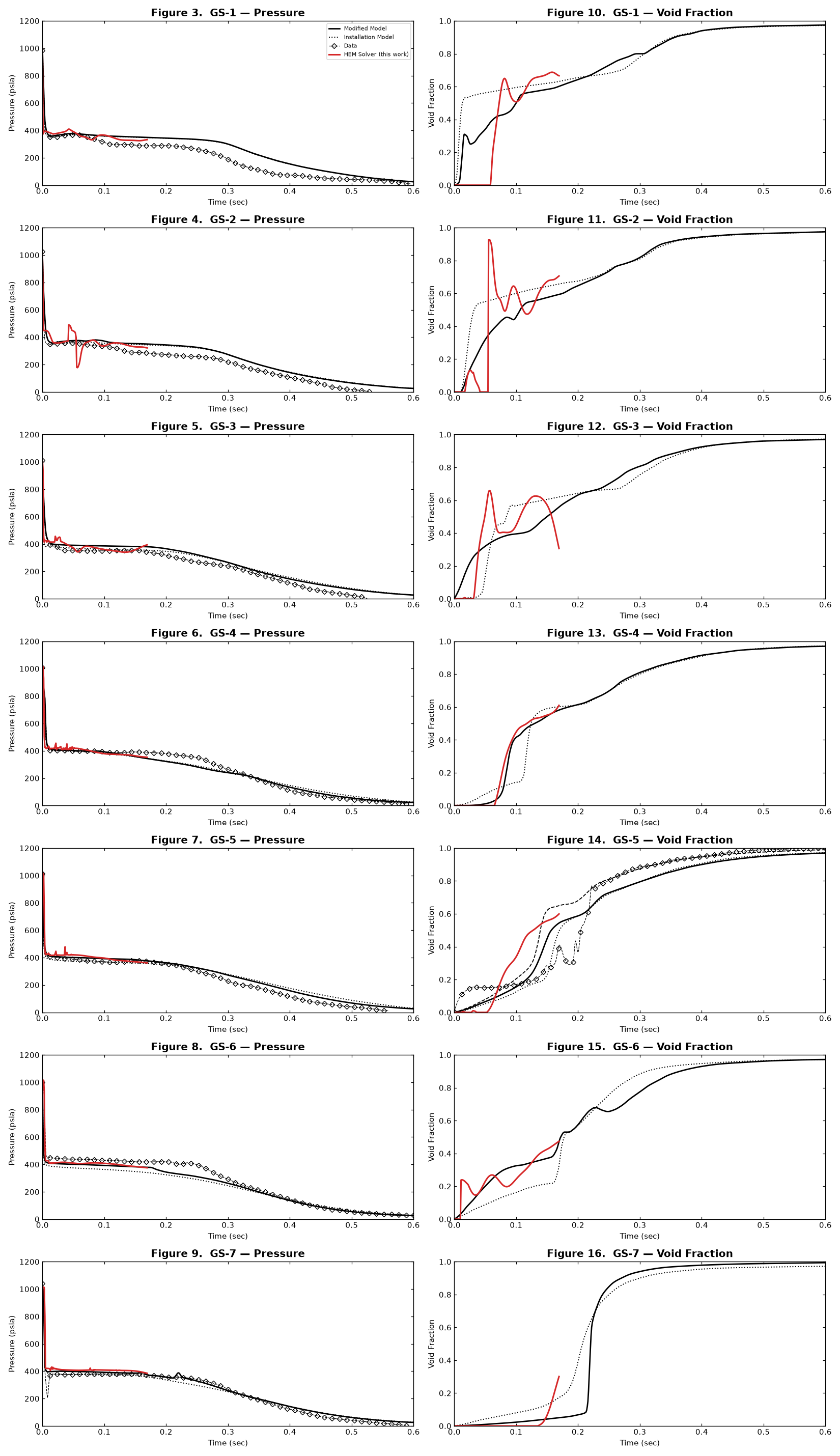}
	\caption{Ringing Fix with rhoCentralFoam introduced new numerical 
	instability}
    \label{fig:edwards-rhocentralfoam-fix}
\end{figure}

To remove the ringing and oscillation and maintain stability, the 
author then decided to use a rhoCentralFoam 
like solver in the shock regime, and rhoPimpleFoam like solver in the 
subsonic regime. This became known as the HybridAllMach mode in 
Table~\ref{tab:edwards-debug-timeline}. Hence, the TampinesSteamArray,
the Rust struct responsible for solving the Edwards Pipe, had two modes,
a rhoPimpleFoam mode and a HybridAllMach mode. This was user selectable.

The Mach numbers used to switch the solver to a rhoCentralFoam like solver 
are based on the current "owner" cell $O$ and neighboughring cell $N$:

\begin{equation*}
	\mathrm{Ma}_O = \frac{\left| \mathbf{u}_O \right|}{c_O},
	\qquad
	\mathrm{Ma}_N = \frac{\left| \mathbf{u}_N \right|}{c_N}
\end{equation*}

\noindent where $c$ is the local speed of sound evaluated from the
homogeneous-equilibrium steam tables at the cell state $(p, h)$. Inside the
saturation dome this is the two-phase equilibrium sound speed, which is far
lower than that of either phase alone. The blending weight is then:

\begin{equation*}
	\beta_f = \mathrm{clamp}\left(
	\frac{\mathrm{Ma}_f - \mathrm{Ma}_\text{lo}}
	{\mathrm{Ma}_\text{hi} - \mathrm{Ma}_\text{lo}},\ 0,\ 1 \right),
	\qquad
	\mathrm{Ma}_f = \min\left( \mathrm{Ma}_O,\ \mathrm{Ma}_N \right)
\end{equation*}

$\mathrm{Ma}_\text{hi} $ and $ \mathrm{Ma}_\text{lo}$ are set at 1.0 and 0.3 
respectively. This makes sense because Mach 1 is sonic, and Mach 0.3 is the 
lower bound where compressibility effects start to matter.

Using this weighting allows us to switch on or off the rhoCentralFoam like 
KNP solver based on Mach number.

Nevertheless, introducing this solver also produced yet another bug.
The solver again overdrains enthalpy and some cells in 
the system went below freezing in the later part of the depressurisation.
This necessitated the final fix.

\paragraph{Tail Stability Fix}

For the last fix, we know that rhoPimpleFoam is stable in all regimes, 
just produces ringing during depressurisation. Since the enthalpy overdrains 
in the low pressure, low density regime where the system is already significantly 
depressurised, it makes sense to turn off the rhoCentralFoam type solver 
and restore the rhoPimpleFoam type. Intuitively, this makes sense.

This is because rhoPimpleFoam is the more numerically stable of the 
solvers and prevents enthalpy divergence. Having the KNP switch off at
50 $\text{kg/m}^3$, close to the density of steam, 
is an elegant way to patch the solver to make it more stable all the 
way up to 600 ms. This is more of a stability patch than anything else,
but it worked!

All in all, the compressibility and $\rho_\text{cont}$ fixes are more 
grounded in thermodynamics, whereas the KNP (rhoCentralFoam) type 
solver blending helps fix ringing and potential numerical instabilities 
associated with it. These are more of stopgap measures and numerical improvements 
to an otherwise stable solver with some ringing. 

This yields the figure shown in Figure~\ref{fig:edwards-final-hybridallmach}:

\begin{figure}[H]
    \centering
	\includegraphics[width=0.65\textwidth]{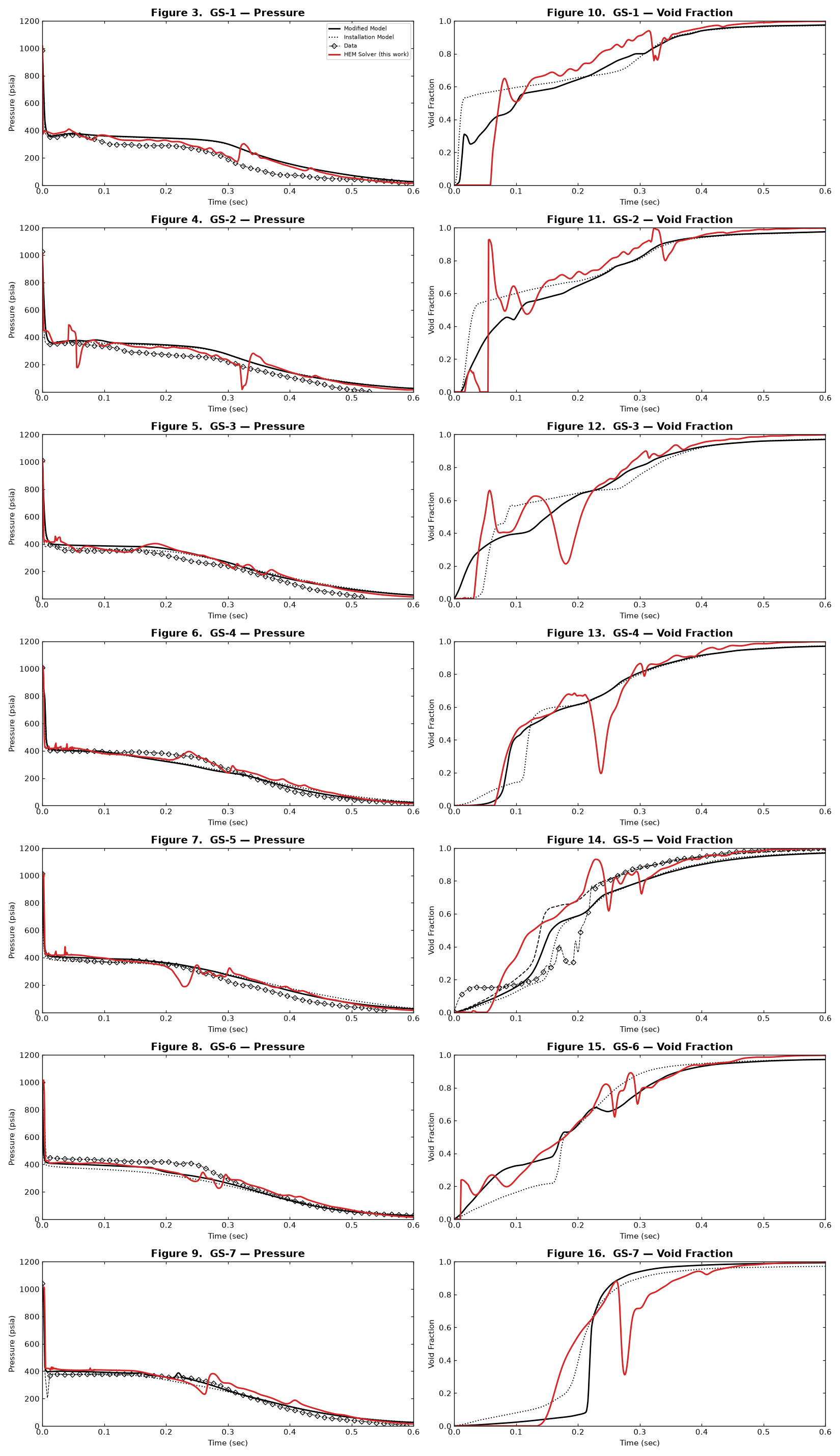}
	\caption{Final performance of HybridAllMach after debugging}
    \label{fig:edwards-final-hybridallmach}
\end{figure}

Significant ringing was still observed, but it was damped compared to before. 
This is an acceptable result for a first attempt, which was after all, a 
HEM solver with limitations. Instabilities notwithstanding, it still performs adequately,
the trend of pressure drop still follows experiment and RELAP code 
somewhat well at a root mean square error of 30.6 psia. 
Though much remains to be desired of the void fraction.

\paragraph{Workflow and future work}

The workflow that emerged for the Edwards case was as follows. Claude Code
proposed and implemented the numerical fixes described above. Because the
first author was not initially familiar with the derivations behind them, he
asked Claude Code to set out each one from first principles, and then checked
that derivation himself against the finite-volume literature
\cite{greenshields2022notes,jasak1996openfoamthesisphd} and against the
conservation arguments reproduced in this section, before accepting the fix.

It is worth being explicit about what this does and does not establish,
because the distinction is the one this research log is arguing for. Asking a
model to justify its own output is not verification: a model that produced a
wrong fix can produce a plausible derivation for it, and the two failures are
correlated. What made the step meaningful was that the derivation was checked
by a human against independent sources, and that it terminated in quantities
--- the continuity-consistent density $\rho^{\mathrm{cont}}$ and the
compressibility $\psi_h$ --- that can be written down from the governing
equations without reference to the code. Where that check could not be closed
independently, the result is reported here as interim rather than verified.

The GS-1 pressure comparison against the Edwards data is accordingly a
validation of the solver's gross behaviour only. A full derivation, a
mesh-convergence study of the kind performed for the cavity in
Section~\ref{sec:cavity}, and quantitative void-fraction validation are
reserved for future work.

\section{Discussion}

\subsection{Initial V\&V Efforts of Outram-Foam and Outram-Foam 
Derived Solvers in Tampines Steam Tables}

Using Claude Code, the path from draft code to V\&V-ed case has been 
significantly shorter as the code produced was able to match solutions 
in literature and experimental data to within reasonable bounds. Moreover,
the development of an initial draft of the HEM solver validated using 
the Edwards Pipe Blowdown test using Claude Code demonstrates its effectiveness 
in producing new thermal-hydraulics code quickly from existing libraries, 
provided there is some guidance from the human in the loop.

This contribution is significant because prior to this, critical choked 
flow for two phase steam and water was scarcely found in open-source 
repositories. From this agentic coding experience, the bottleneck for 
code development now is the human V\&V required to validate and certify 
that the drafts produced by AI agents is useful and safe enough 
for educational and research uses.

\subsection{Speed Increase and Productivity Gain}

Both agentic development and validation efforts 
in this research log required human-in-the-loop development, which should dispel 
the myth that AI is ready to replace human beings, at least in this field.
In this case, it was demonstrated that Agentic AI augmented capabilities of 
human beings to boost productivity.

\subsubsection{Humans are Still Required}
AI-assisted coding is effective, and much faster. From 
Table~\ref{tab:outram_park_timeline}, one month can produce code for 
thermodynamics, PRKE, Monte Carlo, and NJOY translation.

In the span of a few months, the Rust ports were done. 

as of July 23 2026 we have 
- Rust Translation and Refactoring of many codes 
- Verified CFD in Outram-Foam 
- NJOY and Outram MC (not in this research log)
- Work in Progress PFLOTRAN translation
- Preliminarily Verified 1D Homogeneous Equilibrium Solver in TampinesSteamArray

However, this speed increase is only possible with AI for someone with 
background knowledge of:

- Rust Programming Language 
- CFD 
- Reactor Physics 
- Thermal Hydraulics 
- Some experience with NJOY, OpenFOAM and OpenMC

The author has had these experiences from PhD \cite{ong2024digital}, and 
these proved to be extremely valuable in guiding and prompting the AI,
and doing V\&V on the code afterwards, as was done in TUAS \cite{ong2024tuas}. 
These skills are indispensable, and it is still up to humans to verify 
and validate the codes generated. AI will not replace human beings.
Rather, it is a workforce enablement tool, as is the case here.

This is in alignment with the government, as Dr Tan See Leng said

" While an AI-ready workforce offers significant potential to improve 
productivity, we must steer AI adoption to enhance our workers' 
potential, not displace or replace it." \cite{tan_mom_cos_2026} In this case,
AI was used to draft and generate code, but it is human input which then 
checked and validated to make sure it is of use and correct enough for 
teaching and research. Even during development, as mentioned, the authors 
needed to give AI constant guidance to ensure tokens were not wasted chasing 
the wrong solution, when the implemented solution was obviously wrong.

This use of Agentic Coding and translation in Outram Park is exactly a case 
study in AI.

\subsubsection{Ballpark Productivity Gains when using Agentic Coding}

\paragraph{Estimate using Boehm et al's Method}
With this combination however, it is possible to achieve significant savings.
When estimating the value of outram park, it is possible to use several 
estimation techniques. The upper bound is classical software cost 
estimation where cost can be estimated by person months \cite{boehm1981software}:

\begin{equation}
	\text{Person Months} = 2.4 * (KLOC)^{1.05}
\end{equation}

For outram park backend in end july, the repository was in the order of 
350,000 lines of code or 350 (kilo lines of code, KLOC). 
Using Boehm's formula, we obtain about 1125 person 
months of productivity.

Priced at SGD\$10,000 per month, we would have a value of about 
SGD\$11.26 million or about USD\$8.66 million depending on exchange rate.
These are rough estimates that estimate a repository's value based on the 
number of lines of code, which may not reflect the value that this is  
open-source code. Moreover, it does not reflect the use of modern tools 
such as language server protocol, autocompletes and neovim which helped 
speed up development.

\paragraph{Estimate using pre-agentically written Rust Code as Baseline}

To obtain a better estimate, consider the author's pre-agentic repositories: TUAS \cite{ong2024tuasgithubrepo},
,tampines-steam-tables \cite{ong2026tampinessteamtables} and boon lay 
(still in development) \cite{ong2026boonlay} as an example.
During his PhD the author spent roughly two years learning Rust and developing it 
(2022-2023) formerly named "thermal\_hydraulics\_rs"\cite{ong2024thermalhydraulicsrs}, 
and another 2 years in SNRSI (2024-2025) where it evolved into TUAS 
\cite{ong2024tuasgithubrepo}. This 
is 4 years total just to get it out and validate it. Suppose verification and 
validation took about 1.5 years of that effort. With TUAS, there were 
96 KLOC including doc comments pre-AI. With tampines-steam-tables there were 
about 55 KLOC in the same era by roughly end 2025, with a little bit of AI 
assitance from NUS AI-know for root finding solvers. Boon Lay was another 
15 KLOC roughly, partly generated with NUS AI-know, but at a rate nowhere near 
the level of agentic coding Claude Code offered.
A naive computation gives three years to generate roughly 165~KLOC without
validation, allowing that other projects and work were ongoing throughout. This would be about 55 KLOC per year
done at a more sustainable pace. This is shown in detail, using AI 
summarised review of git logs in Table~\ref{tab:preagentic_baseline} 
for work done before Agentic coding:

\input{./results_and_discussion/kloc_accounting/baseline_table.tex}

Two repositories appear in Table~\ref{tab:preagentic_baseline} that were
not named in the discussion above: \texttt{teh-o-prke}
\cite{ong2025tehoprke}, the point reactor kinetics package, and the chemical
engineering real-time process control simulator
\cite{ong2024chemengprocesscontrol}. Both predate Outram Park and are
vendored into it, so both belong in the baseline rather than in the agentic
tally. \texttt{teh-o-prke} is in turn the predecessor of the
\texttt{nee\_soon} coupling layer. The process control simulator dates from
the author's doctoral work and is the one baseline component carrying a
published verification and validation record, having been validated against
Scilab in \cite{ong2024digital}.

The baseline reaches 181.3~KLOC of code by a different route to the running
total in the text. The text sums TUAS at its head (96~KLOC),
\texttt{tampines-steam-tables} (56~KLOC) and \texttt{boon-lay} (15~KLOC); the
table separates \texttt{thermal\_hydraulics\_rs} from the TUAS work that
followed it, counts \texttt{boon-lay} at its 9.6~KLOC of code rather than its
15.8~KLOC total, and adds the two repositories above.

Spread over 367 active
days, the baseline works out at roughly 494 lines of code per working day.
This is the running average for code in Table~\ref{tab:preagentic_baseline},
with some help with NUS AI-know. Ranking the repositories by that rate, as in
Table~\ref{tab:baseline_rates}, is more informative than the average alone:
the pre-agentic repositories span 206 to 572 lines per active day, a factor of
roughly three, which is the ordinary variation between projects of differing
nature. That spread is the yardstick the agentic rate should be read against.

\input{./results_and_discussion/kloc_accounting/rate_table.tex}

Note also that the ranking does not line up with the amount of AI help
received. The fastest repository, \texttt{tampines-steam-tables}, had only
root-finding solvers generated with NUS AI-know and nothing more; the slowest,
\texttt{teh-o-prke}, had none at all, and \texttt{boon-lay}, which had the most
non-agentic code generation of the three, sits second from the bottom. Whatever
sets the pace of pre-agentic work, chat-assisted code generation is not the
dominant term.

Now, if one were to consider the crates built with help from
NUS-AI know via chat and code generation, the average is
525 lines of code per working day compared to 470 lines of code per working
day for the older crates. This is a 12\% increase in productivity. Not all
of it can be attributed to NUS-AI know however, because the author has 
gotten more proficient in Rust programming via neovim and its plugins 
in 2025-2026 as compared to 2023-2024. Hence, this is not a clean comparison.
Moreover, the nature of the crates differ. Therefore, not all lines of 
code are created equal. Hence, this should only be treated as a ballpark 
figure.

Hence, the pre-agentic ballpark for Rust based development was around
\textbf{500 lines of code/active day of coding}. At that rate the agentically
written portion of Outram Park, 175\,997 lines, represents about 352
engineer-days, and the full 181\,298-line baseline about 362; the figure
depends on which body of code one takes as the thing being replaced. Assuming
engineers are paid US\$200 per day (a conservative figure, as lower-bound
estimates are about US\$100,000/year \cite{veseli2025navigating}, or about
USD\$273 per day), the cost to replace \textbf{Outram Park pre-agentically is
in the range USD\$70,000--73,000}. This range, rather than a single figure, is
what the line counts support.
Note also that this assumes all days are used for coding. In reality, a 
software engineer or nuclear numerical simulation engineer would be expected 
to regularly attend meetings and non coding related activites. Sometimes,
rest is also required to ensure coding is sustainable. This means that 
the figure cited here for human cost to replace is lower than what it may
be in real life.

\paragraph{Agentically Developed Source Code}

The comparison becomes clearer in 
Table~\ref{tab:agentic_crates}, which gives the corresponding breakdown for
\texttt{outram-park-backend}, by crate, as of 2026-07-23. The repository holds
349\,541 lines of Rust code across 26 crates; subtracting the 173\,544 lines
of pre-agentic work vendored into it leaves \textbf{175\,997 lines written
agentically}. That subtracted figure is not an independent estimate: it is the
baseline of Table~\ref{tab:preagentic_baseline} less the 7\,754 code lines
that were written in \texttt{thermal\_hydraulics\_rs} and never carried across
at spin-out. Those lines are pre-agentic work and belong in the baseline, but
they are not vendored into the backend and so cannot be subtracted from it.
The two tables otherwise partition the same body of code without overlap or
omission, which is the arithmetic check on both. Classification follows each
crate's own declared provenance in its \texttt{Cargo.toml} manifest.

\input{./results_and_discussion/kloc_accounting/agentic_table.tex}

\begin{figure}[H]
	\centering
	\includegraphics[width=0.95\textwidth]{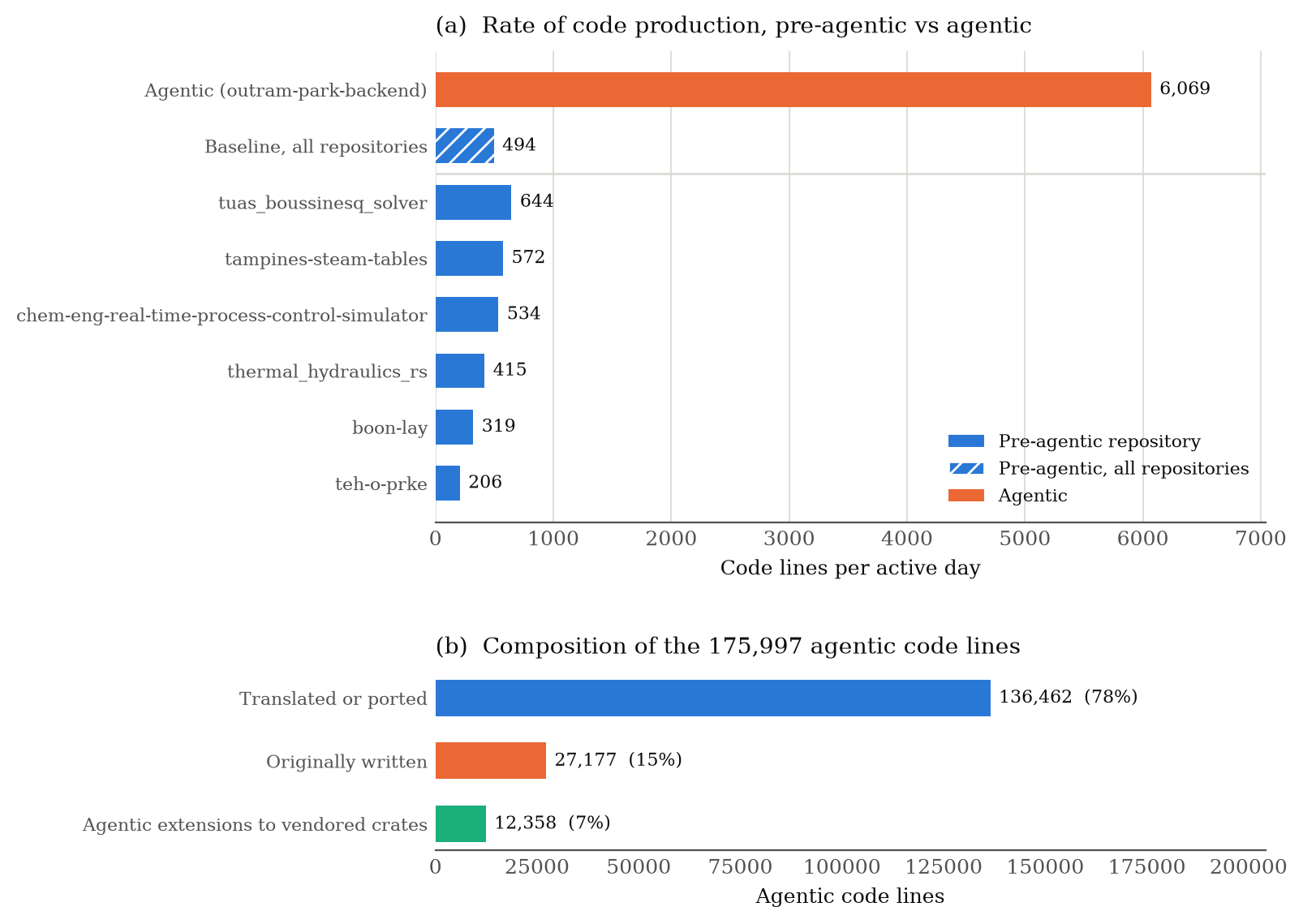}
	\caption{Rate and composition of code production. (a) Code lines per active
	day for each pre-agentic repository, for the pre-agentic baseline as a
	whole, and for the agentic month. The aggregate is separated by a rule
	because its denominator is the union of active dates across overlapping
	projects, not the sum of the rows above it. (b) The agentic output split by
	provenance. Generated from the repositories and their git histories by
	\texttt{kloc\_accounting.py}; the underlying figures are those of
	Tables~\ref{tab:preagentic_baseline} and~\ref{tab:agentic_crates}.}
	\label{fig:kloc_productivity}
\end{figure}

Three features of Table~\ref{tab:agentic_crates} bear on how the productivity
figure should be read. First, \textbf{77~per cent of the agentic output is
translation} rather than original design: porting an existing, already-debugged
numerical implementation into Rust is substantially faster per line than
inventing one, and the headline figure should not be read as 175~KLOC of novel
engineering. Second, the 15~per cent that is original is dominated by tooling
and interface code, chiefly the KOVAN agent-facing utilities and the terminal
interfaces, rather than by new physics. Third, the agentic month produced
175\,997 lines against a pre-agentic baseline of 181\,298 lines accumulated
over roughly three years: \textbf{as much code in one month as in the preceding
three years}, which on a per-working-day basis is a factor of 12.3 --- 6\,069
code lines per active day against 494. Both rates, and the composition
underlying the first two points, are plotted in
Figure~\ref{fig:kloc_productivity}.

Note also that the author is comparatively proficient with neovim and Rust,
which if anything flatters the pre-agentic baseline.

Suppose one were to naively extrapolate this to Outram Park, where 
about 187 KLOC was written within the span of one month. Note that this was not sustainable: the author fell ill repeatedly over the
course of the month. 
(please see health advisory) Suppose we took 2.5 months instead to produce 
agentic code at a sustainable pace. Hence, the productivity with agentic 
coding is about 74.8 KLOC per month. In other words, working agentically 
with Claude Code produced more KLOC than the author would normally produce in a
year. This would almost be 900 KLOC per year or a roughly 23 times 
productivity gain as compared to manual + some AI chat bot assisted coding 
if one were to consider Boon Lay. 

Whether we use the low bound of 12x or high bound of 23x, this reflects 
a one order of magnitude productivity increase over pre-agentic AI coding.

Hence, for agentic translation, the ballpark figure is about 
\textbf{6000 lines of code/active coding day}. Which means for human 
translation of codes (plus some new code generation), 
agentic AI via Claude Code was about one order of 
magnitude faster than human coding. Assuming roughly the same amount
of 500 lines of code/active coding day for human translation, \textbf{this is
the same USD\$70,000--73,000 replacement cost given above, arrived at from the
translation side rather than from the line-count baseline --- not a further
cost on top of it}.
This of course, assumes 
that there is a team with intimate knowledge of Rust, C, C++ and Fortran
which knows each of the nuclear science and engineering domains enough 
to translate such codes and develop applications from it.

\subsubsection{Ballpark Productivity Costs when using Agentic Coding}

The work above was done under a flat-rate subscription (actually two, because 
both Claude Pro and Claude Max were used). This totalled around USD\$400 
listed price, but Anthropic charged SGD\$600 for both.
There is worry that tokens 
will not always be cheap \cite{infographics_show_ai_bill_2026}. Hence, 
it is also of interest to see whether the tokens used even when 
API is used, will outweigh the gains described in earlier sections.

Table~\ref{tab:token_cost} shows the estimated token cost of Outram Park:

\input{./results_and_discussion/discussion_markdowns/token_cost_table.tex}

Two features of the bill also deserve comment, and are set out in
Table~\ref{tab:token_cost_components}. 

\input{./results_and_discussion/discussion_markdowns/token_cost_components_table.tex}

Now, prior to 23 July, no proper token accounting was done. Therefore 
Table~\ref{tab:token_cost} would underestimate the cost of the tokens 
used to construct Outram Park agentically. However, taking the ballpark figure of USD\$1600 and adding a \textbf{10 times buffer}, 
USD\$16000, we would still be producing several times the value with the code 
in Outram Park should it be constructed non-agentically. Comparing 
\textbf{USD\$16,000 for approximate token cost to USD\$70,000--73,000 for cost to replace},
it would have been far cheaper to use agentic coding to perform this 
task even at API costs. Therefore, Agentic Coding even with the higher API 
costs still has its value in longer term development of code bases 
and open-source nuclear simulation frameworks such as Outram Park.

Naturally, we must be judicious about the use of tokens in this manner.
As the price of tokens increase, one has to balance its costs against Manpower
and the use of open source local AI models available in ollama 
\cite{marcondes_using_ollama_2025}.

\subsection{The Role of AI in Software Development}

Many advocate for the use of AI, but how should AI be used? And how 
do humans still have relevance?

Indeed, in the pre-AI era, we have:

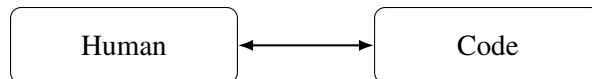
\begin{figure}[H]
\centering
\begin{tikzpicture}
\node (human) [wfbox] {Human};
\node (code)  [wfbox, right=1.8cm of human] {Code};

\draw[wfboth] (human) -- (code);
\end{tikzpicture}
\caption{Traditional software development workflow. The engineer directly develops, tests, and maintains the code.}
\end{figure}

Whereas now, we have a newer code development model:

\begin{figure}[H]
\centering
\begin{tikzpicture}
\node (human)    [wfbox] {Human};
\node (artifact) [wfbox, right=2.6cm of human] {Code\\+\\Documentation};
\coordinate (mid) at ($(human)!0.5!(artifact)$);
\node (agent) [wfbox, below=1.6cm of mid] {AI Agent};

\draw[wfboth] (human) -- (artifact);
\draw[wfaux]  (human) -- (agent);
\draw[wfaux]  (agent) -- (artifact);
\end{tikzpicture}
\caption{Modern human-in-the-loop development. AI agents assist implementation while humans remain directly engaged with the code and documentation.}
\end{figure}
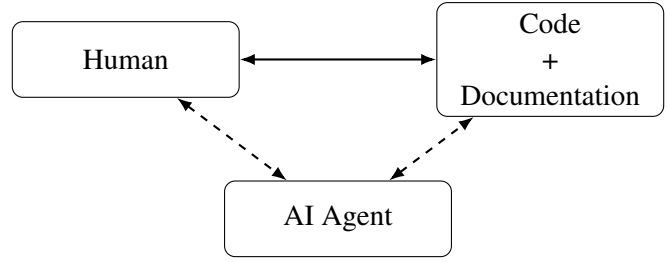

With this new style of development, there have been arguments and questions raised 
on whether AI will replace human beings \cite{andriole2024big,eng2024will}.
Moreover, there is great concern that even when humans use AI, the result 
is cognitive atrophy \cite{trajanov2026cognitive}.
We make the argument AI should augment human intelligence, not replace it 
\cite{de2021ai}. Herein, we describe the proper use of AI, where humans 
should not work in such a way as to experience cognitive decline.
This is the model used for developing Outram Park.

Anecdotally, it was impossible for humans to just dump one prompt into 
Claude Code to result in Outram Park. Outram Park was done with specific 
directions for AI, to translate open-source nuclear codes from Fortran and 
C++ to Rust library by library. For each library, sufficient levels 
of domain knowledge were required to guide the AI. This involvement increases 
where AI is asked to synthesise new codes and solvers such as those 
seen in tampines-steam-tables. Without prior Thermodynamics knowledge 
from the human prompter, AI would have wasted a good amount of tokens using 
the wrong functions in classic hallucination \cite{rawte2023survey}. 
Humans therefore are present and required to mitigate the hallucination risks.

It should also be noted that hallucination risks are lowered when there is 
a codebase of source material. AI is after all a transformer at heart. It is 
able to hold onto large amounts of data and synthesise something new based 
on the data. An interpolation or extrapolation perhaps, or as in literature 
a stochastic parrot \cite{bender2021dangers}. In today's state of AI in 
2024-2026, it is still doubtful whether Artificial General Intelligence (AGI)
exists \cite{feng2024far}. For now, it is the author's opinion to be sceptical 
of agentic AI capabilities until otherwise proven. Therefore, we adopt the 
approach that AI is just (at least) a very good stochastic parrot. 
Of course, a stochastic parrot 
is useful if one wants to do major refactoring of code, where source code 
is verified correct, or to make GUIs of which plenty of working examples are 
already available. However, when it comes to synthesising something new or 
novel, AI in today's state may still be a 
useful stochastic parrot, but still fall short of a human being. It is 
therefore up to the human being to think creatively, innovatively and with 
a noble purpose in mind - to solve problems to make the life of other human 
beings better.

With this in mind, in Outram Park, all code is considered an incomplete draft unless humans validate 
it and check it first. We do not adopt the practice that AI generated code
is considered correct. This is the wrong model:

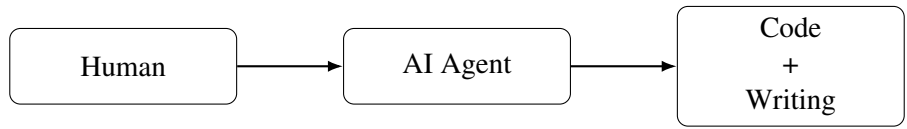
\begin{figure}[H]
\centering
\begin{tikzpicture}
\node (human)    [wfbox] {Human};
\node (ai)       [wfbox, right=1.4cm of human] {AI Agent};
\node (artifact) [wfbox, right=1.4cm of ai] {Code\\+\\Writing};

\draw[wfflow] (human) -- (ai);
\draw[wfflow] (ai) -- (artifact);
\end{tikzpicture}
\caption{Undesirable workflow in which the human delegates both implementation and understanding to an AI system.}
\end{figure}

Instead, for Outram Park, AI was used to speed up the drafting and code 
debugging stage. Thereafter, humans would then manually check the validation 
stage with verified V\&V data:

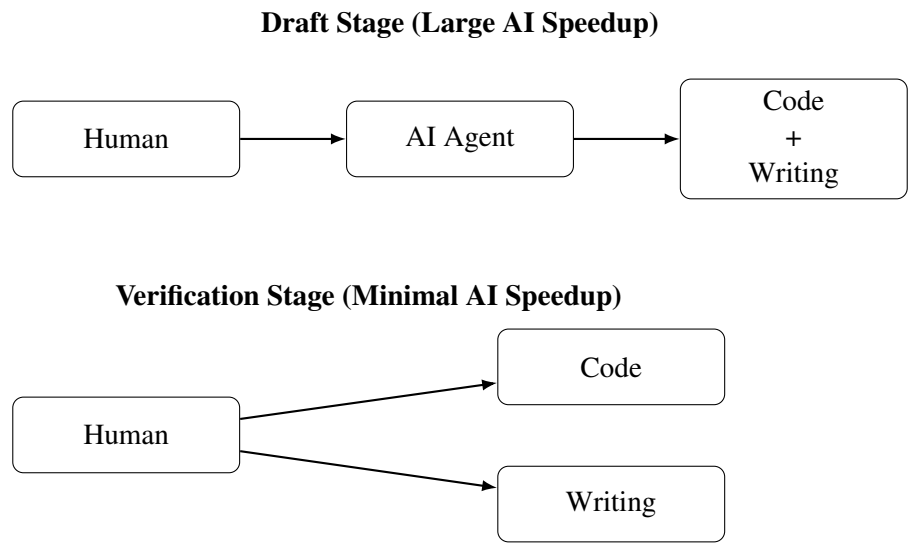
\begin{figure}[H]
\centering

\begin{tikzpicture}
\node (human1)    [wfbox] {Human};
\node (ai1)       [wfbox, right=1.4cm of human1] {AI Agent};
\node (artifact1) [wfbox, right=1.4cm of ai1] {Code\\+\\Writing};

\draw[wfflow] (human1) -- (ai1);
\draw[wfflow] (ai1) -- (artifact1);

\coordinate (top1) at ($(human1.north)!0.5!(artifact1.north)$);
\node (drafttitle) [wflabel, above=0.55cm of top1]
  {Draft Stage (Large AI Speedup)};

\node (human2) [wfbox, below=2.9cm of human1] {Human};
\node (code2)  [wfbox, right=3.4cm of human2, yshift=0.9cm] {Code};
\node (writing2) [wfbox, below=0.8cm of code2] {Writing};

\draw[wfflow] (human2) -- (code2);
\draw[wfflow] (human2) -- (writing2);

\coordinate (top2) at ($(human2.north)!0.5!(code2.north)$);
\node (verifytitle) [wflabel, above=0.55cm of top2]
  {Verification Stage (Minimal AI Speedup)};
\end{tikzpicture}

\caption{
Human-in-the-loop development model adopted by OUTRAM PARK.
AI provides significant acceleration during drafting and implementation,
while verification and validation remain primarily human responsibilities.
}
\end{figure}

Writing and coding were done with AI help, but the concepts were not from AI. As 
mentioned in the Methods section, AI was used to adapt language but 
not come up with new scientific concepts \cite{lin2025supercharging}.
Humans must verify code is correct and can be usable without AI help.

In fact, the humans had to guide the AI throughout the agentic debugging of
the TampinesSteamArray HEM solver. Reaching a stable solution took months spent
reading about pressure-based solvers, density-based solvers and choked flow;
that background is what made it possible to direct the agent effectively rather
than merely to generate more code. For AI-assisted agentic development, the
domain knowledge and engineering experience of the human remain indispensable
to obtaining effective and accurate solutions.

It is useful in the new model to think of AI, deterministic code and humans 
as performing three complementary roles in this process:

\begin{itemize}
	\item Human: manager, responsible owner, and verifier of code and coding agents 
	\item Code: Does deterministic calculations based on human and agent input 
	\item AI Agent: Useful Stochastic Parrot able to draw from large libraries 
		of data and codebases to draft templates and codes for human use
\end{itemize}

AI Agents have limitations, due to their stochastic nature, and in essence 
they are stochastic parrots, only able to regurgitate information with 
some degree of randomness. Humans are best at innovation and creative thinking 
but are not able to hold libraries of information in short term memory.
Codes are accurate as far as they are programmed to be, but reproduce output 
deterministically much better than human beings. Thus they have a capability 
to be more "correct" in calculating a value than a human and AI-LLM, and for 
potentially much lower computational cost for the desired accuracy.

To use AI agents effectively, the human agentic engineer should delegate 
tasks to himself, the AI agent or code based on these strengths. Codes 
can reproduce and calculate cross sections effectively, but should not be
asked to innovate or make value judgements. AI can create new versions 
of codes based on older versions much better than a human, but may not 
innovate on new ideas outside their existing database as well as a human.
Humans should no longer do code refactoring by hand, AI can do better, or do 
calculations by hand, as code is faster and more accurate. But determining 
which problem to solve, how to innovate and what value and moral judgements
should be exclusively within the domain and responsibility of humans. 
As Einstein mentioned, "formulation of a problem is far more essential 
than its solution, which may be merely a matter of \ldots{} skill. 
To raise new questions, new possibilities, to regard old problems 
from a new angle, requires creative imagination and marks real advance." 
\cite{murphy2016problem}

In essence, \textbf{delegate work to the entity best suited to solve it, but 
let the human define the problem well first.}

\subsection{Health Advisory}

Agentic coding can be quite taxing, as seen in Figure~\ref{fig:commit-hrs}:

\begin{figure}[H]
	\centering
	\includegraphics[width=0.95\textwidth]{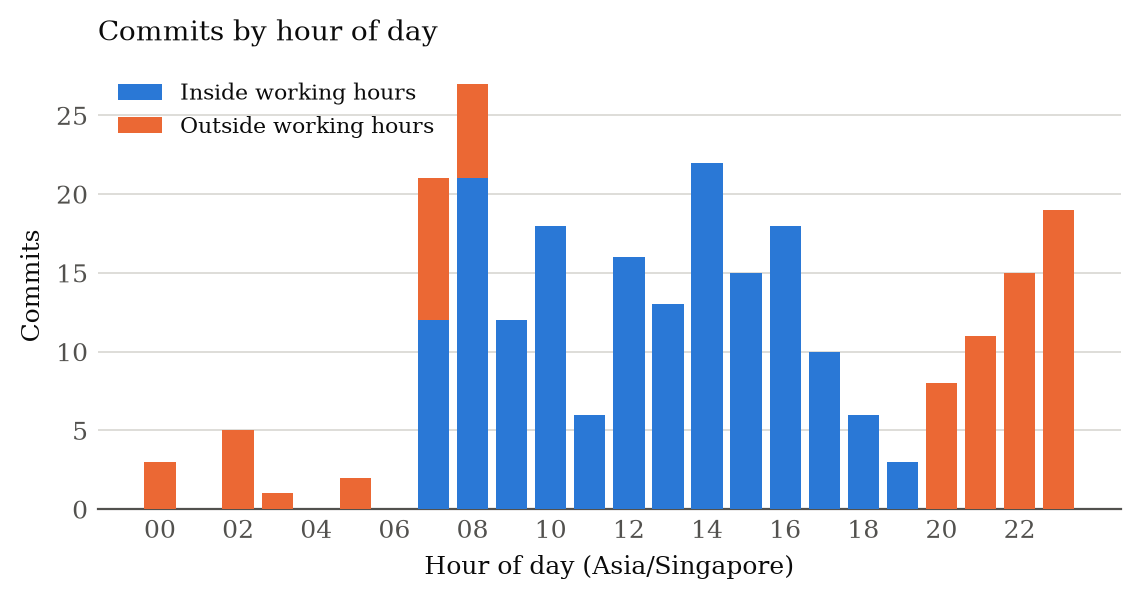}
	\caption{Commit by hours of the day from 19 jun 2026 to 10 july 2026, the 
	period before working hour guardrails were adopted. Times are in GMT+8}
	\label{fig:commit-hrs}
\end{figure}

The author wished to maximise token usage, and consequently became anxious
about drawing fully on the quota so as to extract the full value of the Claude
Code subscription. Sleep quality suffered as a result. This is known in literature 
as technostress \cite{tarafdar2007technostress,tarafdar2019trifecta}. 
With long period of poor sleep quality,
long hours of agentic coding due to hyperfocus 
\cite{dwyer2024monotropism,dupuis2022hyperfocus}, it led to higher levels 
of cognitive fatigue \cite{pessiglione2025fatigue,steward2025neurobiology}
and decision fatigue \cite{choudhury2026decision}. Stresses also led 
to lower immunity \cite{cohen1991commoncold,cohen1995respiratory,cohen1996immunity}
which would likely have contributed to the author's longer recovery periods
from a sore throat.

Of course, we do not suggest AI-assisted coding causes more illness, but 
the AI-assisted developer should do well to ensure that health is taken 
care of so that coding is done sustainably. To ensure this, outram-park-backend 
has instructions and warnings in its repository to ensure agentic coding 
is done only during specified working hours as seen in 
Appendix~\ref{app:health_warning}.

\section{Future Work}

\subsection{Verification and Validation}

The V\&V efforts in this research log are quite preliminary. Claude has helped 
develop some of the cases here for Outram Foam and the reduced order 
Tampines Steam Array HEM solver derived from OutramFoam rhoPimpleFoam and 
rhoCentralFoam. AI assisted debugging has resulted in a deterministic 
solver which was able to reproduce these results with some reduced ringing. 

However, it remains to be seen if such fixes can work outside the domain 
of the Edwards test. If the exact same HEM solver could work for the 
Marviken tests \cite{hall1978marviken}, as well as natural circulation 
tests \cite{guo2025thermal}, then one can know that this solver works 
sufficiently well and did not just receive random AI fixes that made it 
useful for Edwards but useless for other nuclear engineering use cases.
One useful case is that of Zhang and Brooks \cite{zhang2021linear,zhang2024assessment}
where nodalisation and results are given in the paper. 

Moreover, the void fraction validation is not yet complete as there is 
significant ringing even with the HybridAllMach Solver. Complete resolution 
of this problem is reserved for future work, whether using an improved HEM 
solver, or driftFlux or 6 equation model.

\subsection{Thermal Hydraulics}

The next milestone for 1D thermal-hydraulics is the drift-flux and six-equation
model, which is the natural next frontier of experimentation for the Edwards
test.

\subsection{Neutronics and Nuclear Data}

As mentioned earlier, thermal-hydraulics is only part of Outram Park. 
Neutronics is covered in a companion paper to follow shortly after 
the thermal-hydraulics paper is published. Much remains to be done in 
the realm of neutronics. This includes the chord length sampling methods 
used in TRISO pebble beds and methods suitable for liquid fuelled molten 
salt reactors. Depletion and burnup are also important as OpenMC covers 
them. The port known as Outram MC will be discussed in future.

\subsection{Materials}

Outram Park as it stands in July 2026 is lacking much in fuel performance
and materials simulation. These are important for nuclear safety evaluations 
and construction of proper digital twins. Initial ports of Offbeat, the 
foamForNuclear code built along with GeN-Foam \cite{nervi2026foamfornuclear},
have been partially completed but no validation has been done yet. Moreover,
corium interaction studies, which have been done in OpenFOAM 
\cite{mahboob2026corium} have yet to be ported, studied and validated yet.

These are in the realm of future work and will require more manpower.

\subsection{Source Term and Dispersion}

For source term and dispersion studies, these are important for nuclear 
safety analysis. TRISO Atops has already been ported into Outram Park,
but there is little else in this arena as of July 2026. 

\subsection{Digital Twin Work}

Initial FHR simulators have already been done in Outram Park prior to agentic coding. 

Figure~\ref{fig:fhr-sim-v1v2} shows the pre-agentic UI of the simulator:

\begin{figure}[H]
	\centering
	\includegraphics[width=0.95\textwidth]{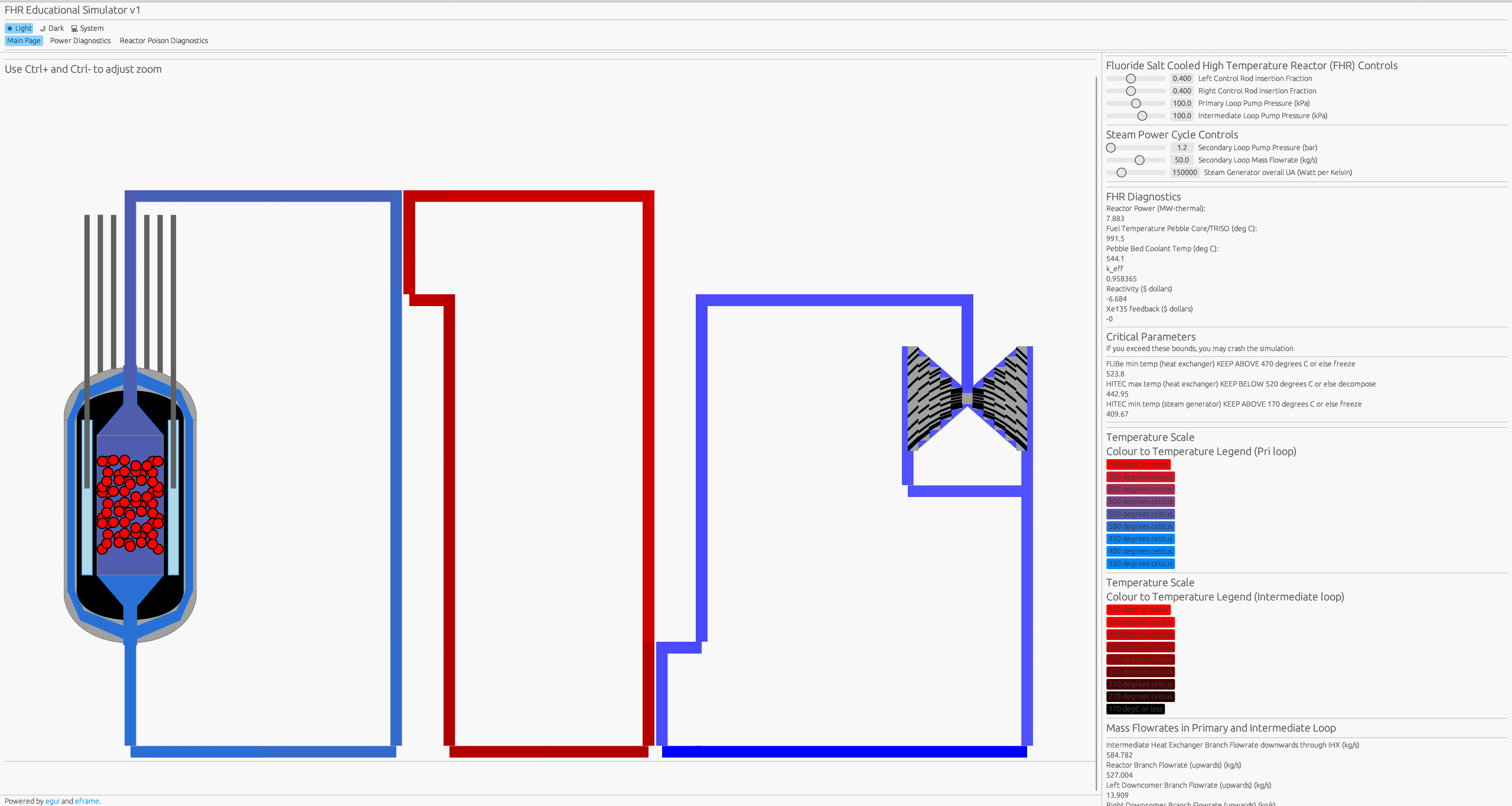}
	\caption{FHR Educational Simulator v1/v2 (still undergoing development)}
	\label{fig:fhr-sim-v1v2}
\end{figure}

The models in figure~\ref{fig:fhr-sim-v1v2}, rudimentary and require improvement.
This is because the steam turbine cycle is merely a steady state Rankine 
cycle loop without critical heat flux (CHF) or integration of 
the transient TampinesSteamArray developed in this work. The aesthetics,
User Interface and physics need to be incorporated into these 
simulators to improve it. Moreover, LWR and HTGR simulators 
need to be developed in order for Outram Park to be 
applicable to these reactors.

\section{Conclusions}

In this work, we present the use of Agentic AI in translating libraries,
performing initial V\&V and its role in solver development. Cost considerations 
and value analysis of Agentic AI have also been discussed in the context 
of Outram Park development in June-July 2026. 

AI-assisted Agentic Development has been greatly accelerated with the use 
of Claude Pro (1 year) and Claude Max (1 month) subscriptions totalling 
SGD\$600, and the value returned is roughly two orders of magnitude that 
amount. Even when considering raw API costs, Agentic AI could 
potentially return value approximately one order of magnitude 
cheaper compared to hiring a full software development team with 
nuclear expertise. 

Agentic development workflows have been shown to greatly improve speed and 
productivity, but we have shown in this research log, in the development 
of tampines-steam-tables and TampinesSteamArray HEM solver, that human 
familiarity with the numerical solution and physics of the problem cannot 
be 100\% substituted by AI. However, AI-assisted agentic translation and 
debugging have served to greatly enhance the speed and efficiency of 
iterative development workflows especially when the humans operating the 
AI Agents have specific domain knowledge and coding knowledge of the systems 
at hand. This is once again in line with what the Singapore government 
has been saying, that AI should not replace human beings, but augment them 
\cite{wong_singapore_budget_2026_ai,tan_mom_cos_2026}.

%
%
\noindent \textbf{ACKNOWLEDGMENTS} \vskip 1em
\noindent This work was performed with funding support from the 
Singapore Nuclear Research and Safety Institute (SNRSI). NRF Funding 
number (A-8002968-00-00).
The author also thanks colleagues at SNRSI for introducing him to 
vibe coding fundamentals and Claude Code, as well as many useful constructive 
conversations. These include, Goh Zhi Zheng, Darryl Foo, Than Yan Ren, Seow Chun 
Yong, Isaac Yap and Vitesh. Thanks to ZheXi Guo for introducing Zhang and 
Brooks validation cases, which are now part of future work.
The author also thanks SNRSI Director
Prof Chung Keng Yeow and SNRSI CEO Low Xin Wei continued support of Outram
Park's various libraries and advocating the use of Artificial Intelligence
in general. The author further thanks SNRSI CEO Low Xin Wei for permitting the
Claude Code subscription, which the author had initially purchased at his own
expense for personal use, to be claimed as a work expense once its value to
Outram Park development had been demonstrated.

\vskip 2em

\section{CRediT Authorship Contribution Statement}
\label{sec:credit}

\textbf{Theodore Kay Chen Ong:} Conceptualization; Methodology; Software;
Validation; Formal analysis; Investigation; Data curation; Visualization;
Writing --- original draft; Writing --- review \& editing; Project
administration.
\textbf{Sicong Xiao:} Supervision; Writing --- review \& editing.

\vskip 2em

\section{Declaration of Generative AI and AI-assisted Technologies in the
Manuscript Preparation Process}
\label{sec:declaration}

During the preparation of this work, the authors used Anthropic Claude Code
(Opus and Sonnet models primarily, with Haiku models for lighter-weight tasks
such as V\&V memo drafting, LaTeX table generation, and case run
orchestration) for agentic porting of OpenFOAM, NJOY, and CoolProp source
code into Rust under a strict provenance-tracking protocol, and for
authoring Rust utility code that
mechanically generates figures and tables from author-verified numerical data.
The authors also used Microsoft Enterprise Copilot, Claude Code
and ChatGPT as a sounding board for
discussing methodology, for literature search assistance, for cross-checking
references, for assistance with \LaTeX{} table syntax and matplotlib plotting
code, and for critique and feedback on hand-written Singlish draft prose. All
references identified with AI assistance were independently verified by the
authors against primary sources. The National University of Singapore's
institutional AI assistant (NUS AI Know), which routes user prompts to
Anthropic Claude, OpenAI ChatGPT, and Google Gemini backends under
institutional data-governance controls, was evaluated but not adopted for
this project: at the time of use it operated as a conversational assistant
only, without the ability to autonomously read, edit, run, and iterate on
a multi-file Rust workspace, which was required for the agentic porting
scale of this work. AI tools were used extensively in the preparation of the
manuscript text: the authors drafted informal engineering notes and
technical brainstorming material, in part in Singlish (Singapore Colloquial
English), and used AI assistance to convert these into publication-ready
academic prose. The underlying technical content, analysis, interpretation, and
conclusions are the authors' own; AI assistance operated on the expression
of that material rather than its substance, and every AI-revised passage
was reviewed and edited by the authors. All
AI-generated code was reviewed, tested, and integrated into the Outram Park
repository under GPLv3 provenance-tracking conventions. The authors reviewed
all AI-assisted output and take full responsibility for the content of the
publication. Further detail on how AI tools were used, including a verbatim
example of the Singlish-to-academic-prose workflow, is given in the methods
section.

The productivity comparison reported in this work is drawn against a baseline
of four earlier repositories by the same author, whose AI-use status differs
and is stated here for transparency. \texttt{thermal\_hydraulics\_rs}
(June 2023 to October 2024) and \texttt{tuas\_boussinesq\_solver} (October 2024
to June 2026) were developed without AI assistance of any kind.
\texttt{tampines-steam-tables} (January 2025 to June 2026) used minimal AI
assistance, confined to root-finding solvers and similar numerical routines.
\texttt{boon-lay} (January to May 2026) used non-agentic AI code generation,
for user interface code and some Monte Carlo solvers, alongside conversational
assistance. Three of \texttt{boon-lay}'s data dependencies ---
\texttt{openmc-endf-8-depletion-lib-a} \cite{ong2026openmcdepletiona},
\texttt{-lib-b} \cite{ong2026openmcdepletionb} and
\texttt{fission-yields-data} \cite{ong2026fissionyieldsdata} --- require
separate comment. Their serialisation parsers were generated largely by OpenAI
ChatGPT-5 (non-agentic, conversational) with author corrections. The bulk of
each repository, however, is ENDF/B-VIII.0 decay-chain and fission-yield data
taken verbatim from the OpenMC project; the author separated this data by
nuclide and wrapped it into Rust functions by hand in Neovim. It is therefore
neither authored logic nor AI-generated code, but reorganised public reference
data, and it is excluded from the productivity accounting on that basis. None of these repositories employed an
agentic coding workflow, in which an AI tool autonomously reads, edits, runs,
and iterates on a multi-file workspace. The comparison reported here is therefore between agentic
development and non-agentic development, where the latter comprises both
unassisted manual programming and conversational AI assistance. It is not a
comparison between AI-assisted and AI-free development, and should not be read
as one. This distinction is deliberate: it isolates the agentic workflow,
rather than the availability of a language model, as the variable under
examination.

\printbibliography

\appendix

\section{Reproducing the Line Counts}
\label{sec:reproducing_the_counts}

The productivity claims above rest entirely on line counts, and a line count is
only as good as the rule used to produce it. Tables~\ref{tab:preagentic_baseline}
and~\ref{tab:agentic_crates} and Figure~\ref{fig:kloc_productivity} are
therefore not compiled by hand. They are emitted by a single script,
\texttt{kloc\_accounting.py}, which is distributed with this manuscript and
which takes its inputs directly from the repositories and their git histories.
Anyone may re-derive every number by running

\begin{center}
\texttt{python3 kloc\_accounting.py -{}-from-github -{}-check}
\end{center}

\noindent which clones the public repositories, measures them, and prints a
row-by-row comparison against the figures printed here. It requires nothing on
the machine beyond \texttt{git}, \texttt{python3} and \texttt{matplotlib}.

Four decisions in that script materially affect the result and are stated here
rather than buried:

\begin{enumerate}
\item \textbf{Comment stripping is lexical, not regular.} A code line is a line
carrying at least one non-whitespace character once comments are removed. The
remover is a scanner that tracks nested block comments, raw strings
(\texttt{r\#"..."\#}), byte strings and character literals, so that a
\texttt{//} inside a string literal is not mistaken for a comment and a
lifetime \texttt{'a} is not mistaken for a character. Doc comments, which in
Rust carry executable doctests, are excluded from the code count and appear
only in the total.
\item \textbf{Measurement is on \texttt{develop}, not \texttt{main}.} These
repositories squash-merge into \texttt{main}. Measuring \texttt{main} reports
nine active days for \texttt{thermal\_hydraulics\_rs} where the development
branch records 144, and would understate the pre-agentic baseline by more than
a factor of two --- which would inflate, not deflate, the productivity claim.
\item \textbf{An emptied repository is measured before it was emptied.}
\texttt{thermal\_hydraulics\_rs} was reduced from 260 Rust files to nine on
2024-10-11, when its code moved into TUAS. It is therefore counted at commit
\texttt{4d534af}, the last commit at full extent. Its branch tip would report
1.6~KLOC and would make the "TUAS net of what it inherited" subtraction
meaningless.
\item \textbf{TUAS is netted against what it imported, not against its
predecessor's size.} TUAS was spun out of \texttt{thermal\_hydraulics\_rs} and
imported it wholesale at its second commit, \texttt{c451c8e}, 42~minutes after
the initial commit. The subtrahend is that imported tree --- 236 files,
52\,054 code lines --- and not the predecessor's own full extent of 260 files
and 59\,808 code lines. The difference, 7\,754 code lines, is work written in
the predecessor and abandoned rather than carried forward. Netting against the
predecessor's full size would delete that work from the baseline, and since
the baseline is the denominator of the productivity ratio, doing so would
inflate the headline figure from 12.3 to 12.8. The narrower reading of
"inherited" is the one that can be defended, and it is the one used here.
\item \textbf{Provenance classification is the one editorial input.} Whether a
crate counts as a translation, as original work, or as an extension of vendored
pre-agentic code is a judgement, not a measurement. Those judgements are listed
explicitly in the script, and each crate's own \texttt{Cargo.toml} description
is carried into the emitted CSV beside the classification so that a reader can
disagree with any individual assignment and recompute the subtotals.
\item \textbf{Classification follows the code, not the package boundary.} The
two terminal interfaces, \texttt{njoy-tui} and \texttt{outram-mc-tui}, ceased to
be workspace crates on 2026-07-23 and became feature-gated binaries inside the
libraries they drive, so that a library consumer no longer sees them as
separate packages. Their code is newly written, but it now sits inside crates
that are ports of an existing upstream. The script therefore splits them out at
the path boundary (\texttt{src/bin/}) and credits them to original work; folding
them into their host crates would have moved 2.5~KLOC of interface code into the
translated column and overstated the translated share.
\end{enumerate}

The script refuses to fail silently in the two ways that would matter. A crate
present in the repository but absent from the classification is reported and
excluded rather than quietly folded into a subtotal, and a classification with
no corresponding crate in the checkout is reported as stale. Both checks earned
their place: the crate reorganisation described above was detected by the second
of them rather than noticed by hand.

Finally, these counts measure volume, not worth. Lines of code are a poor proxy
for engineering value at the best of times, and they are a particularly poor
one here, where 77~per cent of the agentic output is translation of an existing
and already-debugged implementation. The counts are reported because they are
reproducible, not because they are sufficient.

\section{Token Consumption of an Agentic Crate}
\label{app:token_accounting}

Token accounting was not performed properly during the early Rust translations
of Outram-Foam and Outram MC. Token accounting scripts and instructions for
Claude Code were introduced later. 

This makes it easier to see how much an agentic porting costs in terms 
of tokens. This is useful for management decisions, since management may want 
to know a realistic upper limit of API costs in the long run as opposed 
to subscription costs. 

To check, we used Claude Code to translate Offbeat, which is 
the fuel performance sister code of GeN-Foam\cite{nervi2026foamfornuclear} , into 
\texttt{outram-park-fork-offbeat},  one single active day sufficed 
(2026-07-29), got ten commits totalling 13\,863 lines of Rust code. This 
was done after the snapshot used in the main body text. The reader should note 
that agentic development of Outram Park is an ongoing process. By the 
time the papers come out, the figures here will already be outdated. This 
should be read as historical snapshots, and the offbeat translation is an 
addendum to the work discussed in the prose. Do note also that offbeat 
postdates any of the agentic coding work described in the prose.

AI-generated scripts with Claude Code were used to insert token usage counts 
into git commit histories. Then after that, Claude Code was asked to 
summarise token usage between commits for offbeat from the Git histories.
These were done in Tables~\ref{tab:offbeat_cost} and \ref{tab:offbeat_tokens}:

\begin{table}[H]
\centering
\caption{API token consumption for \texttt{outram-park-fork-offbeat}, ten
commits on 2026-07-29. \texttt{total} $=$ \texttt{in} $+$ \texttt{out} $+$
\texttt{cache\_read} $+$ \texttt{cache\_write}.}
\label{tab:offbeat_tokens}
\footnotesize
\setlength{\tabcolsep}{8pt}
\begin{tabular}{@{}l r r@{}}
\toprule
\textbf{Category} & \textbf{Tokens} & \textbf{Share} \\
\midrule
Input           &       2\,867 & 0.0\,\% \\
Output          &     658\,725 & 0.2\,\% \\
Cache write     &   6\,440\,829 & 2.1\,\% \\
Cache read      & 305\,272\,222 & 97.7\,\% \\
\midrule
\textbf{Total}  & \textbf{312\,374\,643} & \textbf{100\,\%} \\
\bottomrule
\end{tabular}
\end{table}

\begin{table}[H]
\centering
\caption{Estimated API cost for \texttt{outram-park-fork-offbeat} at
Claude Opus~4.8 list rates (input \$5.00, output \$25.00, cache read \$0.50, cache
write \$6.25 at the five-minute cache tier per million tokens). The cache-write
tier is not recorded in the trailers; the one-hour tier (\$10.00 per million)
gives the upper figure.}
\label{tab:offbeat_cost}
\footnotesize
\setlength{\tabcolsep}{8pt}
\begin{tabular}{@{}l r r@{}}
\toprule
\textbf{Category} & \textbf{Rate (\$/Mtok)} & \textbf{Cost (\$)} \\
\midrule
Input           & 5.00  & 0.01 \\
Output          & 25.00 & 16.47 \\
Cache read      & 0.50  & 152.64 \\
Cache write (5\,min) & 6.25  & 40.26 \\
\midrule
\textbf{Total (5\,min cache)}  & & \textbf{209.37} \\
Total (1\,h cache)             & & 233.53 \\
\bottomrule
\end{tabular}
\end{table}

We can see that, assuming Opus costs (this is the upper limit), offbeat 
would have cost about USD\$210 for the entire code translation based on API 
costs. Compared to Claude Max subscription for one month, which is USD\$200,
one offbeat library translation would have easily made up for the sunk 
cost of Claude Max.

Assuming a 500 lines/active coding day pre-agentic coding speed, offbeat
would have taken 26 engineer-days to develop. Assuming USD\$200 per working 
day, this would mean a cost of about USD\$5,200 to translate offbeat. In 
reality, it may have taken longer and cost more if such an engineer 
did not have the domain knowledge of the numerical methods and engineering  
knowledge to translate offbeat by hand. In the case of the author, the 
author's main knowledge is thermal-hydraulics and digital twins 
\cite{ong2024digital}, so fuel performance lies outside his expertise. However,
translation of offbeat via AI to produce such a draft demonstrates the
use of AI to translate code outside the author's main domain knowledge 
within nuclear engineering. This illustrates the capability that agentic 
coding offers the user.

Thus, even in this case, using the API costs, it would be more 
economical to translate offbeat using Claude Code as compared to human 
translation.

\section{Development of Tampines Steam Tables}
\label{app:tampines}

\begingroup
\footnotesize
\setlength{\tabcolsep}{4pt}
\begin{longtable}{@{}l>{\raggedright\arraybackslash}p{0.28\linewidth}>{\raggedright\arraybackslash}p{0.40\linewidth}l@{}}
\caption{Major development milestones of \texttt{tampines-steam-tables}, from the standalone repository (v0.0.1--v0.1.8) through migration into the \texttt{outram-park-backend} workspace (v0.2.0--v0.2.4).}
\label{tab:tampines-history}\\
\toprule
\textbf{Date} & \textbf{Milestone} & \textbf{Implementation and verification} & \textbf{Commit} \\
\midrule
\endfirsthead

\multicolumn{4}{@{}l}{\footnotesize\itshape Table \thetable\ (continued)}\\
\toprule
\textbf{Date} & \textbf{Milestone} & \textbf{Implementation and verification} & \textbf{Commit} \\
\midrule
\endhead

\midrule \multicolumn{4}{r@{}}{\footnotesize\itshape continued on next page}\\
\endfoot
\bottomrule
\endlastfoot

\multicolumn{4}{@{}l}{\textbf{Phase I --- Standalone repository: IAPWS-IF97 property library}}\\
\addlinespace[2pt]
2025-01 & Project bootstrap & Physical constants, triple and boiling points, region 1--5 coefficient tables & \texttt{3e266d5} \\
2025-01 & Region 1 forward equations (v0.0.1) & Gibbs energy and derivatives for subcooled liquid; dimensioned with \texttt{uom}. Verified vs.\ IST Table 2.5 sets A--C & \texttt{25d2a67} \\
2025-01 & Region 2 and metastable vapour (v0.0.2) & Ideal-gas and residual $\gamma$ terms, $\kappa$, $\alpha$, $\kappa_T$; separate metastable subregion. Verified vs.\ IST region 2 sets A--C & \texttt{8ba86e5} \\
2025-01 & Region 3, B23 boundary, saturation line (v0.0.3--4) & Helmholtz $\phi$ derivatives; auxiliary B23 equation; region 4 saturation $p$--$T$. Verified vs.\ IST region 3 sets A--C & \texttt{f9d18ec} \\
2025-01 & Region 5 high-temperature steam (v0.0.5) & Forward equations above \SI{800}{\celsius}. Verified vs.\ IST region 5 sets A--C & \texttt{eb77b17} \\
2025-01 & Backward $(p,h)$ equations (v0.0.6) & $T(p,h)$ regions 1--2, subregion 2b--2c boundary, $v(p,h)$ and $T(p,h)$ for 3a/3b. Verified vs.\ IST Tables 2.38, 2.42 & \texttt{4cf0793} \\
2025-01/02 & Region dispatch and $(p,T)$ flash & Functional \texttt{interfaces/} API, validity-range checks, boundary logic across all five regions and the two-phase dome & \texttt{3663628} \\
2025-01/02 & Full $(p,h)$ flash verification & $v$, $T$, $s$, $x$, $c_p$, $c_v$, $w$ over the whole IF97 surface, \SIrange{0.00611}{1000}{bar}, \SIrange{0}{800}{\celsius}; accuracy loss documented near the critical point & \texttt{46fd78c} \\
2025-02 & Two-phase $(p,T)$ flash & Multiphase flash; revised $\kappa$/$\alpha$ in region 3 near saturation; near-critical enthalpy fix & \texttt{d9cc877} \\
2025-02 & Dynamic viscosity (IAPWS 2008) & $\psi_0$, $\psi_1$ terms with $(\rho,T)$, $(p,T)$, $(p,h)$ entry points; agreement within \SI{2}{\percent} & \texttt{3f60ab4} \\
2025-02 & Thermal conductivity (IAPWS 2011) (v0.0.7--8) & $\lambda_0$, $\lambda_1$, $\lambda_2$ critical enhancement; larger uncertainty documented at 160/\SI{220}{bar} and supercritical & \texttt{6978dab} \\
2025-02 & Backward $(p,s)$ equations and flash (v0.0.9) & $T(p,s)$ regions 1, 2a--2c; $v(p,s)$, $T(p,s)$ for 3a/3b; $B_{ps3}$ boundary. Verified \SIrange{10}{800}{bar} & \texttt{e1208ba} \\
2025-02 & Surface tension and dielectric constant (v0.1.0) & IAPWS surface-tension correlation and Harvey--Fern\'andez-Prini dielectric constant & \texttt{a483cfd} \\
2025-03/04 & Backward $(h,s)$ equations (v0.1.1) & $p(h,s)$ regions 1, 2a--2c, 3a/3b; $T_\mathrm{sat}(h,s)$ region 4; B13 and $T_{B23}$ boundaries. Partial: inaccurate below \SI{10}{bar} & \texttt{3703731} \\
2025-03--2026-03 & OpenFOAM algorithm ports & Rust translations of \texttt{simpleFoam}, \texttt{rhoPimpleFoam}, \texttt{driftFluxFoam}, \texttt{chtMultiRegionTwoPhaseEulerFoam} with \LaTeX{} derivations & \texttt{fecd17d} \\
2025-06 & FHR educational simulator v1 (v0.1.2, tagged) & \texttt{egui} GUI; steam generator and Rankine loop coupled to salt loops via \texttt{tuas\_boussinesq\_solver} and \texttt{teh-o-prke} & \texttt{33bfa9b} \\
2026-02/03 & Object-oriented control-volume API (v0.1.4) & \texttt{TampinesSteamTableCV} wrapper, transient mass balance, \texttt{serde} support & \texttt{df9edfc} \\
2026-03 & Steam turbine and generator models & Euler/impulse turbine equations, velocity triangles, torque and EMF braking, discretised three-phase generator (\SI{250}{MW_e}) & \texttt{e9a8824} \\
2026-03 & Isentropic nozzle solver & Force-balance residual, entropy iteration after $(h,s)$/$(p,s)$ disagreement, iterative $(p,h)$-based $(h,s)$ flash for low quality & \texttt{5c3043d} \\
2026-03/04 & Choked-flow foundations & Critical pressure ratio, HEM sound speed, nozzle/diffuser $\mathrm{d}p$--$\mathrm{d}A$--$\mathrm{d}v$ relations, Rayleigh-line balance. Verified vs.\ \c{C}engel worked examples & \texttt{01cc3e7} \\
2026-04/05 & Converging--diverging nozzle regimes (v0.1.6--7) & Bisection/secant solvers for subsonic, perfectly-expanded and overexpanded branches; Mach-number and Joule--Thomson tests & \texttt{8b1c313} \\
2026-05 & Marviken critical-flow tests (v0.1.7) & Blowdown harness after the Marviken full-scale vessel series (NUREG/CR-2671) & \texttt{8ad8819} \\
2026-05/06 & Moody critical mass flux (HEM) & Critical mass flux from stagnation $(p_0,h_0)$; isobar verification vs.\ Moody (1975), NEDO-21052, $p_0/p_\mathrm{ref}=0.5$--$30$ & \texttt{0b5c72f} \\
2026-06 & Wood--Wallis and equilibrium sound speed & \texttt{w\_ps\_eqm} via finite-difference volume derivatives; verified vs.\ Kieffer (1977) eq.~28, except 100/\SI{200}{bar} near saturation & \texttt{2329742} \\
2026-06 & Zaloudek dataset and round-trip validation & Digitised critical-pressure curves for $x=0$--$0.75$; forward and backward throat$\to$stagnation tests; bubble-point edge cases. Data per Saha (1978), NUREG/CR-0417 & \texttt{aaed718} \\
2026-06 & Choked-flow module split (v0.1.8) & Separate single-phase, in-dome and outside-dome stagnation solvers; in-dome passing, near-bubble-point artifact unresolved at migration & \texttt{e7b5097} \\
\addlinespace[4pt]
\multicolumn{4}{@{}l}{\textbf{Phase II --- Migrated into \texttt{outram-park-backend}: dependency refresh and new functionality}}\\
\addlinespace[2pt]
2026-06-19 & Migration into workspace & Vendored with three sibling crates (813 files, +291k lines); v0.1.8$\to$v0.2.0; dependencies centralised in \texttt{[workspace.dependencies]} & \texttt{8a048d3} \\
2026-06-22 & Dependency upgrade & \texttt{uom} 0.36$\to$0.38, \texttt{ndarray} 0.15$\to$0.17, \texttt{thiserror} 1.0$\to$2.0, \texttt{egui} 0.29$\to$0.34; breaking-change fixes across the crate & \texttt{0148c75} \\
2026-06-24 & BLAS decoupling & \texttt{ndarray-linalg} and \texttt{openblas-src} demoted then removed; library reduced to four dependencies & \texttt{85ac0d8} \\
2026-06-26 & Near-bubble-point artifact resolved & Root-caused as numerical, not physical; quality discriminator $x<0.03$ plus precomputed sonic mass flux along the saturated-liquid line. Zaloudek $x_t=0$--$1.0$ within $\pm0.04$ in $\log_{10}G$ & \texttt{06e33fb} \\
2026-06-26 & Superheated/supercritical HEM solver & \texttt{get\_critical\_pressure\_and\_mass\_flux\_superheated\_vapour\_ph} for stagnation states outside the dome & \texttt{3dbbdd9} \\
2026-06-29 & Unified HEM dispatcher & Routes $(p_0,h_0)$ by \texttt{ph\_flash\_region} to in-dome, subcooled and superheated solvers; handles region-3 isentropes re-entering the dome. All 13 stagnation tests re-enabled; region~3 $<\SI{0.5}{\percent}$, region~1 $<\SI{1.4}{\percent}$ & \texttt{50632f0} \\
2026-06-29 & Transient mass and energy balance API & \texttt{CvMassEnthalpyChanges} ledger and \texttt{advance\_timestep()}; regula-falsi $(p,h)$ inversion on target specific volume & \texttt{0ffbe91} \\
2026-06-30 & Moody isobars pass (v0.2.1) & Region-filtered tests across 13 isobars at \num{0.06} $\log_{10}G$ tolerance; subcooled Moody--Zaloudek regime reconciliation documented & \texttt{7173480} \\
2026-07-01 & \texttt{TampinesSteamArray} scaffolded & One-dimensional compressible PIMPLE pipe solver on an \texttt{FvMesh}; vendored OpenFOAM field, FV-operator, LDU-matrix and thermophysics sources & \texttt{01bda7c} \\
2026-07-06 & TUAS dependency dropped & Property core made fully standalone; \texttt{tuas\_boussinesq\_solver} removed as a library dependency & \texttt{18d7a6f} \\
2026-07-13 & \texttt{tampines} framework crate & Facade composing steam tables, CoolProp fork and DWSIM equipment models; balance-of-plant and cooling-tower modules; hosts \texttt{fhr\_sim\_v2} & \texttt{5bca434} \\
2026-07-13 & Real IF97 flash in \texttt{correct\_thermo()} & Placeholder EOS replaced by a genuine two-phase $(p,h)$ flash in the array solver & \texttt{356daff} \\
2026-07-14 & Steam-generator tube integration & Array wired into \texttt{fhr\_sim\_v2}; OpenFOAM-style pressure bounding; \texttt{lambda\_ph\_eqm} critical-enhancement term made two-phase-safe & \texttt{032e95f} \\
2026-07-15 & Android portability (v0.2.2) & $(p,h)$ flash hardening for regions 4--5; dev-dependencies gated off \texttt{aarch64-linux-android} & \texttt{bbde28a} \\
2026-07-16 & Edwards--O'Brien blowdown V\&V (v0.2.3) & 24-CV RELAP nodalisation, Hendrie (1973) axial IC, HEM break BC, \SI{600}{ms} at $\Delta t=\SI{1e-4}{s}$; conservative energy derivative and two-phase $\psi=\partial\rho/\partial p|_h$ by finite difference. GS-1 RMSE \SI{60}{psia}, down from \num{276} & \texttt{8380513} \\
2026-07-16 & \texttt{HybridAllMach} solver stabilised (v0.2.4) & KNP central-upwind flux with Kieffer HEM sound speed plus rarefied-tail blend taper. GS-1 RMSE \SI{30.6}{psia} vs.\ \num{58.6} for PIMPLE; 920 library tests pass & \texttt{46b75a6} \\
\end{longtable}
\endgroup

\section{Developer Health Warning: Agentic Software Development}
\label{app:health_warning}

The following reproduces, near-verbatim, the project's
\texttt{DEVELOPER\_HEALTH\_WARNING.md}, authored by the maintainer 
(agentically) after roughly
one month of intensive agentic development. The prose is unchanged; three
presentational changes have been made for this manuscript. The in-text
citations, written in APA style in the original, are rendered here in the numeric
style of this research log. The original document's separate reference list has been
folded into the manuscript bibliography, where its ten sources appear via the
citations below. And the two flow sketches, drawn as ASCII art in the original,
have been redrawn as diagrams with their node text preserved --- these are
therefore \emph{not} verbatim reproductions of the original figures.

\begin{tcolorbox}[breakable, colback=red!5!white, colframe=red!60!black,
  fonttitle=\bfseries, title={Warning}]
Outram Park was developed using highly agentic software engineering workflows
involving Claude Code, Microsoft Copilot, ChatGPT, Gemini, and other AI-assisted
development tools. While these tools can dramatically increase engineering
productivity, they may also increase the risk of technostress, cognitive
fatigue, decision fatigue, hyperfocus, insomnia, burnout, and prolonged illness.

Developers are encouraged to treat wellbeing, recovery, and sustainable working
practices as first-class engineering requirements.
\end{tcolorbox}

\subsection*{Why This Document Exists}

This document was written after approximately one month of intensive agentic
software development on the Outram Park project.

During this period, the maintainer experienced:

\begin{itemize}
\item Persistent fatigue.
\item Decision fatigue.
\item Difficulty disengaging from development activities.
\item Hyperfocus lasting many consecutive hours.
\item Sleep disruption and insomnia.
\item Reduced awareness of accumulating exhaustion.
\item A prolonged sore throat lasting approximately one month.
\item Continued development activity despite recognized fatigue.
\end{itemize}

These experiences motivated a review of:

\begin{itemize}
\item Repository commit history.
\item Working-hour patterns.
\item Existing literature on technostress.
\item Existing literature on cognitive fatigue.
\item Existing literature on hyperfocus.
\item Existing literature on psychological stress and immune function.
\end{itemize}

The goal of this document is to help future contributors avoid repeating the
same mistakes.

\subsection*{1. Anecdotal Evidence from the Outram Park Project}

\subsubsection*{Commit History Analysis}

A review of repository history identified:

\begin{table}[H]
\centering
\small
\begin{tabular}{@{}lr@{}}
\toprule
\textbf{Metric} & \textbf{Value} \\
\midrule
Total commits analysed           & 282 \\
Commits inside working hours     & 201 \\
Commits outside working hours    & 81 \\
Percentage outside working hours & 28.7\% \\
\bottomrule
\end{tabular}
\end{table}

This means:

\begin{itemize}
\item Nearly \textbf{29\% of all commits} occurred outside working hours.
\item Roughly \textbf{1 in every 3.5 commits} was made outside the eventual
safety window.
\item Almost one-third of repository activity occurred during periods later
designated as recovery or off-work time.
\end{itemize}

\paragraph{Breakdown of Outside-Hours Activity}

\begin{table}[H]
\centering
\small
\begin{tabular}{@{}lr@{}}
\toprule
\textbf{Category} & \textbf{Count} \\
\midrule
Weekday commits before 07:30 or after 20:00 & 52 \\
Sunday commits outside permitted hours      & 13 \\
Saturday commits (full rest day)            & 16 \\
\midrule
\textbf{Total outside-hours commits}        & \textbf{81} \\
\bottomrule
\end{tabular}
\end{table}

Observed patterns included:

\begin{itemize}
\item Late-night development sessions.
\item Activity extending beyond 22:00.
\item Occasional development after midnight.
\item Multiple weekend coding sessions.
\item Repeated Sunday evening work.
\item Several full Saturdays spent developing software.
\end{itemize}

Importantly, the working-hours policy discussed later did \textbf{not} exist
during most of this period. The majority of these commits therefore do not
represent policy violations; rather, they represent historical behaviour that
preceded the creation of project-level health safeguards.

Nevertheless, the commit history provides an objective record of sustained
after-hours development activity.

\subsubsection*{The Productivity Trap}

One particularly important observation was that excessive work was not primarily
driven by project deadlines.

A major motivating factor was:

\begin{quote}
Maximizing utilization of a Claude Pro subscription.
\end{quote}

The thought process often looked like:

\begin{center}
\begin{tikzpicture}[node distance=5mm,
  fbox/.style={draw, rounded corners, align=center, font=\small,
               minimum width=6.6cm, minimum height=0.66cm, inner sep=3pt}]
\node (a) [fbox] {Available token budget};
\node (b) [fbox, below=of a] {Unused AI capacity};
\node (c) [fbox, below=of b] {``I should use what I am paying for''};
\node (d) [fbox, below=of c] {``One more feature''};
\node (e) [fbox, below=of d] {``One more benchmark''};
\node (f) [fbox, below=of e] {``One more refactor''};
\node (g) [fbox, below=of f] {Several additional hours of work};
\draw[wfflow] (a) -- (b);
\draw[wfflow] (b) -- (c);
\draw[wfflow] (c) -- (d);
\draw[wfflow] (d) -- (e);
\draw[wfflow] (e) -- (f);
\draw[wfflow] (f) -- (g);
\end{tikzpicture}

\smallskip
{\footnotesize\itshape Redrawn as a diagram; the original document rendered this
as an ASCII sketch, so this figure is not verbatim.}
\end{center}

This behaviour was reinforced by:

\begin{itemize}
\item Immediate implementation availability.
\item Extremely short development feedback loops.
\item Rapid visible project progress.
\item The feeling that unused model capacity represented wasted value.
\end{itemize}

In retrospect, this created a novel productivity trap:

\begin{quote}
Because software development had become unusually inexpensive, there was
increased pressure to continue generating more work.
\end{quote}

The AI model never became tired.

The developer did.

\subsubsection*{The One-Month Illness}

During the same period, the maintainer experienced:

\begin{itemize}
\item Persistent fatigue.
\item Sleep disruption.
\item Decision fatigue.
\item A sore throat lasting approximately one month.
\end{itemize}

This document does \textbf{not} claim that AI tools directly caused illness.

However, these observations motivated a review of literature exploring the
relationship between:

\begin{itemize}
\item Technostress,
\item Hyperfocus,
\item Sleep disruption,
\item Cognitive fatigue,
\item Psychological stress,
\item Immune function.
\end{itemize}

\subsection*{2. Engineering Safety Controls}

\subsubsection*{Working-Hour Restrictions}

Following this period, formal safeguards were introduced into \texttt{CLAUDE.md}.

Agentic development is restricted to:

\begin{table}[H]
\centering
\small
\begin{tabular}{@{}ll@{}}
\toprule
\textbf{Day} & \textbf{Working Hours} \\
\midrule
Monday--Friday & 07:30--20:00 \\
Saturday       & No agentic development \\
Sunday         & 12:00--19:00 \\
\bottomrule
\end{tabular}
\end{table}

Outside these periods:

\begin{itemize}
\item Claude Code must not perform agentic development work.
\item Claude Code may assist only with planning activities.
\item Tasks should instead be recorded using Beads or TODO systems.
\end{itemize}

\subsubsection*{Purpose of the Policy}

The purpose of this policy is not to reduce productivity.

The purpose is to:

\begin{itemize}
\item Protect sleep.
\item Protect weekends.
\item Protect long-term sustainability.
\item Reduce burnout risk.
\item Preserve engineering judgement.
\item Prevent recurrence of prolonged illness.
\end{itemize}

\subsection*{3. Literature Review}

\subsubsection*{Technostress}

Technostress refers to stress arising from technology use and technology-enabled
work.

Tarafdar, Tu, and Ragu-Nathan \cite{tarafdar2007technostress} introduced several
major dimensions of technostress:

\begin{itemize}
\item Techno-overload.
\item Techno-invasion.
\item Techno-complexity.
\item Techno-insecurity.
\item Techno-uncertainty.
\end{itemize}

\paragraph{Techno-Overload}

Techno-overload occurs when technology increases the amount of work an individual
can perform and therefore increases expectations and workload
\cite{tarafdar2007technostress}.

Agentic development workflows may amplify techno-overload because:

\begin{itemize}
\item Code generation becomes dramatically faster.
\item Refactoring becomes inexpensive.
\item Architectural experimentation becomes easier.
\end{itemize}

The bottleneck therefore shifts from implementation toward:

\begin{itemize}
\item Review,
\item Verification,
\item Validation,
\item Decision making.
\end{itemize}

\paragraph{Techno-Invasion}

Techno-invasion occurs when technology erodes the boundary between work and
personal life \cite{tarafdar2007technostress}.

Examples include:

\begin{itemize}
\item Evening coding sessions.
\item Weekend work.
\item Difficulty disengaging from projects.
\item The feeling that productive work remains continuously available.
\end{itemize}

Tarafdar, Cooper, and Stich \cite{tarafdar2019trifecta} note that technology can
simultaneously generate benefits (``techno-eustress'') and harms
(``techno-distress''), highlighting the need for deliberate safeguards.

\subsubsection*{Hyperfocus}

Hyperfocus refers to periods of unusually intense attentional engagement.

Research has documented hyperfocus across autistic, ADHD, and general
populations \cite{dwyer2024monotropism}.

Dwyer et al. \cite{dwyer2024monotropism} reported associations between hyperfocus
and:

\begin{itemize}
\item Reduced quality of life.
\item Repetitive thinking.
\item Greater anxiety symptoms.
\item Attentional regulation difficulties.
\end{itemize}

At the same time, hyperfocus is not inherently negative.

Dupuis et al. \cite{dupuis2022hyperfocus} found that intense attentional
engagement may also function as a genuine cognitive strength that contributes to
productivity and deep work.

Consequently:

\begin{quote}
Hyperfocus may facilitate exceptional progress while simultaneously increasing
the risk of overwork.
\end{quote}

Warning signs include:

\begin{itemize}
\item Skipping meals.
\item Missing breaks.
\item Delaying sleep.
\item Losing awareness of time.
\item Continuing work despite obvious fatigue.
\end{itemize}

\noindent\cite{dwyer2024monotropism, dupuis2022hyperfocus}

\subsubsection*{Cognitive Fatigue}

Cognitive fatigue emerges following prolonged periods of demanding mental effort.

Pessiglione et al. \cite{pessiglione2025fatigue} reviewed evidence suggesting
that cognitive fatigue:

\begin{itemize}
\item Accumulates over time.
\item Is often poorly recognized by the individual experiencing it.
\item Alters subsequent decision making.
\item Reduces willingness to engage in demanding cognitive tasks.
\end{itemize}

Steward et al. \cite{steward2025neurobiology} further demonstrated that repeated
cognitive exertion alters subsequent effort-based decision making.

Their findings suggest that fatigue affects not only performance but also future
willingness to exert effort \cite{steward2025neurobiology}.

In agentic development:

\begin{itemize}
\item Coding effort decreases.
\item Review effort increases.
\item Validation effort increases.
\item Architectural decision making increases.
\end{itemize}

Consequently, significant cognitive fatigue may develop despite reduced manual
programming effort.

\subsubsection*{Decision Fatigue}

Decision fatigue refers to deterioration in decision quality following prolonged
decision making.

Choudhury and Saravanan \cite{choudhury2026decision} identified several
consequences:

\begin{itemize}
\item Reduced efficiency.
\item Lower-quality decisions.
\item Increased cognitive burden.
\item Choice avoidance.
\end{itemize}

Agentic systems frequently generate:

\begin{itemize}
\item Multiple architectures.
\item Multiple valid implementations.
\item Multiple debugging approaches.
\item Multiple optimization strategies.
\end{itemize}

The developer therefore becomes responsible for choosing among many plausible
alternatives.

AI systems may therefore reduce implementation effort while simultaneously
increasing decision-making effort.

\subsubsection*{Stress, Immunity, and Illness}

The relationship between psychological stress and immune function is one of the
strongest empirical foundations relevant to this document.

Cohen, Tyrrell, and Smith \cite{cohen1991commoncold} conducted a landmark viral
challenge study in which healthy volunteers were intentionally exposed to
respiratory viruses.

They observed a dose-dependent relationship between psychological stress and
susceptibility to respiratory illness \cite{cohen1991commoncold}.

Participants experiencing higher levels of psychological stress were more likely
to:

\begin{itemize}
\item Become infected.
\item Develop clinical cold symptoms.
\end{itemize}

\noindent\cite{cohen1991commoncold}

Subsequent reviews concluded that psychological stress is associated with
measurable changes in immune functioning and increased susceptibility to upper
respiratory infections \cite{cohen1995respiratory, cohen1996immunity}.

The literature therefore supports the plausibility of the following pathway:

\begin{center}
\begin{tikzpicture}[node distance=5mm,
  fbox/.style={draw, rounded corners, align=center, font=\small,
               minimum width=6.8cm, minimum height=0.66cm, inner sep=3pt}]
\node (a) [fbox] {Agentic development};
\node (b) [fbox, below=of a] {Technostress};
\node (c) [fbox, below=of b] {Hyperfocus};
\node (d) [fbox, below=of c] {Extended working hours};
\node (e) [fbox, below=of d] {Sleep disruption};
\node (f) [fbox, below=of e] {Chronic psychological stress};
\node (g) [fbox, below=of f] {Immune system effects};
\node (h) [fbox, below=of g] {Increased susceptibility to illness};
\draw[wfflow] (a) -- (b);
\draw[wfflow] (b) -- (c);
\draw[wfflow] (c) -- (d);
\draw[wfflow] (d) -- (e);
\draw[wfflow] (e) -- (f);
\draw[wfflow] (f) -- (g);
\draw[wfflow] (g) -- (h);
\end{tikzpicture}

\smallskip
{\footnotesize\itshape Redrawn as a diagram; the original document rendered this
as an ASCII sketch, so this figure is not verbatim.}
\end{center}

Importantly, the literature does \textbf{not} demonstrate that AI tools directly
cause illness.

Instead, it demonstrates that prolonged stress, fatigue, poor recovery, and
sleep disruption may contribute to physiological states associated with
increased illness susceptibility
\cite{cohen1991commoncold, cohen1995respiratory, cohen1996immunity}.

\subsubsection*{Key Takeaway}

The current scientific literature does not yet contain extensive studies
specifically examining long-duration agentic coding workflows.

However, converging evidence from:

\begin{itemize}
\item Technostress research \cite{tarafdar2007technostress, tarafdar2019trifecta},
\item Hyperfocus research \cite{dwyer2024monotropism, dupuis2022hyperfocus},
\item Cognitive fatigue research \cite{pessiglione2025fatigue, steward2025neurobiology},
\item Decision fatigue research \cite{choudhury2026decision}, and
\item Stress-immunity research \cite{cohen1991commoncold, cohen1995respiratory, cohen1996immunity}
\end{itemize}

suggests that highly productive AI-assisted development workflows should be
treated as potential occupational health risks when recovery, boundaries, and
sustainable work practices are neglected.

\subsection*{4. Recommendations}

\subsubsection*{Protect Sleep}

Sleep is an engineering dependency.

Treat insufficient sleep as seriously as:

\begin{itemize}
\item Failing tests.
\item Broken builds.
\item Invalid validation results.
\end{itemize}

\subsubsection*{Record Decisions Instead of Continuing Work}

When fatigued:

\begin{itemize}
\item Write notes.
\item Create TODO items.
\item Create Beads tasks.
\end{itemize}

Do not force major architectural decisions.

\subsubsection*{Respect Stopping Points}

Establish explicit end-of-day criteria.

Do not rely on:

\begin{quote}
``Just one more task.''
\end{quote}

\subsubsection*{Preserve Recovery Time}

Protect:

\begin{itemize}
\item Sleep.
\item Meals.
\item Exercise.
\item Weekends.
\item Annual leave.
\item Family time.
\end{itemize}

\subsubsection*{Remember the Human Bottleneck}

\begin{quote}
AI systems do not become tired.

Developers do.
\end{quote}

A highly productive project that burns out its developers has failed one of its
most important engineering constraints.

\end{document}

%% file: results_and_discussion/kloc_accounting/baseline_table.tex
\begin{table}[H]
\centering
\caption{Pre-agentic repositories forming the productivity baseline, measured
2026-06-22. Code lines exclude blank and
comment-only lines; total lines include both, so Rust doc comments, which carry
executable doctests, appear only in the total. An active day is any calendar
date carrying at least one commit; the total is a union of distinct dates
rather than a column sum, because these projects overlapped in time. TUAS is
reported net of the code it inherited from \texttt{thermal\_hydraulics\_rs} at
spin-out --- specifically net of the tree TUAS imported wholesale at its second
commit, \texttt{c451c8e}, rather than net of the predecessor's own full extent,
so that what is removed is what actually came across. At its head TUAS stands at
171\,618 total and 96\,507 code lines, the 96~KLOC
quoted in the text. All figures are measured on each repository's
\texttt{develop} branch, which is where the development history lives ---
these repositories squash-merge into \texttt{main}, so measuring \texttt{main}
would collapse years of work into a handful of commits.
\texttt{thermal\_hydraulics\_rs} is measured at commit \texttt{4d534af}, the
last before its code was moved into TUAS on 2024-10-11 and the repository
emptied. None of these repositories used an agentic coding workflow.
\emph{Every figure in this table is produced by \texttt{kloc\_accounting.py},
which walks the repositories and their git histories directly; see
Section~\ref{sec:reproducing_the_counts}}}
\label{tab:preagentic_baseline}
\footnotesize
\setlength{\tabcolsep}{5pt}
\begin{tabular}{@{}
  >{\raggedright\arraybackslash}p{0.33\linewidth}
  >{\raggedright\arraybackslash}p{0.19\linewidth}
  r r r@{}}
\toprule
\textbf{Repository} & \textbf{Period} & \textbf{Total} & \textbf{Code} &
\textbf{Active} \\
 & & \textbf{lines} & \textbf{lines} & \textbf{days} \\
\midrule
\texttt{thermal\_hydraulics\_rs} (predecessor to TUAS)$^{a}$ \cite{ong2024thermalhydraulicsrs} & Jun 2023 -- Oct 2024 & 111\,437 & 59\,808 & 144 \\
\texttt{chem-eng-\ldots-simulator}$^{a}$ \cite{ong2024chemengprocesscontrol} & Oct 2023 -- Apr 2024 & 8\,848 & 4\,802 & 9 \\
\texttt{teh-o-prke}$^{a}$ \cite{ong2025tehoprke} & Apr 2024 -- Jun 2025 & 10\,751 & 6\,579 & 32 \\
\texttt{tuas\_boussinesq\_solver} (net-new)$^{a}$ \cite{ong2024tuasgithubrepo} & Oct 2024 -- Jun 2026 & 76\,672 & 44\,453 & 69 \\
\texttt{tampines-steam-tables}$^{b}$ \cite{ong2026tampinessteamtables} & Jan 2025 -- Jun 2026 & 79\,990 & 56\,082 & 98 \\
\texttt{boon-lay}$^{c}$ \cite{ong2026boonlay} & Jan 2026 -- May 2026 & 15\,765 & 9\,574 & 30 \\
\midrule
\textbf{Baseline total} & \textbf{2023 -- 2026} & \textbf{303\,463} & \textbf{181\,298} & \textbf{367} \\
\bottomrule
\end{tabular}

\vspace{3pt}
{\footnotesize
\begin{minipage}{\linewidth}
$^{a}$~No AI assistance of any kind.
$^{b}$~Minimal AI assistance, confined to root-finding solvers, via NUS AI-know.
$^{c}$~Non-agentic AI code generation for user interfaces and some Monte Carlo solvers, via NUS AI-know.
\end{minipage}}
\end{table}

%% file: results_and_discussion/kloc_accounting/rate_table.tex
\begin{table}[H]
\centering
\caption{Code lines per active day, each pre-agentic repository ranked fastest
first, against the agentic month. The spread among the pre-agentic
repositories --- roughly a factor of three, from 206 to 572 --- is the useful
context for the agentic figure: it is the ordinary variation between projects
of differing nature, and it is far smaller than the gap to the agentic rate.
The \emph{AI?} column records only non-agentic assistance; none of the
pre-agentic repositories used an agentic workflow.
\emph{Produced by \texttt{kloc\_accounting.py}; see
Section~\ref{sec:reproducing_the_counts}}}
\label{tab:baseline_rates}
\footnotesize
\begin{tabular}{@{}r
  >{\raggedright\arraybackslash}p{0.42\linewidth}
  >{\raggedright\arraybackslash}p{0.26\linewidth}@{}}
\toprule
\textbf{Code lines} & \textbf{Repository} & \textbf{Non-agentic} \\
\textbf{per active day} & & \textbf{AI assistance} \\
\midrule
644 & \texttt{tuas\_boussinesq\_solver} (net-new) & none \\
572 & \texttt{tampines-steam-tables} & root-finding solvers only \\
534 & \texttt{chem-eng-\ldots-simulator} & none \\
415 & \texttt{thermal\_hydraulics\_rs} (predecessor to TUAS) & none \\
319 & \texttt{boon-lay} & code generation \\
206 & \texttt{teh-o-prke} & none \\
\midrule
\textbf{494} & \textbf{Pre-agentic baseline, all repositories} & \textbf{mixed} \\
\textbf{6\,069} & \textbf{\texttt{outram-park-backend}} & \textbf{agentic} \\
\bottomrule
\end{tabular}
\end{table}

%% file: results_and_discussion/kloc_accounting/agentic_table.tex
\begin{table}[H]
\centering
\caption{Agentic lines of code in \texttt{outram-park-backend} by crate,
2026-06-19 -- 2026-07-23. Code lines exclude blank
and comment-only lines. Translated crates are pure-Rust forks or ports of the
named upstream project and are independent works, not the official software.
Extensions are the excess of a vendored crate over its standalone pre-agentic
original, and are credited here rather than to the baseline.
\emph{Every figure in this table is produced by \texttt{kloc\_accounting.py}
from the repository pinned at commit \texttt{3130b38}, because
\texttt{develop} moves daily and a tip measurement could not be reproduced
later; the classification of each
crate is the one editorial input, and is carried in the script beside each
crate's own \texttt{Cargo.toml} description so it can be audited. See
Section~\ref{sec:reproducing_the_counts}.}}
\label{tab:agentic_crates}
\footnotesize
\setlength{\tabcolsep}{5pt}
\begin{tabular}{@{}
  >{\raggedright\arraybackslash}p{0.42\linewidth}
  >{\raggedright\arraybackslash}p{0.22\linewidth}
  r@{}}
\toprule
\textbf{Crate} & \textbf{Upstream} & \textbf{Code lines} \\
\midrule
\multicolumn{3}{@{}l}{\textit{Translated or ported from an existing codebase}} \\
\texttt{njoy-outram-park-fork} & NJOY2016 & 32\,481 \\
\texttt{outram-park-fork-coolprop} & CoolProp & 28\,707 \\
\texttt{outram-foam-appbuilder-lib} & OpenFOAM & 20\,495 \\
\texttt{outram-foam-basic-lib} & OpenFOAM & 15\,691 \\
\texttt{outram-mc-libs} & OpenMC & 15\,406 \\
\texttt{outram-park-fork-pflotran} & PFLOTRAN & 7\,939 \\
\texttt{outram-foam-mesh} & OpenFOAM & 6\,218 \\
\texttt{outram-blender} & Blender & 4\,769 \\
\texttt{outram-park-fork-dwsim-libs} & DWSIM & 2\,079 \\
\texttt{outram-foam-turbulence-lib} & OpenFOAM & 1\,717 \\
\texttt{outram-foam-cli} & OpenFOAM & 960 \\
\cmidrule(r){1-3}
\textbf{Subtotal, translated} & & \textbf{136\,462} \\
\midrule
\multicolumn{3}{@{}l}{\textit{Originally written}} \\
\texttt{outram-park-digital-twin-engine} & --- & 8\,651 \\
\texttt{tampines} & --- & 7\,306 \\
\texttt{kovan-tui} & --- & 1\,716 \\
\texttt{kovan-cli} & --- & 1\,354 \\
\texttt{njoy-tui (bin in njoy-outram-park-fork)} & --- & 1\,304 \\
\texttt{kovan-codegen} & --- & 1\,267 \\
\texttt{kovan-semantics} & --- & 1\,222 \\
\texttt{outram-mc-tui (bin in outram-mc-libs)} & --- & 1\,202 \\
\texttt{kovan-literature} & --- & 1\,114 \\
\texttt{kovan-discovery} & --- & 934 \\
\texttt{kovan-common} & --- & 617 \\
\texttt{nee\_soon}$^{\dagger}$ & --- & 490 \\
\cmidrule(r){1-3}
\textbf{Subtotal, original} & & \textbf{27\,177} \\
\midrule
\multicolumn{3}{@{}l}{\textit{Agentic extensions to vendored pre-agentic crates}} \\
\texttt{tampines-steam-tables} & (own, pre-agentic) & 7\,150 \\
\texttt{boon-lay} & (own, pre-agentic) & 3\,680 \\
\texttt{tuas\_boussinesq\_solver} & (own, pre-agentic) & 836 \\
\texttt{teh-o-prke} & (own, pre-agentic) & 692 \\
\texttt{chem-eng-\ldots-simulator} & (own, pre-agentic) & 0 \\
\cmidrule(r){1-3}
\textbf{Subtotal, extensions} & & \textbf{12\,358} \\
\midrule
\textbf{Total agentic} & & \textbf{175\,997} \\
\bottomrule
\end{tabular}

\vspace{3pt}
{\footnotesize
\begin{minipage}{\linewidth}
$^{\dagger}$~\texttt{nee\_soon} succeeds \texttt{teh-o-prke} in naming --- the
former Teh-O package --- but its code is newly written; the point kinetics
module itself remains vendored as \texttt{teh-o-prke}.
\end{minipage}}
\end{table}

%% file: results_and_discussion/discussion_markdowns/token_cost_table.tex
\begin{table}[H]
\centering
\caption{Token consumption and Anthropic API list-price cost of the agentic
coding work on this project, measured on one machine, as of 2026-07-23. The
work was carried out under a flat-rate subscription, so this is a
\emph{counterfactual} --- what a reader reproducing it through the metered API
would be charged --- and not an amount that was paid. Counts are taken from the
Claude Code session transcripts, which record the token usage the API itself
reported for each call, deduplicated by message id; no figure here is estimated.
List prices were retrieved 2026-07-17. The total is a
\textbf{lower bound} in two independent ways. Transcripts are stored per machine
with no synchronisation, so any machine not scanned is missing; and Claude Code
prunes them on a rolling retention window (\texttt{cleanupPeriodDays}, thirty
days by default), so a later rerun returns a \emph{smaller} total as old
sessions expire. The transcripts counted span 2026-06-23 to 2026-07-23. Every cache write on record
carries the 1-hour time-to-live and is therefore billed at $2.00\times$ the
input rate; none used the cheaper $1.25\times$ five-minute tier.
\emph{Produced by \texttt{token\_accounting.py}, which reads the transcripts
directly and emits this table; re-run it to update the numbers.}}
\label{tab:token_cost}
\footnotesize
\setlength{\tabcolsep}{5pt}
\begin{tabular}{@{}
  >{\raggedright\arraybackslash}p{0.34\linewidth}
  r r r@{}}
\toprule
\textbf{Model} & \textbf{Billed} & \textbf{Billable} & \textbf{List-price} \\
 & \textbf{API calls} & \textbf{tokens} & \textbf{cost (USD)} \\
\midrule
\texttt{claude-opus-4-8} & 3\,778 & 1\,396\,968\,434 & \$1\,026.49 \\
\texttt{claude-sonnet-5} & 3\,012 & 1\,430\,185\,326 & \$512.10 \\
\texttt{claude-sonnet-4-6} & 813 & 84\,970\,645 & \$56.84 \\
\texttt{claude-haiku-4-5-20251001} & 54 & 2\,810\,324 & \$0.62 \\
\cmidrule(r){1-4}
\textbf{Total} & \textbf{7\,657} & \textbf{2\,914\,934\,729} & \textbf{\$1\,596.05} \\
\midrule
\multicolumn{3}{@{}l}{Same work, Sonnet~5 introductory pricing
(\$2/\$10 per Mtok)} & \$1\,425.35 \\
\multicolumn{3}{@{}l}{Same work, counterfactual with no prompt caching} & \$11\,670.78 \\
\bottomrule
\end{tabular}
\end{table}

%% file: results_and_discussion/discussion_markdowns/token_cost_components_table.tex
\begin{table}[H]
\centering
\caption{The same spend of Table~\ref{tab:token_cost} split by what was
actually billed, measured on one machine, as of 2026-07-23. Token volume and
cost share disagree sharply, which is the characteristic shape of a long agentic
session: the context is re-read on nearly every turn, so cache reads dominate
the token count, and only the $0.10\times$ read multiplier keeps them from
dominating the bill by the same margin. Generated output, a rounding error by
volume, is a substantial fraction of the cost. Note that the five-minute cache
tier is unused --- every write on record carries the 1-hour time-to-live and is
billed at $2.00\times$ the input rate rather than $1.25\times$. Prices are
Anthropic API list, retrieved 2026-07-17; the work itself was
done under a flat-rate subscription, so these are counterfactual charges rather
than money paid, and they are a lower bound both across machines and in time ---
Claude Code prunes transcripts on a thirty-day rolling window. The transcripts counted span 2026-06-23 to 2026-07-23.
\emph{Produced by \texttt{token\_accounting.py}; see
Table~\ref{tab:token_cost} for the per-model totals.}}
\label{tab:token_cost_components}
\footnotesize
\setlength{\tabcolsep}{5pt}
\begin{tabular}{@{}
  >{\raggedright\arraybackslash}p{0.26\linewidth}
  >{\raggedright\arraybackslash}p{0.15\linewidth}
  r r r r@{}}
\toprule
\textbf{Billed as} & \textbf{Priced at} & \textbf{Tokens} & \textbf{\% of} &
\textbf{Cost} & \textbf{\% of} \\
 & & & \textbf{tokens} & \textbf{(USD)} & \textbf{cost} \\
\midrule
Cache read & $0.10\times$ input & 2\,872\,615\,552 & 98.5\% & \$1\,135.21 & 71.1\% \\
Cache write, 1\,h TTL & $2.00\times$ input & 33\,044\,451 & 1.1\% & \$284.26 & 17.8\% \\
Cache write, 5\,min TTL & $1.25\times$ input & 0 & 0.0\% & \$0.00 & 0.0\% \\
Fresh input & $1.00\times$ input & 1\,039\,602 & 0.0\% & \$4.51 & 0.3\% \\
Output & output rate & 8\,235\,124 & 0.3\% & \$172.08 & 10.8\% \\
\cmidrule(r){1-6}
\textbf{Total} & & \textbf{2\,914\,934\,729} &
\textbf{100.0\%} & \textbf{\$1\,596.05} &
\textbf{100.0\%} \\
\bottomrule
\end{tabular}
\end{table}